\documentclass[iop]{emulateapj}
\defcitealias{kempner2004}{KD04}

\usepackage{epsfig}
\usepackage{subfigure}
\usepackage{graphics}

\newcommand{\chan}{{\it Chandra}}
\newcommand{\ciao}{{\it CIAO}}

\newcommand{\rf}{${r}_{\rm{F}}$}

\newcommand{\kms}{$\,\rm{km\,s^{-1}}$}

\shorttitle{}
\shortauthors{Owers et al.}
\begin{document}
\title{The Dissection of Abell~2744: a rich cluster growing through major and minor mergers.}
\author{Matt S. Owers\altaffilmark{1}, Scott W. Randall\altaffilmark{2}, Paul E.J. Nulsen\altaffilmark{2}, Warrick J. Couch\altaffilmark{1}, Laurence P. David\altaffilmark{2}, Joshua C. Kempner}
\altaffiltext{1}{Center for Astrophysics and Supercomputing, Swinburne University of Technology, Hawthorn, VIC 3122, Australia; mowers@astro.swin.edu.au}
\altaffiltext{2}{Harvard Smithsonian Center for Astrophysics, 60 Garden Street, Cambridge, MA 02138, USA}

\begin{abstract}
New \chan\ X-ray data and extensive optical spectroscopy,
obtained with AAOmega on the 3.9 m Anglo-Australian Telescope, are
used to study the complex merger taking place in the galaxy cluster
Abell~2744.  Combining our spectra with data from the literature
provides a catalog of 1237 redshifts for extragalactic objects lying
within 15\arcmin\ of the cluster center.  From these, we confirm 343 cluster
members projected within 3 Mpc of the cluster center.  Combining
positions and velocities, we identify two major substructures,
corresponding to the remnants of two major subclusters.  The new data
are consistent with a post core passage, major merger taking place
along an axis that is tilted well out of the plane of the sky,
together with an interloping minor merger.  Supporting this
interpretation, the new X-ray data reveal enriched, low entropy gas
from the core of the approaching, major subcluster, lying $\sim 2$\arcmin\ north
of the cluster center, and a shock front to the southeast of the
previously known bright, compact core associated with the receding
subcluster.  The X-ray morphology of the compact core is consistent
with a Bullet-like cluster viewed from within $\sim 45^\circ$ of the
merger axis.  An X-ray peak $\sim 3$\arcmin\ northwest of the cluster center, 
with an associated cold front to the northeast and a trail of low entropy
gas to the south, is interpreted as the remnant of an interloping
minor merger taking place roughly in the plane of the sky.  We infer
approximate paths for the three merging components.

\end{abstract}


\section{Introduction}

In a universe where structure grows hierarchically, clusters of galaxies are the
latest structures to collapse and virialize \citep[e.g.,][]{springel2006}. There
are three main modes of mass accretion onto rich clusters of galaxies: 
the steady infall of matter from the surrounding filamentary large scale 
structures, the discrete accretion of group-sized objects, and the extreme event
of a major cluster-cluster merger. The latter are the most energetic events 
known in the Universe \citep{markevitch1998} and result in the violent 
reassembly of the cluster. Such a dramatic reconfiguration of the cluster 
results in a rapid change in the environment of its member galaxies, although 
what effect this has the galaxies themselves is yet to be fully understood. This
is in part due to the complex nature of cluster mergers and, hence, the 
difficulty in obtaining a detailed picture of their properties.

The well known Butcher-Oemler effect 
\citep[hereafter BO-effect;][]{butcher1978,butcher1984}, that the fraction 
of blue galaxies in the cores of rich clusters is significantly lower at the 
present epoch than it was $\sim$2.5\,Gyrs or more ago, reveals 
that a significant fraction of cluster galaxies have undergone rapid 
transformation in star-forming properties over the intervening period. 
Furthermore, the observed increase with redshift in the fraction of spiral 
galaxies in clusters at the expense of a commensurate decline in the S0
fraction \citep{dressler1997, fasano2000, desai2007,just2010} reveals a 
corresponding rapid evolution in galaxy morphology. While the BO effect appears 
to be widespread, the scatter in the blue galaxy fraction is large at all 
redshifts ($z>0.2$) and exceeds the uncertainties in the measurements 
\citep{butcher1984}. This indicates some internal cluster-specific mechanism is 
responsible for the scatter and \citet{margoniner2001} found that a large 
portion of this scatter can be attributed to the richness of a cluster, where 
less rich clusters have a higher blue fraction. Another intriguing 
possibility is that the scatter is driven by the hierarchical formation of 
clusters, in particular cluster mergers, which become more common with 
increasing redshift. Here it is quite conceivable that they would lead to an 
increased blue galaxy fraction through triggering star formation in the member 
galaxies \citep{kauffmann1995, metevier2000, miller2006}. 

In this context, spectroscopic observations of the Coma cluster have revealed a 
tantalizing correlation between the spatial distribution of post-starburst 
galaxies and regions involved in merger activity 
\citep{caldwell1993,caldwell1997,poggianti2004}. These observations strongly 
suggest that the merging process can affect the star-forming properties of the 
cluster galaxies. This spurred \citet{caldwell1997} to conduct a further study 
to search for ``abnormal'' spectrum galaxies, similar to those BO galaxies
found at higher redshift, in four additional nearby clusters, three of which 
were selected on the basis of harboring clear substructure. Although their
study was limited to galaxies of early-type morphology, the results indicated 
that there is a significant fraction of abnormal-type galaxies in nearby rich 
clusters and, most intriguingly, the triggering of starbursts leading to 
abnormal spectra occurs during the core passage phase of a merger. Recently, 
\citet{hwang2009} have provided further evidence for this scenario by comparing 
the galaxy properties in the post-core passage merger Abell~168 and the pre-core
passage merger Abell~1750. Furthermore, radio observations have revealed that 
clusters which harbor evidence for major merger activity show an increase in 
radio activity amongst their galaxies 
\citep{miller2003,miller2006,venturi2001,venturi2000,johnstonhollitt2008}. 
However, there are counter-examples \citep{venturi2002}, suggesting that the 
merger phase is an important factor. According to the simulations of 
\citet{bekki2010}, this evolution may be driven by the significant increase in
ICM pressure that a galaxy is exposed to during a cluster merger, in particular
when the galaxy passes through regions affected by shocks during the core 
passage stage of the merger \citep[see also][]{roettiger1996}.
{\it Therefore, an understanding of the merger dynamics and merging history of a
cluster is critical when attempting to disentangle the effects leading to 
the triggering and halting of star formation in cluster galaxies.}

The rich X-ray luminous cluster Abell~2744 at $z=0.3$ 
\citep[also known as AC118;][]{couch1984} is an interesting test case, being 
both a well known merging cluster and one that exhibits a significant BO effect.
Two previous studies have addressed the 
dynamical state of the merger occurring in the core of Abell~2744. The 
first was the $Chandra$-based study of \citet[][hereafter KD04]{kempner2004}, 
who showed that Abell~2744 is undergoing a major merger in approximately the 
north-south direction. The second was conducted by \citet{boschin2006} who 
used 85 cluster member spectra to show that the merger has a significant 
velocity component along the line of sight, as suggested by 
\citetalias{kempner2004} \citep[see also; ][]{girardi2001}, and estimated the 
mass ratio of the merging subclusters to be $\sim 3:1$.
This merger also quite naturally explains why Abell~2744 hosts one of the most 
luminous known radio haloes covering its central 1.8\,Mpc, as well as a large 
radio relic at a distance of about 2\,Mpc from the cluster center 
\citep{giovannini1999,govoni2001a,govoni2001b}. The presence of a 
significant blue galaxy excess---a blue fraction that was $2.2\pm0.3$ 
times that seen in the same core regions of nearby clusters---was first 
measured photometrically and confirmed spectroscopically by \citet{couch1987}, 
who found the cluster to have a blue galaxy fraction of $\sim 25$\%. Moreover, 
the spectroscopic study showed that the blue galaxy population in Abell~2744 was
dominated by starburst and post-starburst galaxies. Abell~2744 therefore 
provides an excellent laboratory for studying the link between major merger 
activity and star formation activity in cluster galaxies. This paper reports 
the first step of this study, which is to determine more precisely the 
{\it phase} and {\it history} of Abell~2744's merger, in particular whether it 
is in a pre-core or post-core passage phase. 

As has been shown previously 
\citep[e.g.,][]{zabludoff1995,barrena2007,maurogordato2008,owers2009a,ma2009,ma2010}, 
the combination of high quality X-ray imaging and spectroscopy with
comprehensive optical spectroscopy provides a powerful toolkit for 
disentangling the complex dynamics of cluster mergers.
In this paper, we combine new \chan\ observations, which provide an additional 
100ks of data to the existing 25ks analyzed in \citetalias{kempner2004}, with 
new AAT/AAOmega optical multi-object spectroscopy (MOS), which roughly 
doubles the number of spectroscopically confirmed cluster members over a 3\,Mpc 
radius region, and also doubles the number of cluster member spectra within the 
central regions studied by \citet{boschin2006}. These data sets are used both to
detect substructure in Abell~2744, which is related to the merger activity, and 
to provide a more up-to-date interpretation of the merging history.

The outline of this paper is as follows. In Section~\ref{chandra_data} we 
present the reduction and analysis of the new \chan\ data. In 
Section~\ref{optical_data} present details of the MOS observations and data 
reduction, the redshift catalog, and the cluster membership allocation 
procedure. In Section~\ref{substructure_detection} we present the methods used
for substructure detection. In Section~\ref{SS_nat} we present our 
interpretations of the nature of the structures detected in the optical and 
X-ray data. In Section~\ref{merger_scen} we present our merger scenario based
on the observations and interpretations presented in the paper. Finally, in
Section~\ref{summary} we summarize our results.

Throughout the paper, we assume a standard $\Lambda$CDM cosmology where 
$H_0 =70$\kms, $\Omega_m=0.3$ and $\Omega_{\Lambda}=0.7$. For the assumed 
cosmology and at the cluster redshift (z=0.3064), 1\arcsec=4.52\,kpc.

\section{Chandra data}\label{chandra_data}

\subsection{Observations and Data Reduction}

Abell~2744 has been observed with \chan\ using both the ACIS-I and ACIS-S 
chip arrays and we summarize the dates, exposure times and \chan\ ObsID's in 
Table~\ref{chandra_obs}. The data were reprocessed starting with the level 1 
event files and using \ciao\ version 3.4 with the latest gain and
calibration files applied (CalDB version 3.5.5). 
All observations were performed in Very Faint (VF) data mode, allowing 
$5 \times 5$ pixel islands to be used in identifying cosmic ray events, 
significantly helping to reduce the particle
background. Observation specific bad pixel files were produced and applied
and events flagged with {\it ASCA} grades 1, 5 and 7 were excluded. The 
data were filtered for periods of anomalously high background associated 
with flares. Standard procedures are followed in doing this, i.e., for the
ACIS-I observations, point sources and the region containing the diffuse 
cluster emission were excluded, and a light curve was extracted in the
$0.3-10$\,keV range binned in 259s intervals. The {\it Sherpa} 
{\sf lc\_clean.sl} routine was then used to identify for removal time
intervals when the count rate deviated by more than $20 \%$ from the
mean. No significant flares 
were detected, and the cleaned exposure times are listed in 
Table~\ref{chandra_obs}. 

\begin{deluxetable*}{ccccc}
\tabletypesize{\scriptsize}
\tablecolumns{3}
\tablewidth{0pc}
\tablecaption{Summary of the \chan\ observations.\label{chandra_obs}}
\tablehead {\colhead{Array}& \colhead{ObsId} & \colhead{Date} &\colhead{Exposure Time (ks)} & \colhead{Cleaned Exposure Time (ks)}} 
\startdata
             ACIS-S        &     2212        & 2001 September 3   &   25.14            &  23.79\\
             ACIS-I        &     7712        & 2007 September 10  &   8.17             &  8.07\\
             ACIS-I        &     7915        & 2006 November 8    &   18.86            &  18.62\\
             ACIS-I        &     8477        & 2007 June 10       &   46.5             &  45.91\\
             ACIS-I        &     8557        & 2007 June 14       &   28.17            &  27.75\\
\enddata
\end{deluxetable*}

For background subtraction during spectral and imaging analyses, we use 
the blank sky 
observations\footnote{See http://cxc.harvard.edu/contrib/maxim/acisbg}
appropriate for the chip and epoch of the observation of interest. The
background files were processed in the same manner as the observations,
using the same background filtering, bad pixel and gain files, and were 
reprojected to match the observations. The backgrounds were normalized so that 
the source and background counts in the $10-12$\,keV range matched. We checked 
for excess soft Galactic X-ray emission in the observations by extracting 
spectra from source-free regions and comparing to spectra extracted from the 
background files, finding no significant difference. 

\subsection{Image Analysis}\label{image}

\begin{figure*}
\vspace{10pt}
{\includegraphics[width=0.45\textwidth]{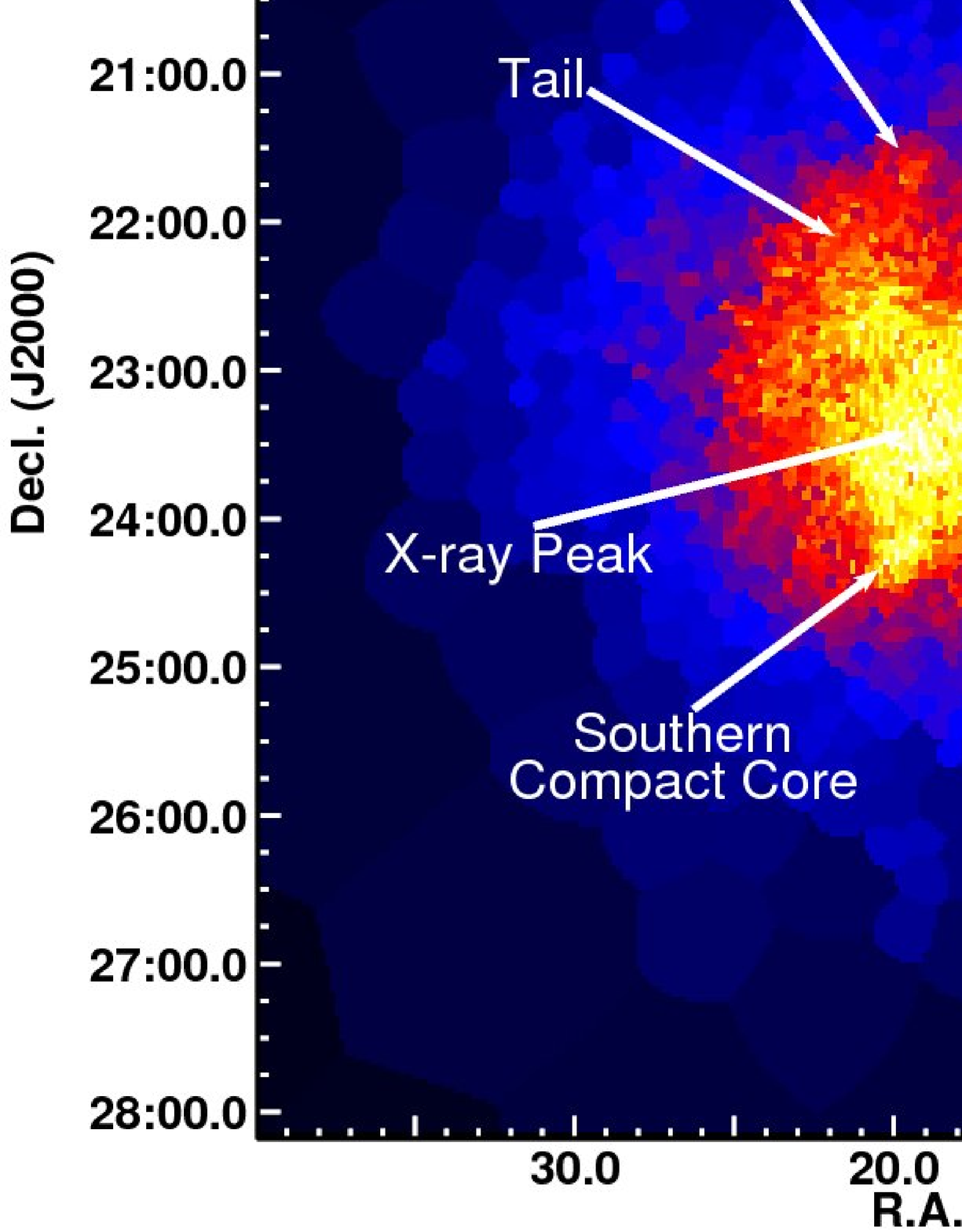}}
{\includegraphics[width=0.45\textwidth]{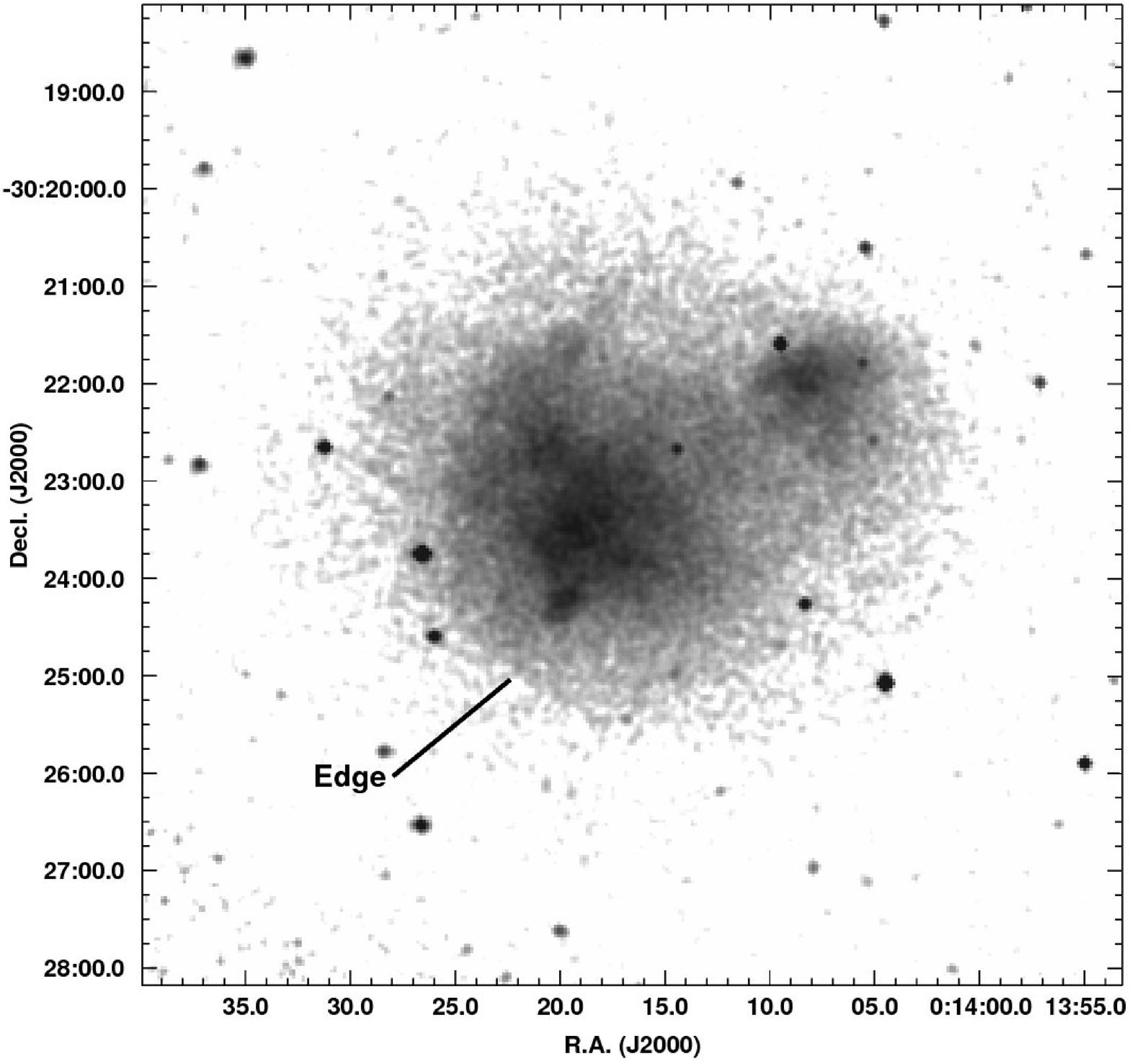}}
\caption{Background-subtracted, exposure-corrected \chan\ images. The image
in the {\it left panel} is binned to S/N=5 using the WVT algorithm and 
the point sources have been removed and filled using {\sf dmfilth}.
The image in the {\it right panel} shows the image with point sources, binned 
to 1\arcsec.968 pixels and smoothed with a Gaussian kernel with 
FWHM=$4\arcsec$.}
\label{wvtimage}
\end{figure*}

The left panel of Figure~\ref{wvtimage} shows a background-subtracted, 
$0.5-7$\,keV, exposure-corrected image of Abell~2744 which has been binned using
 the WVT binning algorithm of \citet{Diehl2006}, which is a generalization of  
\citet{Cappellari2003}'s Voronoi binning algorithm. The bin size is determined 
by the constraint that the signal-to-noise ratio must be $\sim 5$. Prior to 
binning, point sources detected with {\sf wavdetect} were removed and the 
regions were filled via interpolation from a surrounding background region 
using {\sf dmfilth}. The exposure map was generated using standard \ciao\ 
procedures\footnote{see http://cxc.harvard.edu/ciao3.4/threads/expmap\_acis\_multi/} and accounts for the effects of vignetting, quantum efficiency (QE), QE
nonuniformity, bad pixels, dithering, and effective area. The energy dependence
of the effective area is accounted for by computing a weighted instrument map
where the weights are computed from an absorbed MEKAL spectral model with 
Galactic absorption, temperature, redshift and abundance set to the average
cluster values obtained in Section~\ref{tmaps}. 
The right panel of Figure~\ref{wvtimage} shows the 
background subtracted, exposure corrected image of Abell~2744 with point sources
included, and with a light Gaussian smooth (FWHM=4\arcsec) applied.  
Abell~2744 is clearly a disturbed system, with a structure 
$\sim 750$\,kpc to the northwest of the X-ray surface brightness peak (labelled northwestern interloper). 
Within the main X-ray structure, there are a further three substructures: 
one $\sim 190$\,kpc south (labelled southern compact core), one $\sim 480$\,kpc to the 
north (labelled northern core) and another $\sim 220$\,kpc to the north of 
the X-ray surface brightness peak. The northern 
core furthest from the X-ray peak has a tail, which includes the second 
structure north of the X-ray peak, extending southward and curving westward 
towards the peak in the X-ray emission. The southern compact core has a 
blunted cone morphology with a prominent head and a fan-like tail of 
emission extending back towards the peak in the X-ray emission. There is also an
edge in the surface brightness $\sim 250$\, kpc to the southeast of the
southern substructure.

The northwestern interloper appears asymmetric, with an edge just east of 
north and an extended tail toward the south. There also
appears to be a bridge of emission connecting it to the main cluster emission.
Based on a shorter \chan\ observation, \citetalias{kempner2004} tentatively 
identified a cold front and bow shock on the eastern side of the northwestern 
interloper and also a fan-like feature extending towards the west. They
concluded the northwestern interloper was moving from west to east on its way towards the 
cluster core. However, the deeper observations presented in 
Figure~\ref{wvtimage} do not confirm the existence of a bow shock and indicate
the cold front on the eastern side is part of a larger northern edge, meaning 
the interpretation of \citetalias{kempner2004} may not be the correct one. 

\begin{figure*}
\vspace{10pt}
{\includegraphics[angle=-0,width=0.48\textwidth]{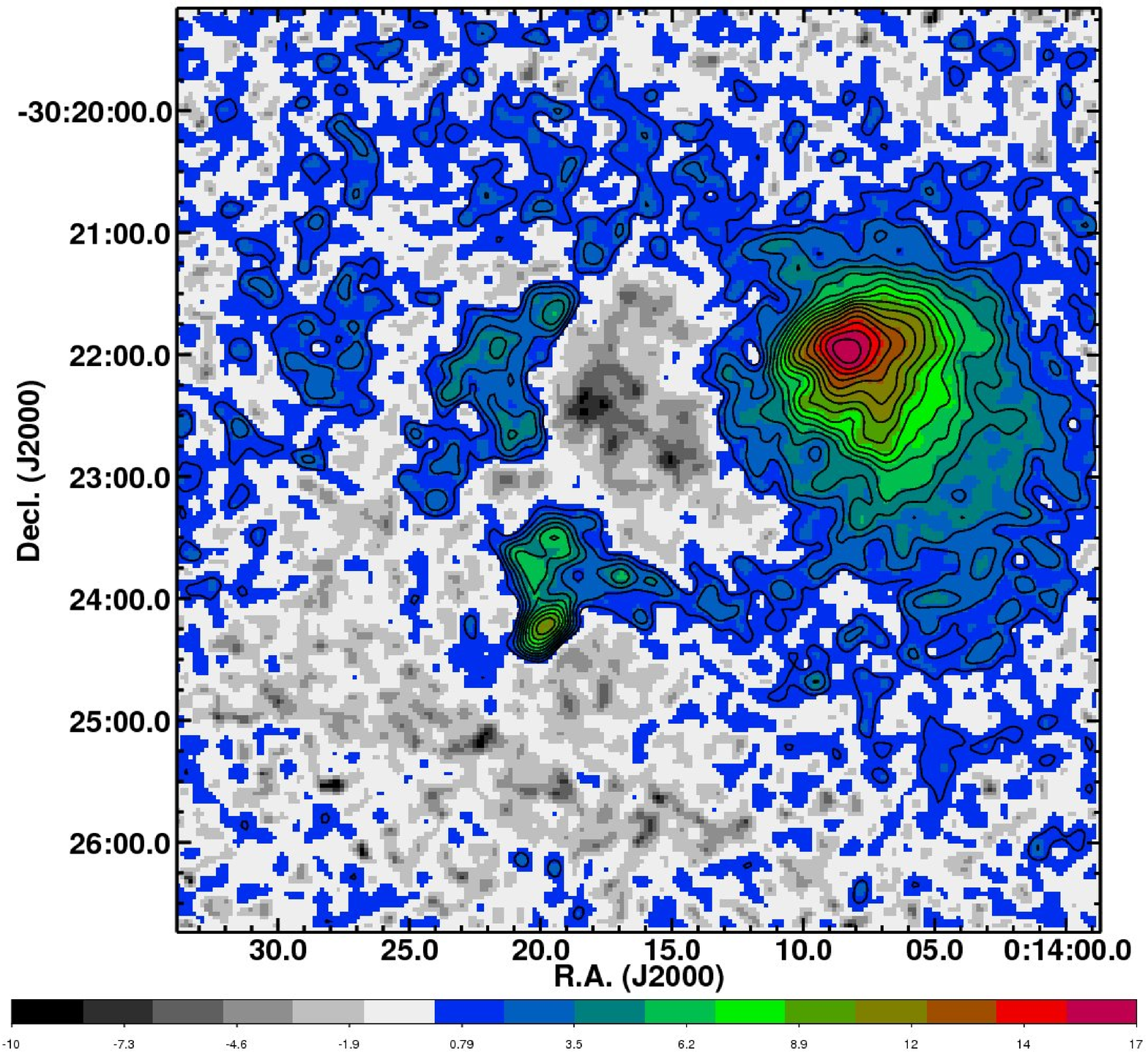}}
{\includegraphics[angle=-0,width=0.48\textwidth]{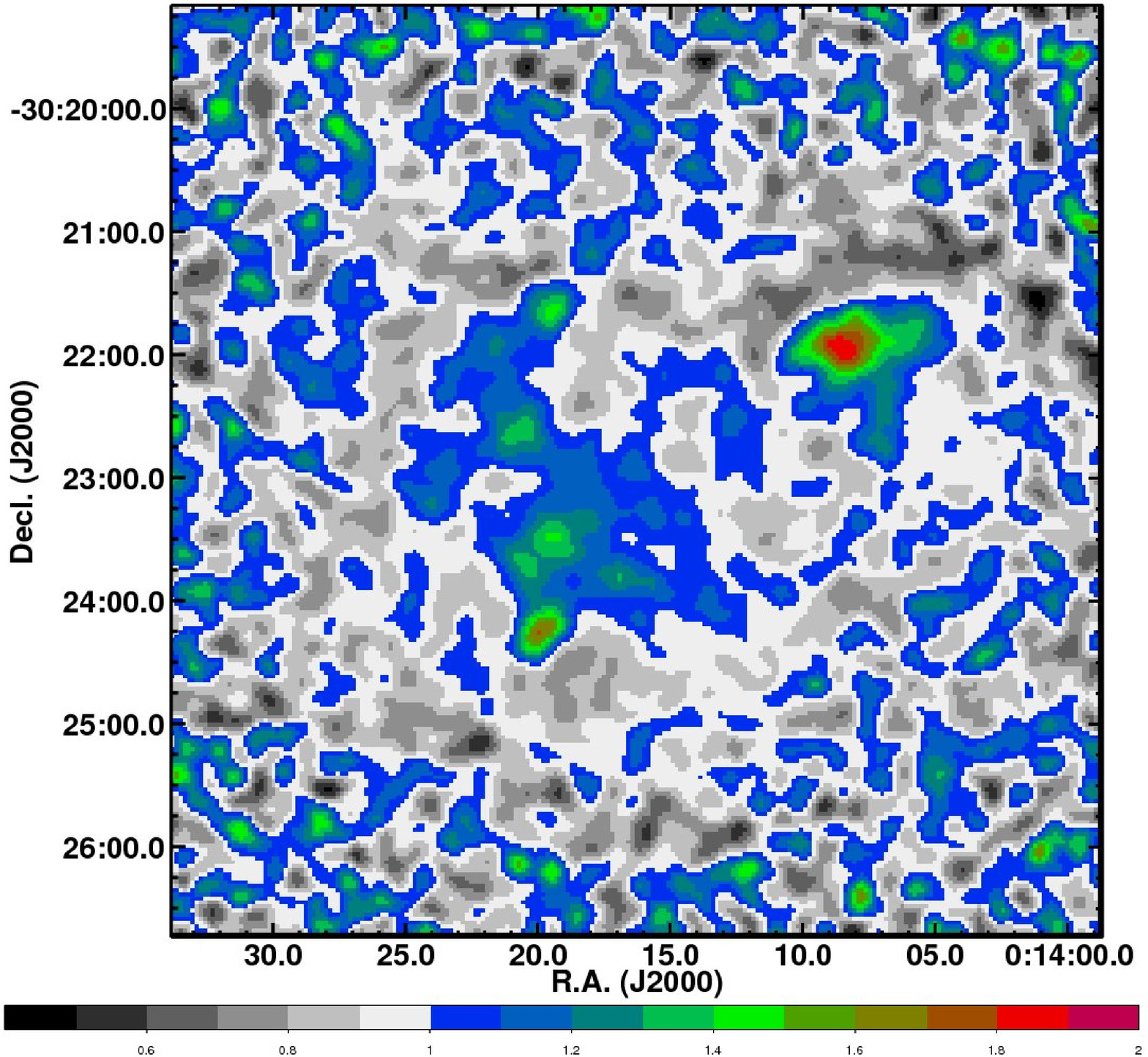}}
\caption{{\it Left panel:} Residual significance image produced using a similar 
method as in \citet{owers2009a}, i.e., an azimuthally symmetric $\beta$-model 
is fitted to the data (excluding the northwestern interloper and point sources) and 
then subtracted to produce a residual image. The residual image is then smoothed
and divided by an error map to produce the residual significance map. Contours 
are spaced linearly in steps of $1 \sigma$ in the range $1-17 \sigma$. 
{\it Right panel:}Unsharp masked image created by dividing an exposure and 
background corrected image smoothed using a Gaussian kernel with 
$\sigma=4.4\arcsec$ (20 kpc) by one smoothed by $\sigma=22\arcsec$ (100 kpc).}
\label{residimage}
\end{figure*}

To emphasize the structures revealed in Figure~\ref{wvtimage} we present the
residual significance map, which highlights low surface brightness structures,
and the unsharp masked image, which is useful for revealing sharp structures 
such shocks and cold fronts, in Figure~\ref{residimage}. To produce the 
significance map, we first fit the \chan\ image with an azimuthally symmetric 
$\beta$-model with best fitting values of $r_c=471\pm8$\,kpc, 
$\beta=0.96\pm0.13$, R.A.=$00^{\rm h}14^{\rm m}18.4^{\rm s}$ and 
decl.=$-30^{\circ}23\arcmin 14\arcsec.0$. During the fitting of the 
$\beta$-model, the northwest substructure was excluded, along with point 
sources. This $\beta$-model accounts for the main cluster emission and is 
subtracted from \chan\ image to produce a residual image which is then smoothed 
with a Gaussian kernel with $\sigma =3\arcsec$. The resulting smoothed 
residual map is then divided by an error map to produce the residual 
significance map \citep[see][for more details]{neumann1997,owers2009a}. The 
left panel in Figure~\ref{residimage} shows that the northwest interloper and 
its extended tail to the south are 
detected with high significance, while the northern core and southern compact core, 
along with the tail near the northern core, are also detected. 
The unsharp masked image is produced by dividing a background subtracted,
exposure corrected image that was smoothed with a 4.4\arcsec\ Gaussian
by the same image smoothed by 22\arcsec, corresponding
to physical scales of 20 kpc and 100 kpc at the cluster redshift. The 
unsharp masked image is shown in the right panel of Figure~\ref{residimage}, 
where it can be seen that the structure in the central cluster region is very 
similar to that seen in the residual significance map. However, the unsharp 
masked image accentuates the edge to the north of the northwestern interloper
and also a finger of emission extending from the northwestern  interloper
towards the south.

\subsection{Thermodynamic maps}\label{thermomaps}
\subsubsection{Temperature maps}\label{tmaps}
\begin{figure*}
\vspace{10pt}
{\includegraphics[width=0.42\textwidth]{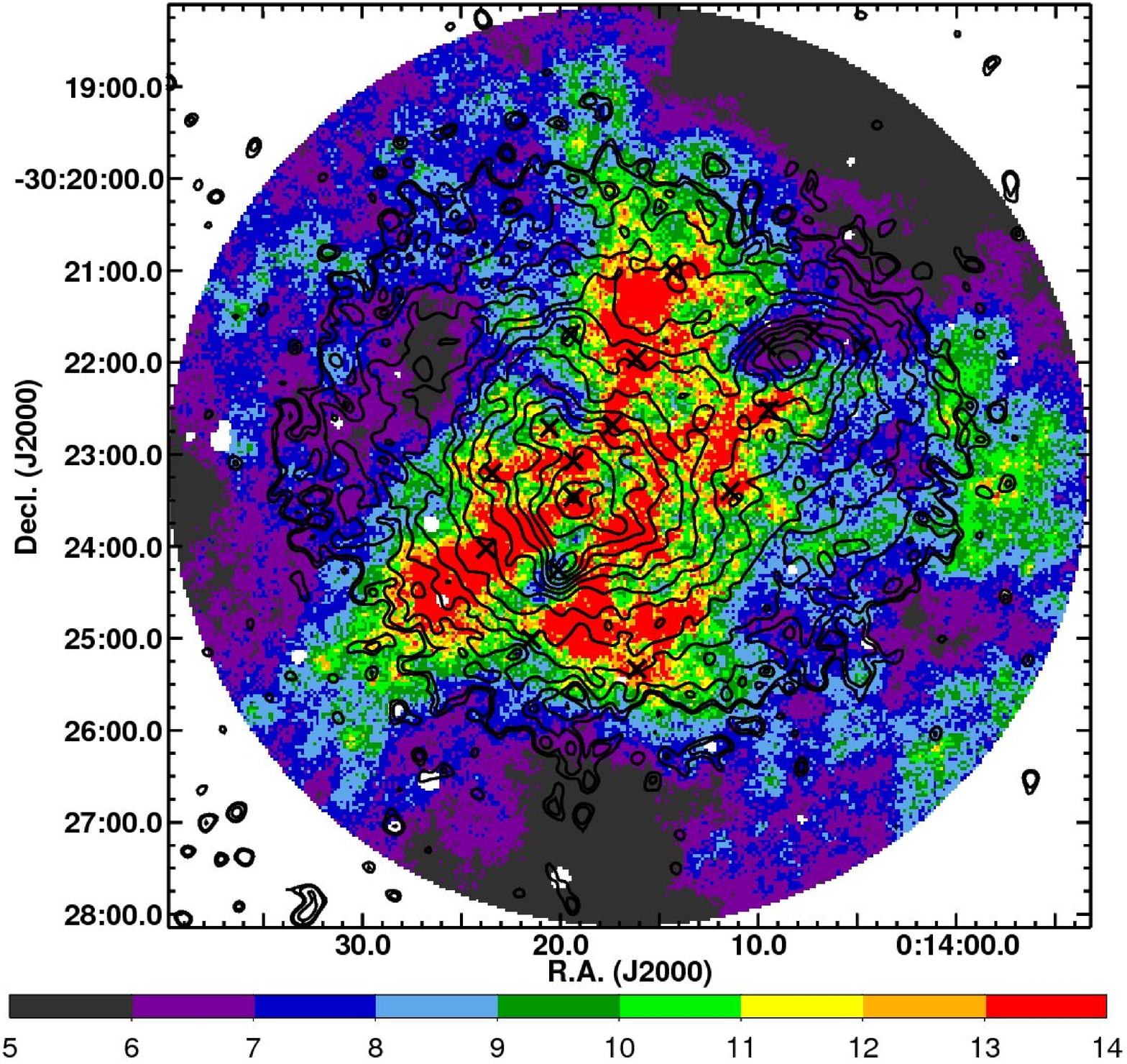}}
{\includegraphics[width=0.42\textwidth]{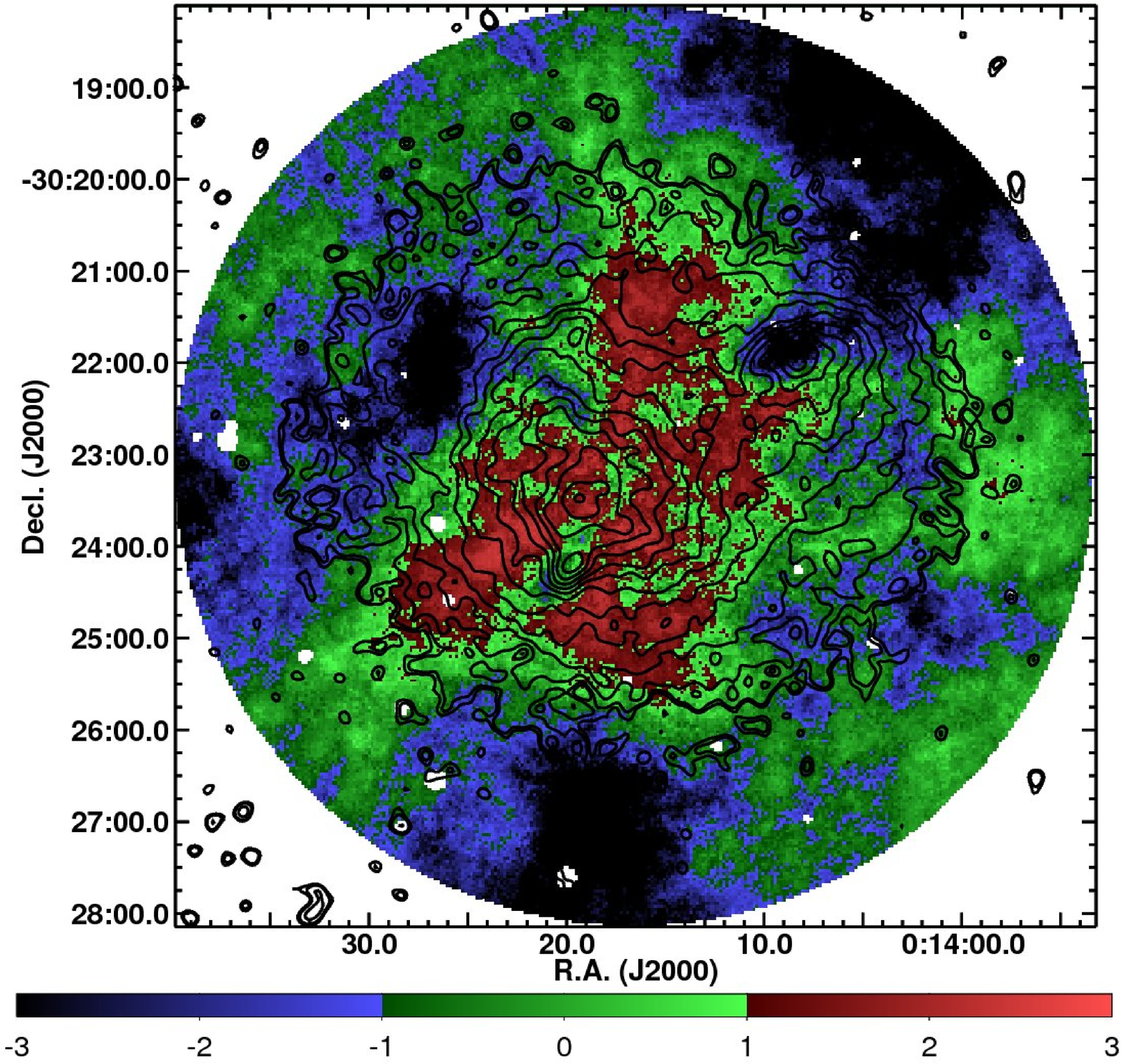}}\\
{\includegraphics[width=0.42\textwidth]{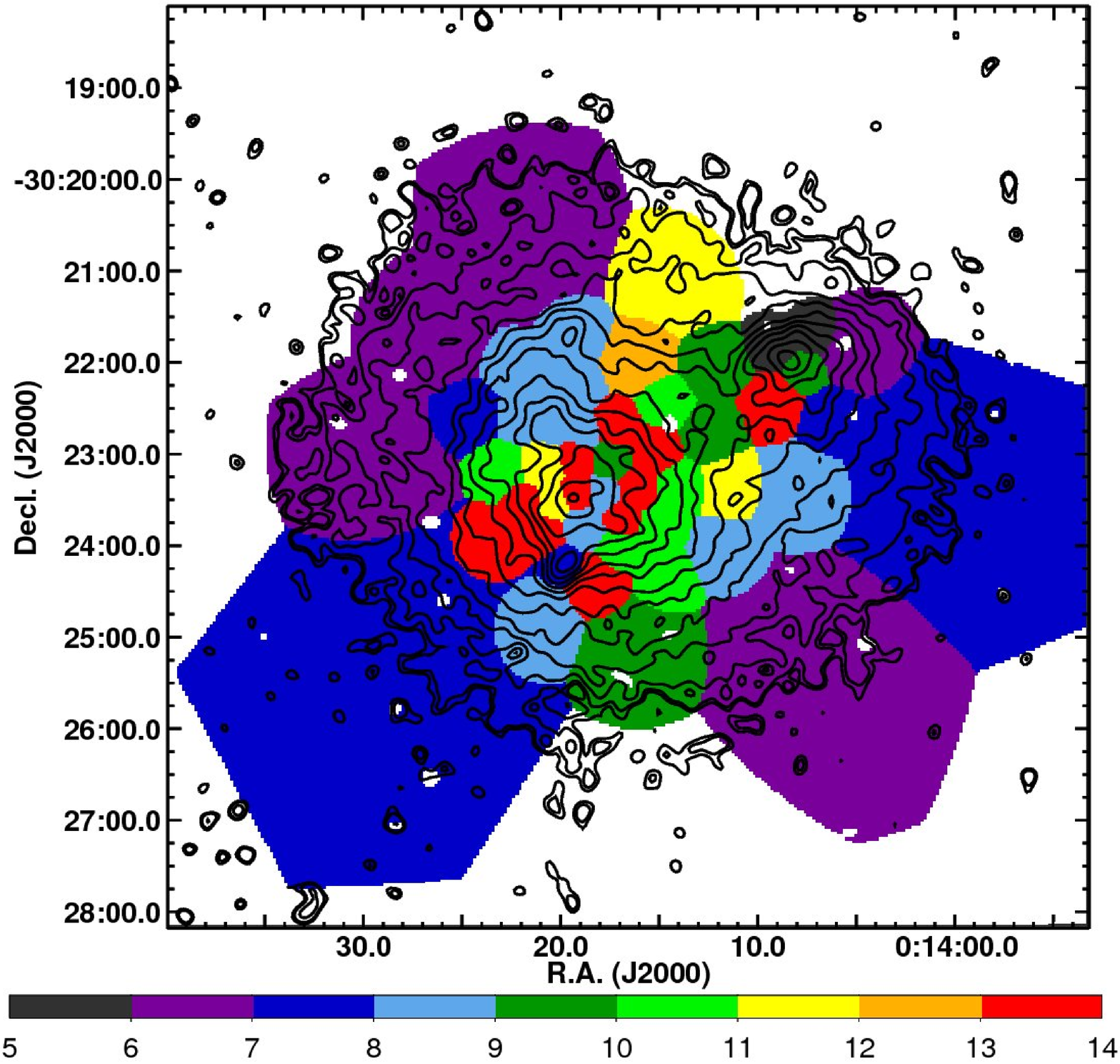}}
{\includegraphics[width=0.42\textwidth]{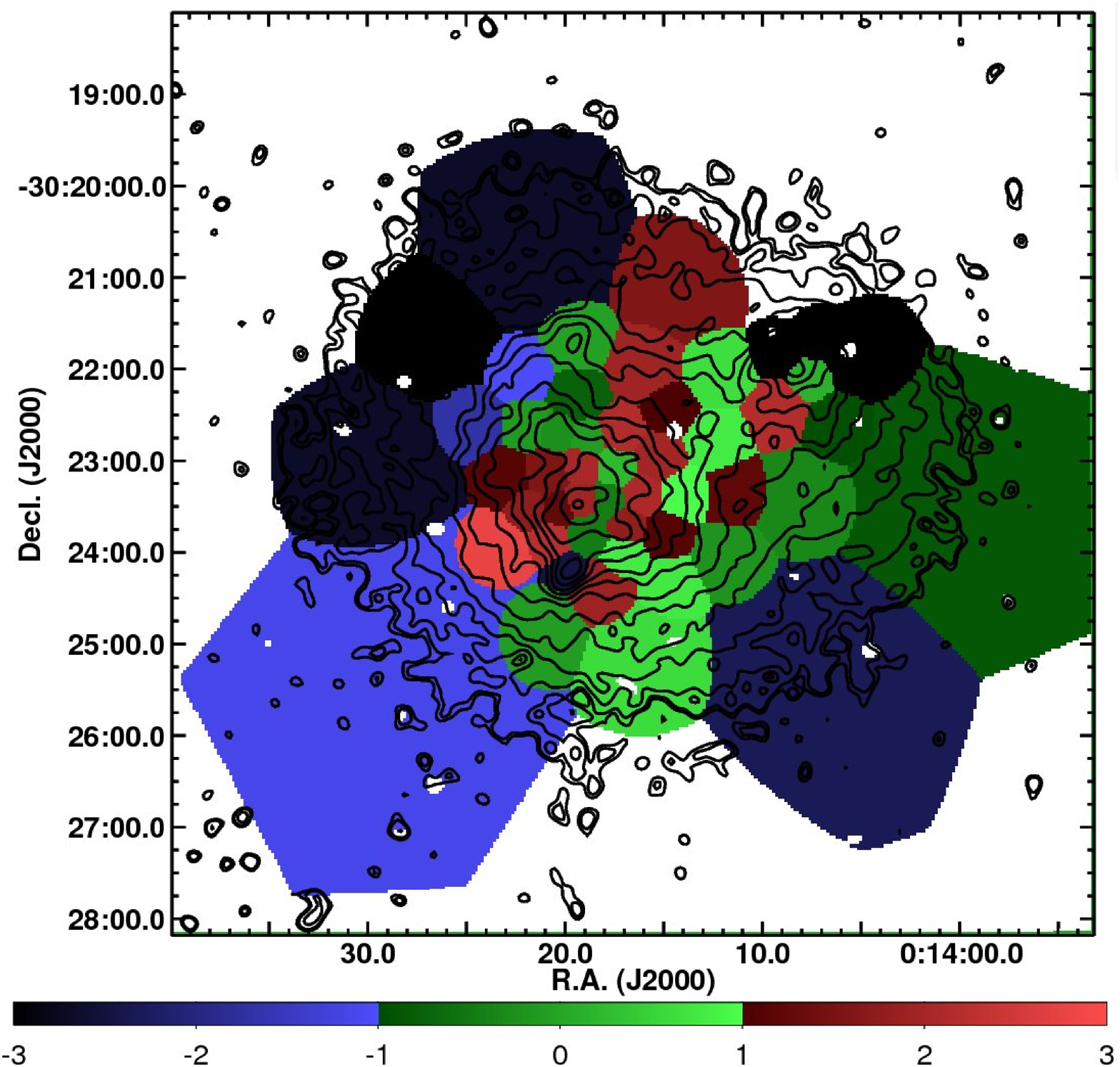}}\\
{\includegraphics[width=0.42\textwidth]{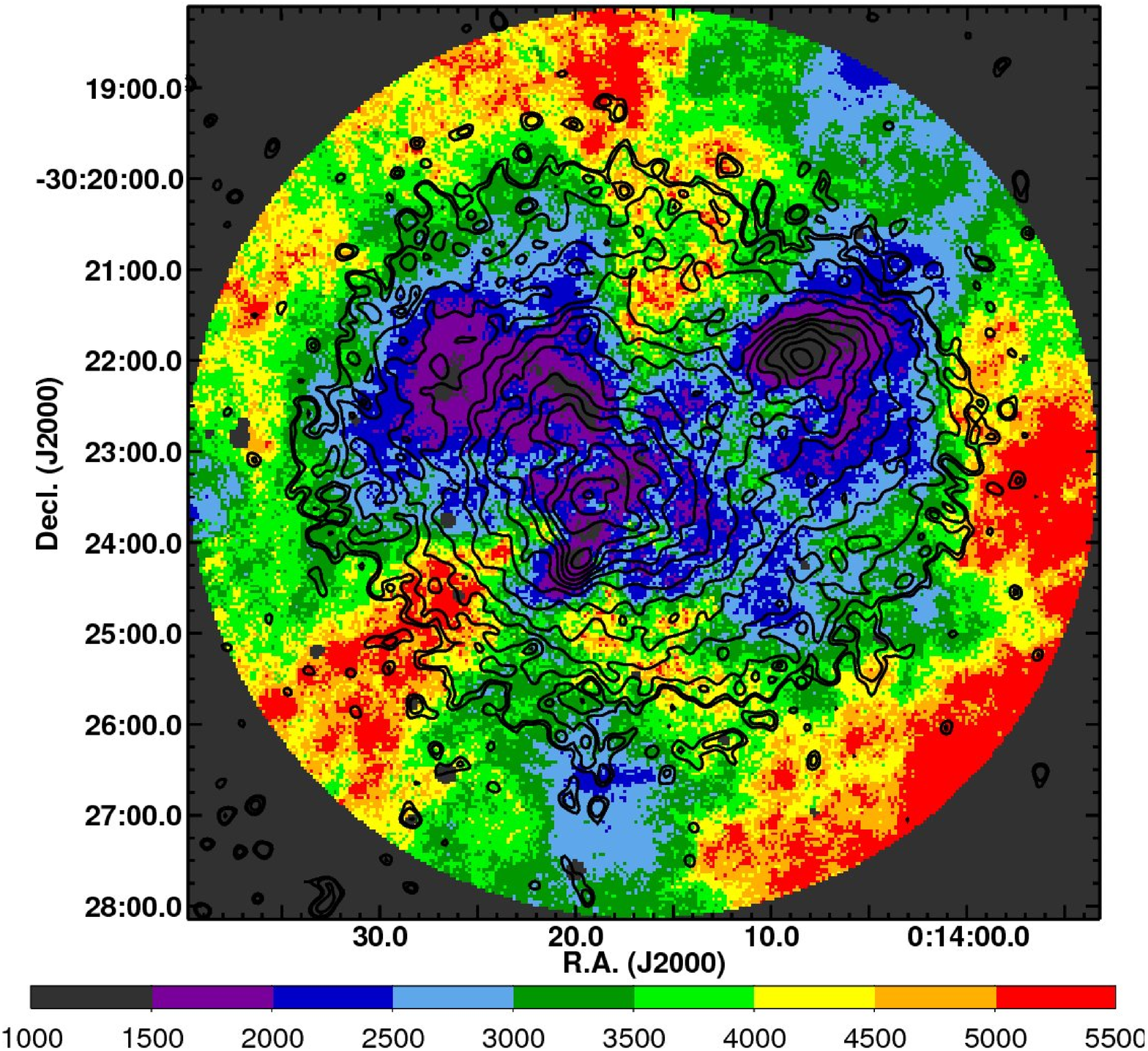}}
{\includegraphics[width=0.42\textwidth]{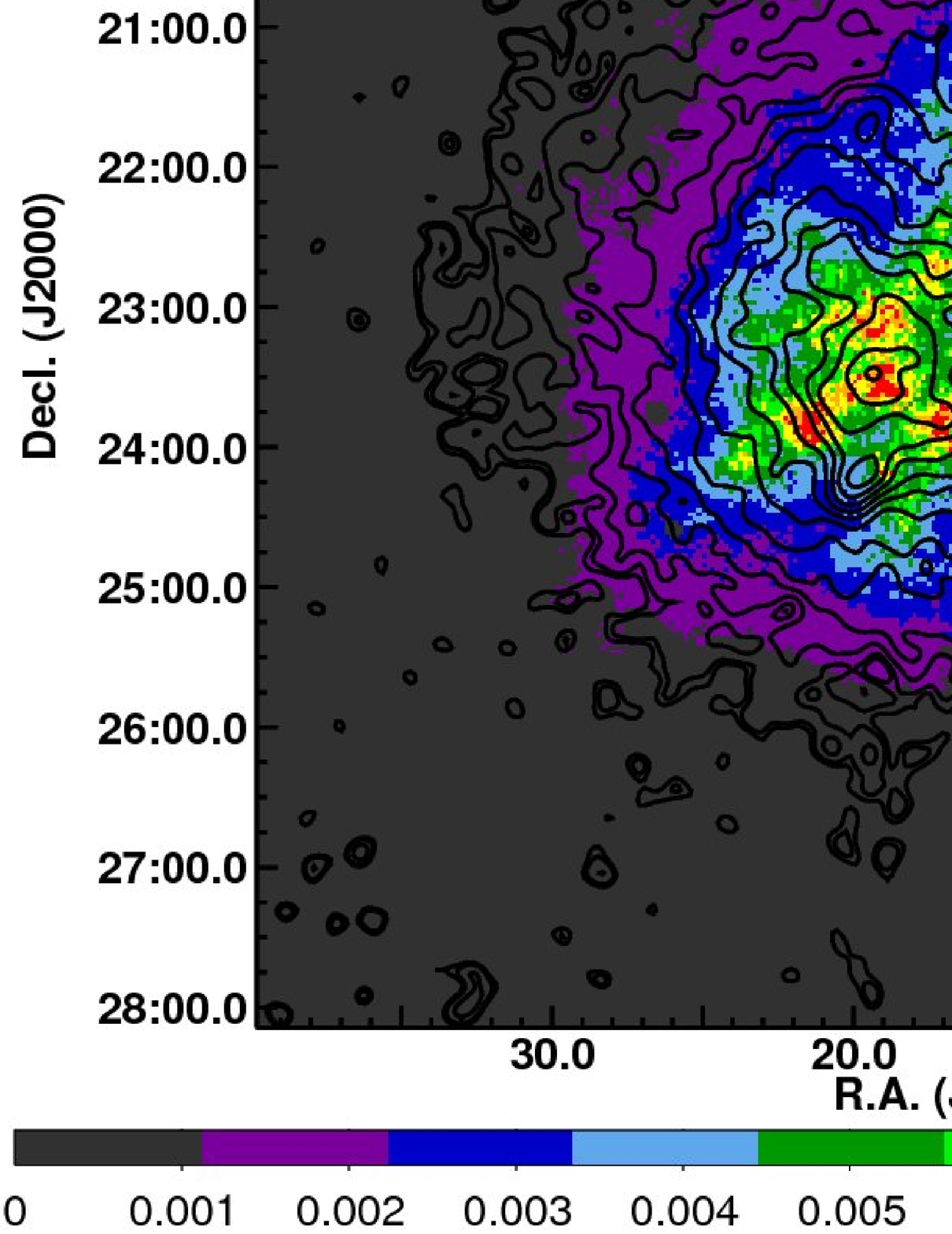}}
\caption{Thermodynamics maps derived from the \chan\ data. All maps have \chan\ 
surface brightness contours overlaid. {\it Top left:} Temperature map using the 
method outlined in \citet{randall2008}. The cross points indicate regions where 
we require a WVT bin to be centered (see text). {\it Middle left:} Temperature
map determined using the WVT binning algorithm to define regions. The color bars
on the {\it top} and {\it middle left} panels show the temperature scale in keV.
{\it Top and middle right:} Significance of the difference in temperature at 
each pixel compared to the global temperature where positive (negative) values 
indicate a higher (lower) temperature. The color bars show the 
significance scale which has units of the number of $\sigma$s deviation from 
the global temperature. {\it Bottom left
and right:} Pseudo-entropy {\it left} and Pseudo-pressure {\it right} maps 
derived from the temperature and normalization. The units of the pseudo-entropy 
and -pressure maps are arbitrary.}
\label{scottstmap}
\end{figure*}

To better determine the nature of the features seen in Figures~\ref{wvtimage} 
and \ref{residimage} and to identify other regions affected by the merger 
activity, we produce two maps of the projected temperature distribution for 
Abell~2744 using complementary techniques. The first temperature map is produced
using a similar method to that developed by
\citet{osullivan2005} and \citet{maughan2006}, which is outlined in full in 
\citet{randall2008}. Briefly, the X-ray image is binned to a
$1\arcsec.968\times 1\arcsec.968$ pixel size and at the position of each pixel, 
source and background spectra are extracted in a circular region surrounding 
the pixel. The radius of this region is defined adaptively so that the extracted
region contains 1500 background subtracted counts. The process is expedited by 
extracting the response files from a coarser grid. For each spectrum, an 
absorbed MEKAL model is fitted in the energy range $0.6-9.5$\,keV, with the 
absorption column density set at the Galactic value, 
$N_{\rm H}=1.61\times10^{20}\, {\rm cm}^{-2}$ \citep{dickey1990}, the abundance set
to the global value of $Z_{ave}=0.262$ (see below) and the temperature 
free to vary. This temperature map is presented in the top left panel of 
Figure~\ref{scottstmap}. We note that this method produces a temperature map 
with correlated pixels and the scale over which the pixels are correlated is 
not clear from the temperature map alone. Therefore, as a check we 
produced a second temperature map using the WVT algorithm to bin the image such 
that each (uncorrelated) spatial bin contains $\sim 1500$ background subtracted 
$0.5-7$\,keV counts. We select a number of regions of interest from the 
temperature map shown in the top left panel of Figure~\ref{scottstmap} (marked 
with crosses) and we require the WVT binning algorithm to place a bin 
centered on each of these regions. For each bin we extract observation-specific 
source spectra, corresponding count-weighted responses, and background spectra 
from the blank sky observations. The background spectra had their exposure times
normalized so that the 9.5-12\,KeV count rates match the observed rates. The 
spectra were re-binned so that there was at least 1 count per energy bin. The 
re-binned observation-specific spectra were fitted simultaneously over the
0.5--7 keV energy range within the XSPEC package \citep{arnaud1996} using an 
absorbed MEKAL model \citep{kaastra1992,liedahl1995} with abundance, column 
density and redshift fixed to those values used for the non-tessellated map. 
The Cash statistic was minimized during fitting. This tessellated temperature 
map is shown in the middle left panel of Figure~\ref{scottstmap}. The 
tessellated and non-tessellated temperature maps are in excellent agreement.

We have produced significance maps for both the tessellated and non-tessellated 
temperature maps which show how significant the spatial variations in 
temperature are when compared to the global average temperature, $kT_{ave}$. The
significance, $\Delta (kT)$, is defined at each pixel as
\begin{equation}
  \Delta (kT) = \left\{
  \begin{array}{ll}
    {(kT_{\rm pix} - kT_{\rm ave})} \over {\sqrt{\sigma (kT_{\rm pix, hi})^2 + 
        \sigma(kT_{\rm ave, lo})^2}}, &kT_{\rm pix} < kT_{\rm ave}, \\
    {(kT_{\rm pix} - kT_{\rm ave})} \over {\sqrt{\sigma(kT_{\rm pix, lo})^2 + 
        \sigma(kT_{\rm ave, hi})^2}}, &kT_{\rm pix} \geq kT_{\rm ave},
  \end{array}
  \right.
\end{equation}
where $kT_{\rm pix}$ is the temperature map value for the pixel of interest, 
$\sigma (kT_{\rm pix, hi}) \,\rm{and}\, \sigma(kT_{\rm pix, lo})$ are the upper
and lower $68\%$ confidence limits for $kT_{\rm pix}$, respectively, and 
$\sigma (kT_{\rm ave, hi}) \,\rm{and}\, \sigma(kT_{\rm ave, lo})$ are the upper and 
lower $68\%$ confidence limits for $kT_{\rm ave}$, respectively. The significance
maps are shown in the top right and middle right panels of 
Figure~\ref{scottstmap} for the non-tessellated and tessellated maps, 
respectively. The global average temperature, $kT_{\rm ave}$, was determined from
fitting spectra extracted from a $\sim 900$\,kpc radius region containing the 
majority of the cluster emission with an absorbed MEKAL model, as outlined 
above for the tessellated maps, with the exception that the
abundance was free to be fitted. We find $kT_{\rm ave}=9.07\pm0.15$\,keV, while 
the global average abundance $Z_{\rm ave}=0.262\pm0.025$. These values are 
consistent with the values presented in \citet{zhang2004}. 

The temperature maps reveal a rich diversity of patchy temperature structures, 
the most significant of which include: (i) A prominent cool $\sim 5$\,keV region
associated with the northwest interloper where the coolest gas is coincident 
with the compressed X-ray isophotes, indicating there is a cold front there. We
also note a hint of a cool swirl heading to the south 
of the northwestern interloper and coincident with the asymmetry seen in the X-ray 
brightness (Figure~\ref{residimage}). (ii) A cool $\sim 7.5$\,keV region 
associated with the southern compact core which is surrounded by hot $> 12$\,keV
gas, the hottest of which resides just to the east of the substructure. (iii) A 
$\sim 700$\,kpc ridge of hot $> 11$\,keV gas running towards the north which
separates the northwestern interloper and northern core. (iv) While the northern 
core and its tail do not appear to have a significantly different 
temperature to the average temperature, they do appear to be significantly 
cooler than the regions directly to the west and hotter than the region directly
to the east. This is especially clear when considering the tessellated map 
(middle left and right panels of Figure~\ref{scottstmap}). (v) The peak in the 
X-ray surface brightness does not host the coolest gas, as would be expected in 
a relaxed, cool core cluster.

\subsubsection{Pseudo-Entropy, Pseudo-Pressure and Abundance maps}

From the non-tessellated temperature map, we have generated projected
pseudo-pressure and pseudo-entropy maps, similar to those presented in 
\citet{finoguenov2005}. These are shown in the bottom left 
and bottom right panels of Figure~\ref{scottstmap}, respectively. The 
pseudo-entropy at each pixel is $K=kT_{\rm pix}A_{\rm pix}^{-1/3}$, while the 
corresponding pseudo-pressure is $P=kT_{\rm pix}A_{\rm pix}^{1/2}$ where $A_{\rm pix}$ is the
normalization determined during the fitting of the spectra corrected for
exposure and normalized by the area of the extraction region. The northwestern
interloper stands out as having the lowest pseudo-entropy in the
cluster complex, while the swirl of cool gas, hinted at in the temperature maps, 
is clearly seen as low entropy gas extending towards
the south. The non-hydrostatic nature of Abell~2744 is further highlighted
by the asymmetric, patchy pressure distribution which peaks at the X-ray surface
brightness peak and has two high pressure fingers trailing off to the southeast 
and northeast.

\begin{figure*}
\vspace{10pt}
{\includegraphics[width=0.45\textwidth]{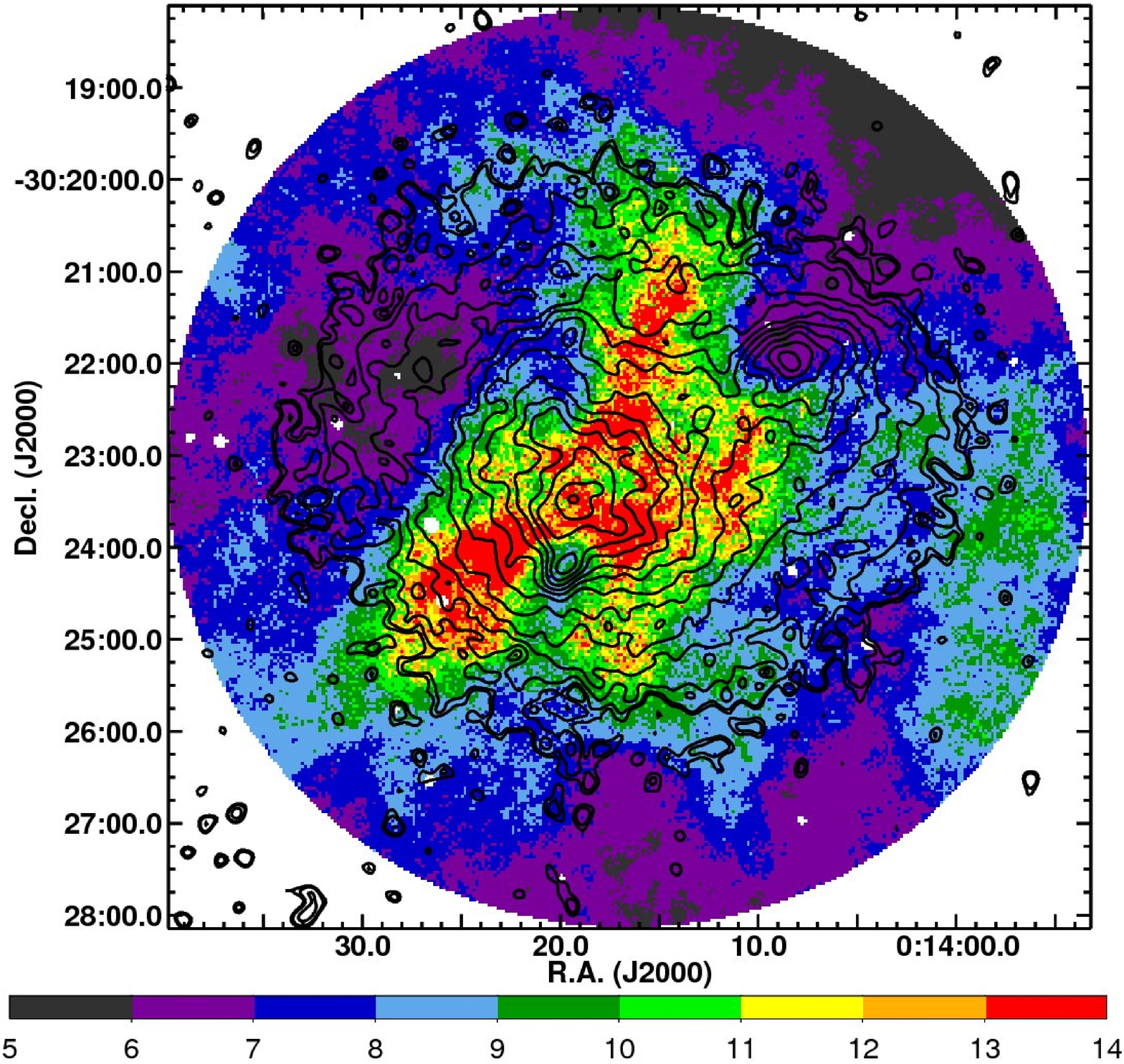}}
{\includegraphics[width=0.45\textwidth]{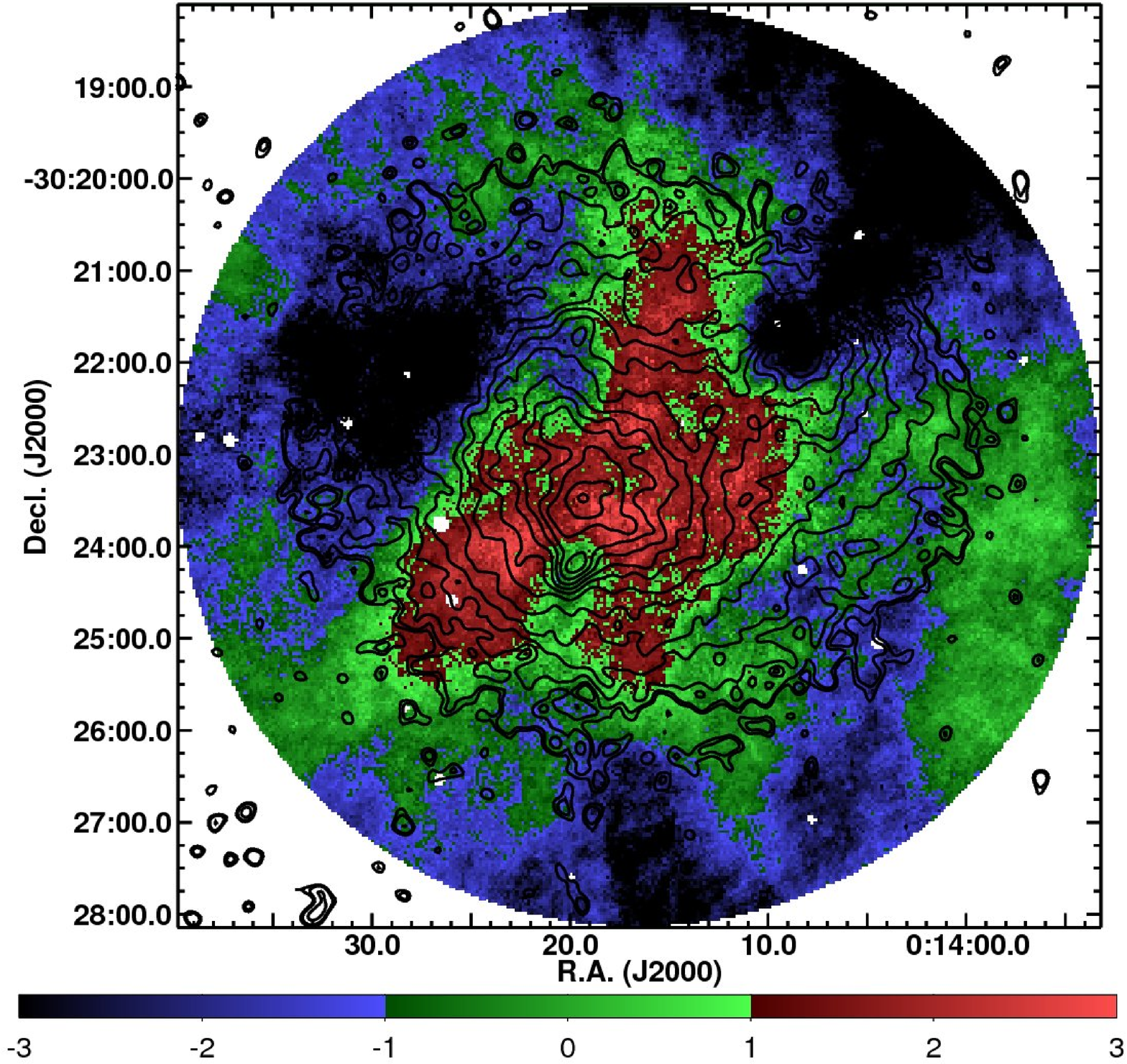}}\\
{\includegraphics[width=0.45\textwidth]{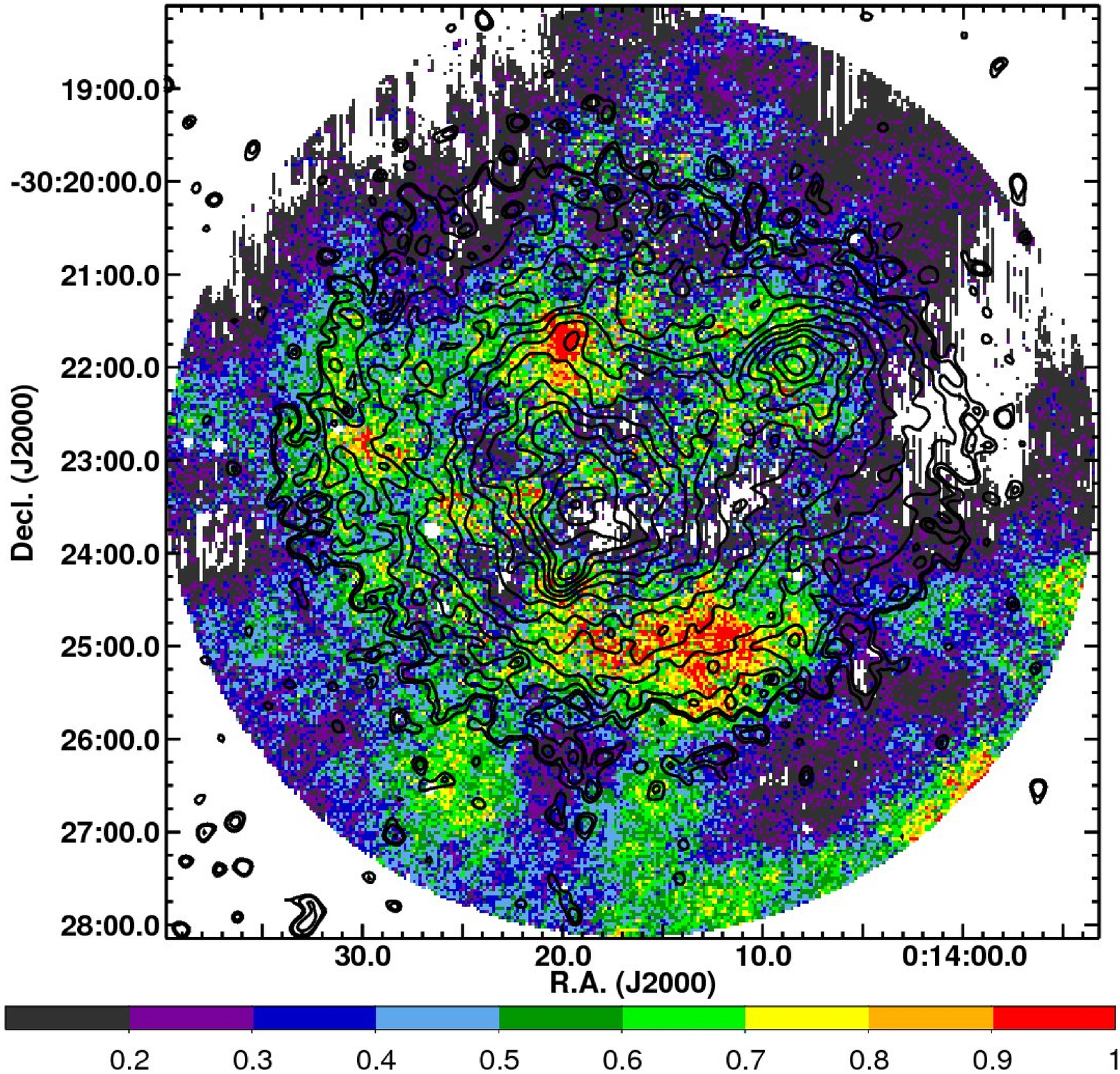}}
{\includegraphics[width=0.45\textwidth]{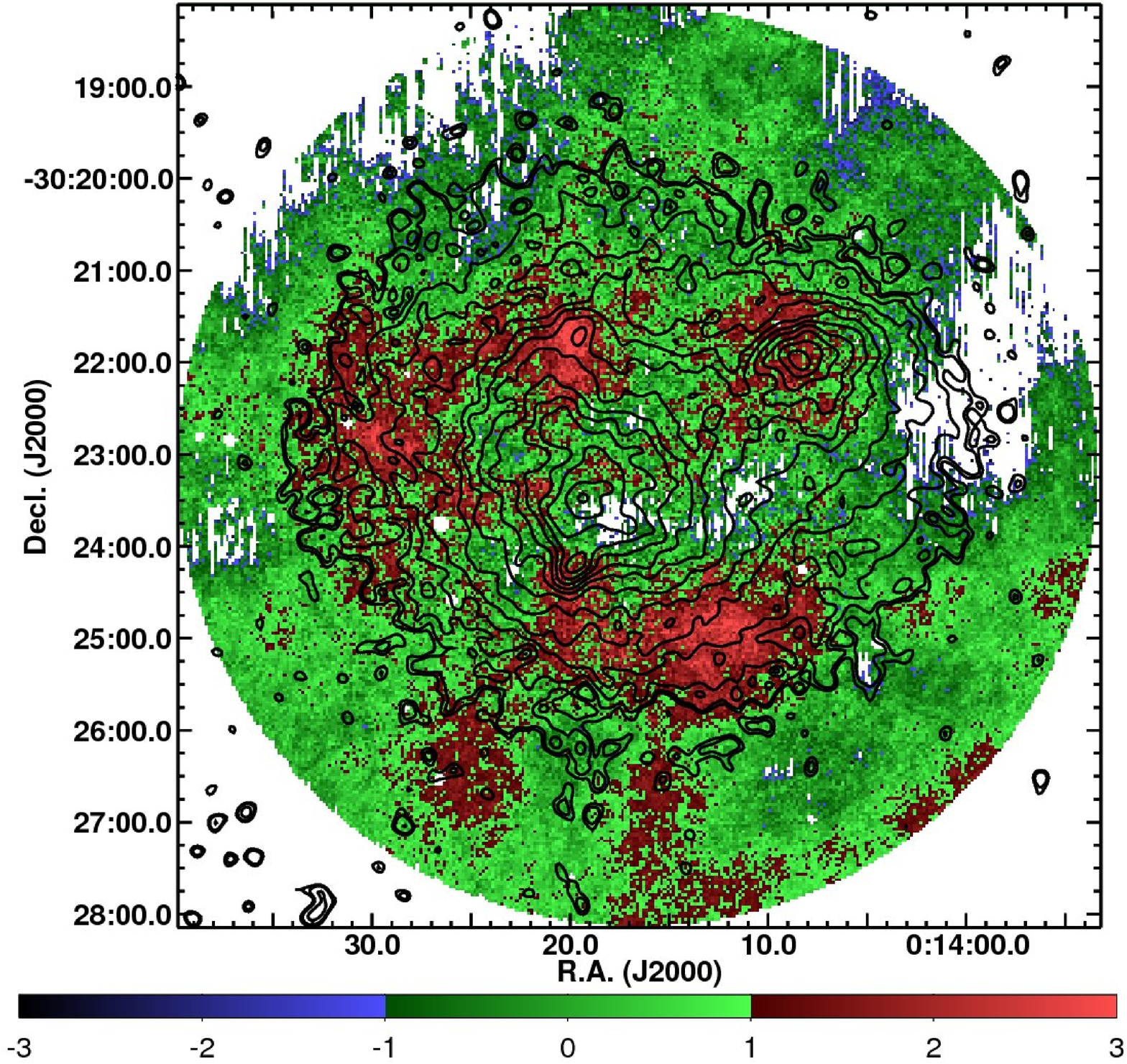}}
\caption{{\it Top left and right} Same as the {\it top left and right} panels in 
Figure~\ref{scottstmap}, but using an extraction region which contains 3000 
background subtracted counts and allowing the abundance to vary during the fit.
{\it Bottom left:} The abundance map where the color bar shows the corresponding
abundance scale relative to solar. {\it Bottom right:} The significance of the
difference of the abundance at each pixel relative to the global abundance 
measure. The color bar gives the scale in units of $\sigma$.}
\label{Zmap}
\end{figure*}

Finally, we present an abundance map in the lower left panel of 
Figure~\ref{Zmap}. The technique outlined in Section~\ref{tmaps} used 
to generate the non-tessellated temperature map was again applied. However, the 
abundance was free to vary during the fitting while, in order to allow more
robust abundance determinations, the spectra were extracted from larger regions 
which contain $\sim 3000$ counts. The abundances were measured relative to the
solar photospheric values of \citet{anders1989}. Included in the top left panel 
of Figure~\ref{Zmap} is the corresponding temperature map which is in broad 
agreement with the temperature maps presented in the top and middle left panels 
of Figure~\ref{scottstmap}. The most notable enhancements in abundance seen in 
the maps presented in the lower panels of Figure~\ref{Zmap} are coincident with 
the southern compact core, the northern core and the northwestern interloper. We identify the northern
core as the highest metallicity region within Abell~2744. Furthermore, 
there is an anti-correlation between the abundance and the temperature insomuch 
as the hottest regions also appear to have the lowest abundances.

\subsection{Regions of Interest}\label{roi}

\begin{figure*}
\vspace{10pt}
{\includegraphics[width=0.48\textwidth]{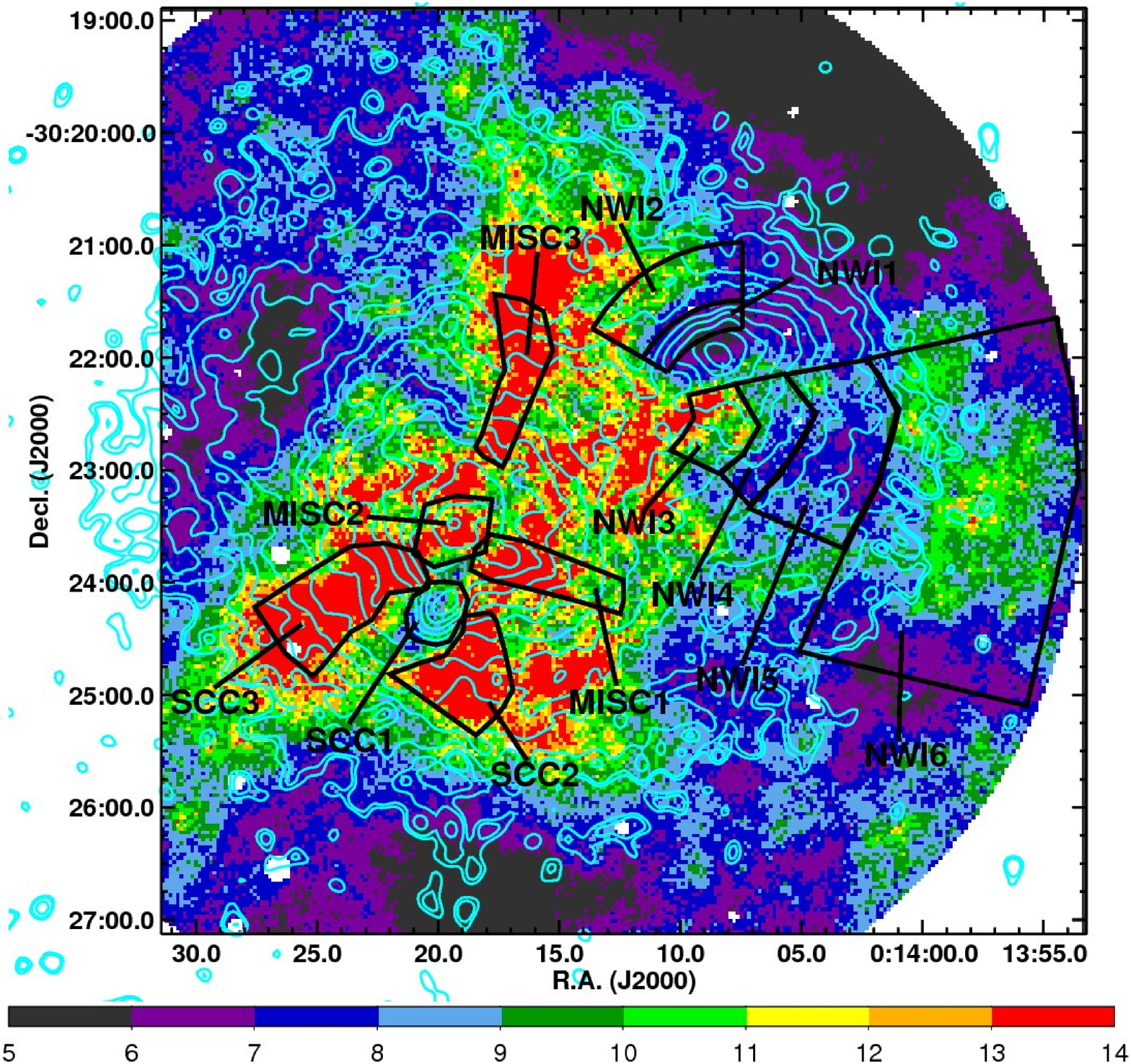}}
{\includegraphics[width=0.48\textwidth]{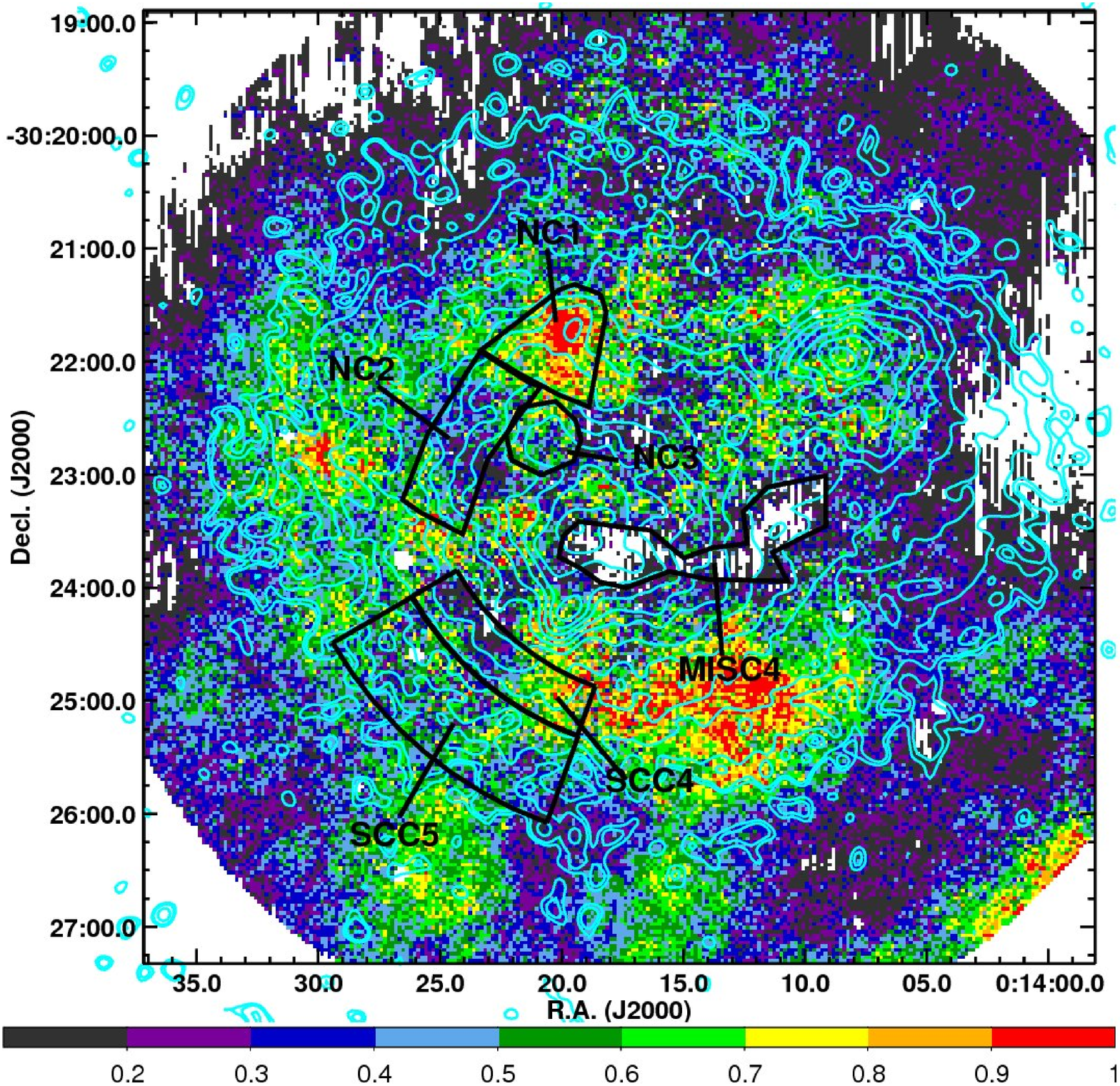}}
\caption{Regions of interest labelled as in Table~\ref{reg_interest} shown overlaid on the temperature map ({\it left panel})
and abundance map ({\it right panel}).}
\label{regint_fig}
\end{figure*}

We now explore the nature of several regions of interest which have been 
identified from Figures~\ref{wvtimage}, \ref{residimage} and \ref{scottstmap}. 
The selected regions are overlaid onto the temperature and abundance maps in 
Figure~\ref{regint_fig}. For each region of interest the source spectra, along 
with the corresponding backgrounds and responses, are extracted and fitted 
in XSPEC with an absorbed MEKAL model. During the fitting, the column density
is fixed to the Galactic value while the temperature, abundance, normalization,
and, where the data are of a high enough quality, the redshift are allowed to
free to vary during the fit. The best-fitting parameters for 
each of the regions are presented in Table~\ref{reg_interest}.

\begin{deluxetable*}{cccccc}
\tabletypesize{\scriptsize}
\tablecolumns{3}
\tablewidth{0pc}
\tablecaption{Summary of fits to spectra extracted from the regions of interest. \label{reg_interest}}
\tablehead {\colhead{Region} & \colhead{Description} &\colhead{$kT$ (keV)} & \colhead{Abundance (solar)} & \colhead{$z_{gas}$}& \colhead{Net Counts}}
\startdata
\cutinhead{Main Cluster}
R1&central 900kpc & $9.07_{-0.18}^{+0.14}$&$0.262_{-0.025}^{+0.026}$&$0.3033_{-0.0030}^{+0.0016}$& 94206\\
\cutinhead{Northern Core Surrounding Structures}
NC1&NC head& $7.36^{+0.75}_{-0.50}$ & $0.480^{+0.130}_{-0.126}$ & $0.3035^{+0.0087}_{-0.0083}$ &3103\\
NC2& NC Tail & $8.69^{+0.67}_{-0.67}$ & $0.424^{+0.125}_{-0.115}$ & $0.2846^{+0.0099}_{-0.0129}$ &4447\\
NC3&structure south of NC & $9.05_{-0.79}^{+0.92}$ &$0.404_{-0.132}^{+0.113}$ & $0.2967_{-0.0114}^{+0.0068}$& 2962\\
\cutinhead{Southern Compact Core and Surrounding Structures}
SCC1&SE cool core & $7.70^{+0.97}_{-0.67}$ & $0.392^{+0.154}_{-0.139}$ &$0.2950^{+0.0088}_{-0.0109}$& 2059\\
SCC2&Outside2 SE cool core & $9.74^{+1.94}_{-1.37}$ & $0.329^{+0.234}_{-0.213}$ &$0.3064$&1551\\
SCC3&Hot reg north of SE cool core & $15.56^{+2.81}_{-1.90}$ & $0.0^{+0.152}_{-0.252}$ &$0.3064$&2986\\
SCC4&Inside edge&$15.83^{+3.80}_{-2.70}$ & $0.143^{+0.292}_{-0.267}$ &$0.3064$&1800\\
SCC5&Outside edge&$8.61^{+1.82}_{-1.42}$ & $0.607^{+0.360}_{-0.291}$ &$0.3064$&1260\\
\cutinhead{Northwestern Interloper and Surrounding Structures}
NW1&NW inside front &$6.33_{-0.70}^{+0.83}$&$0.233_{-0.148}^{+0.173}$ & $0.3086_{-0.0170}^{+0.0232}$ & 1270\\
NW2&NW outside front&$10.45_{-1.80}^{+2.52}$&$0.104_{--}^{+0.259}$ & $0.3064$ & 1263\\
NW1&NW inside front deprojected&$5.09_{-0.71}^{+0.88}$&$0.293_{-0.210}^{+0.259}$ & $0.3095_{-0.0194}^{+0.0240}$ & 1270\\
NW3&NWtail adjacent east & $10.44_{-1.42}^{+1.76}$&$0.300_{-0.201}^{+0.207}$ & $0.2935_{-0.0143}^{0.0159}$ & 1642\\
NW4&NWtail & $8.12_{-0.99}^{+1.10}$&$0.383_{-0.168}^{+0.194}$ & $0.3215_{-0.0107}^{0.0101}$ & 1686\\
NW5&NWtail adjacent West& $8.28_{-1.08}^{+1.21}$&$0.000_{-0.155}^{+0.144}$ & 0.3064 & 1723\\
NW4+NW5&Comb NW4 and NW5 & $8.34_{-0.80}^{+0.83}$&$0.190_{-0.110}^{+0.1146}$ & $0.3189_{-0.0119}^{+0.0161}$ & 1723+1686\\
NW6&NWtail adjacent West 2& $11.28_{-2.37}^{+3.42}$&$0.007_{-0.240}^{+0.366}$ & 0.3064 & 1635\\
\cutinhead{Miscellaneous Regions}
MISC1& ridge c&$12.46^{+1.55}_{-1.14}$ & $0.097^{+0.128}_{-0.119}$ & $0.3064$ &4258\\
MISC2&SB peak&$10.33^{+0.95}_{-0.89}$ & $0.215^{+0.101}_{-0.104}$ & $0.3189^{+0.0092}_{-0.0110}$ &4793\\
MISC3&Hot ridge north&$13.14^{+2.28}_{-1.85}$ & $0.353^{+0.225}_{-0.203}$ & $0.2813^{+0.0127}_{-0.0132}$ &2470\\
MISC4& low abundance &$11.46^{+0.82}_{-0.77}$ & $0.180^{+0.087}_{-0.090}$ & $0.3112^{+0.0096}_{-0.0102}$ &7897\\
\enddata
\end{deluxetable*}

\subsubsection{Northern Core and Surrounding Structures}

The abundance map (bottom left panel of Figure~\ref{Zmap}) reveals the northern 
core has significantly higher metallicity than the surrounding gas,
while the temperature difference is not significant. Furthermore, both the
residual significance and unsharp masked maps (Figure~\ref{residimage})
reveal a curved tail of emission trailing to the south of the substructure,
as well as a second component $\sim 270$\,kpc to the south. We extract and 
fit spectra from regions surrounding these structures (NC1, NC2 and NC3 in 
the right panel of Figure~\ref{regint_fig}). We find that the northern substructure 
(NC1) does in fact have a significantly lower temperature than the global average, 
while the NC2 and NC3 regions have temperatures which are consistent with the 
global average within the quoted confidence levels. Also, we find that at the 
$68\%$ confidence level all three regions harbor gas with higher abundances 
than the global value.

\subsubsection{Southern Compact Core and Surrounding Structures}\label{southern_ss}

There are several regions of interest surrounding the southern compact core
including the compact core itself (SCC1 in the left panel of 
Figure~\ref{regint_fig}) and two hot regions identified in the temperature 
maps which straddle the core on the northeastern and southwestern sides 
(SCC2 and SCC3 in the left panel of Figure~\ref{regint_fig}). While
the detailed spectra extracted for the SCC2 region reveal the 
gas there is not significantly hotter than the global average temperature,
the region to the northeast (SCC3) does contain significantly hotter gas
which also has a very low abundance. We confirm that the
southern compact core harbors gas which is both cooler and of a higher
abundance than the global averages.

\begin{figure}
\includegraphics[angle=-90,width=.48\textwidth]{{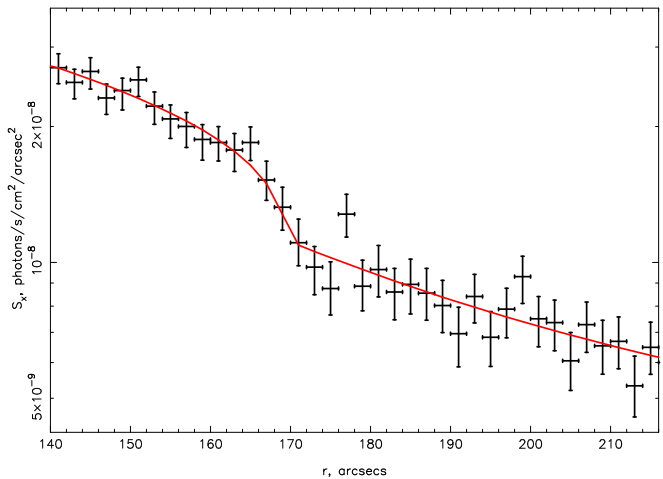}}
\caption{Surface brightness profile across a sector containing the edge to the
southeast of the southern compact core (see regions SCC4 and SCC5 in 
Figure~\ref{regint_fig}). The red curve shows the surface brightness 
profile for the best fitting broken powerlaw density model.}
\label{shock_sb_prof}
\end{figure}

We attempt to quantify the strength of the surface brightness edge just 
southeast of the southern compact core (Figure~\ref{wvtimage}) by fitting 
a spherically symmetric broken power law density model centered at the center of
curvature of the edge \citep[using the same method as][]{owers2009c}. The 
resulting fit gives a jump in density at the edge of $1.60^{+0.15}_{-0.11}$. The
surface brightness profile and the best fitting broken powerlaw density model
are shown in Figure~\ref{shock_sb_prof}. We 
note that in ObsIds 7712 and 7915, the chip gap lies very close to the 
position of the edge. Therefore we repeat the fit using only ObsIds 8477 
and 8557. We find that the best fitting surface brightness model gives a 
density jump of $1.51_{-0.15}^{+0.17}$, consistent with the measurement obtained 
using the full dataset. The temperature measured just inside the edge (SCC4) 
is kT=$15.83_{-2.7}^{+3.80}$\,keV, while kT=$8.61_{-1.42}^{1.82}$\,keV in the
SCC5 region just outside the edge. The change in pressure across this edge is 
$(n_{e,in}kT_{in})/ (n_{e,out}kT_{out})=2.94^{+0.98}_{-0.73}$, which indicates a 
significant pressure jump and thus that the edge is likely a part of
a shock front.

\subsubsection{Northwestern Interloper and Surrounding Structures}\label{NW_edge}

\begin{figure}
\includegraphics[angle=-90,width=.48\textwidth]{{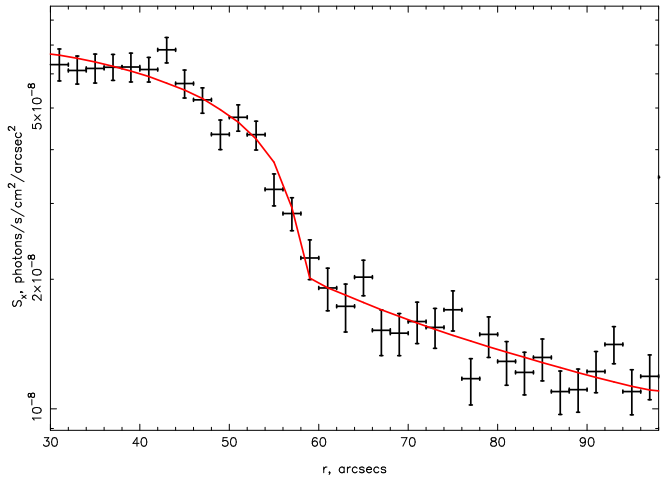}}
\caption{Surface brightness profile across a sector containing the NW 
substructure cold front (see regions NWI1 and NWI2 in 
Figure~\ref{regint_fig}. The red curve shows the surface brightness 
profile for the best fitting broken powerlaw density model.}
\label{sb_prof}
\end{figure}

As noted in Section~\ref{image}, the northwestern interloper contains a sharp
surface brightness edge to the north. The temperatures just inside and just 
outside the front (regions NWI1 and NWI2 in the left panel of Figure~\ref{regint_fig}) 
are $6.33_{-0.70}^{+0.83}$ and $10.45_{-1.80}^{+2.52}$ keV, respectively. 
Figure~\ref{sb_prof} shows the surface brightness profile across the edge, 
which we have fitted with a spherically symmetric broken power law density model
centered at the center of curvature of the edge 
\citep[using the same method as][]{owers2009c} 
with a best fitting density jump of $2.13^{+0.20}_{-0.17}$ (uncertainties are 
$68\%$ confidence limits). Providing the geometry assumed in the density model
accurately represents the physical geometry of the edge, the temperature just 
inside the edge will be artificially boosted by hot gas from outside the
edge which lies along the line of sight. To account for this, we determine
the deprojected temperature inside the edge by including a second MEKAL
component in fitting the spectrum extracted from the NWI1 region. 
This second MEKAL component has temperature, normalizations and 
abundance frozen to the values determined for the NWI2 region, with 
the normalizations multiplied by a constant which accounts for
the different emission measures due to the different volumes probed 
\citep[see][for a detailed explanation of the procedure]{owers2009c}.
The deprojected temperature in the NWI1 region is 
$5.09_{-0.71}^{+0.88}$ and combining this with the measured density jump
and the temperature measured in the NWI2 region, we find that the pressure 
across the front, given by $(n_{e,in}kT_{in})/ (n_{e,out}kT_{out})=1.04^{+0.32}_{-0.25}$ is 
continuous and, thus, the edge is a cold front.

We extract spectra for 4 regions running from east to west across  
the northwestern interloper's low entropy tail 
(NWI3, NWI4, NWI5, NWI6 in the left panel of Figure~\ref{regint_fig}). 
The temperature map (top left panel in Figure~\ref{scottstmap}) indicated that the low entropy
tail may harbor cooler gas than the regions immediately to the east and west.
Region NWI4 coincides with the lowest projected entropy, while region NWI5
coincides with the lowest temperature gas seen in the temperature maps in
the vicinity of the low entropy tail. Regions NWI3 and NWI6 are the regions
immediately east and west of the tail, respectively. The temperature and
abundance measurements are listed in Table~\ref{reg_interest} and we find that
the temperatures in NWI4 and NWI5 are indeed lower than those measured in 
NWI3 and NWI6, but only at the $68\%$ confidence level.

\subsubsection{Miscellaneous regions with high temperature and/or low abundance.}

There are several regions in Figures~\ref{regint_fig} which have either high 
temperature (MISC1, MISC2 and MISC3) and and/or an anomalously low abundance 
(MISC4). The first of the hot regions (MISC1) is roughly coincident with a 
finger of excess emission clearly seen in the unsharp masked and residual 
significance images (Figure~\ref{residimage}) which extends $\sim 450$\,kpc 
from the X-ray surface brightness peak and towards the north-northeast 
\citepalias[labelled Ridge C in][]{kempner2004}. The second region (MISC2) is 
coincident with the peak in 
the X-ray surface brightness while the third region (MISC3) is a streak of hot 
gas which extends $\sim 700$\,kpc from the peak in the X-ray surface brightness 
towards the north. We also extract a spectrum for the region which encompasses 
the low abundance finger running from the peak in the X-ray surface brightness 
towards the north-northwest (MISC4).

We find that the MISC2 region is marginally hotter than the cluster mean 
temperature and has an abundance which is statistically consistent with the 
mean value. The two other regions selected based on their high temperature, 
MISC1 and MISC3 are both significantly hotter than the mean temperature, 
consistent with the temperature map results. The abundance of the MISC3 region 
is consistent, within the uncertainties, with the mean cluster value, while the 
abundance of the MISC1 region is low when compared to the cluster mean value. 
The MISC4 region, selected based on its low metallicity, does have a best 
fitting abundance which is lower than the mean cluster value, but this is not 
statistically significant.

\section{Optical Data}\label{optical_data}

In this section, we outline the new spectroscopic measurements taken using the 
AAOmega multiobject spectrograph (MOS) which, when combined with the available 
redshifts from the literature, provide a sample of 343 spectroscopically 
confirmed cluster members which we utilize to detect substructures and to 
constrain the line-of-sight dynamics of the merger.

\subsection{Sample selection, Observations and Completeness}

Here we outline the selection of galaxies targeted for spectroscopic 
observations, the observation and reduction of the spectra, the precision of 
our redshift measurements and the spectroscopic completeness of our sample.

\subsubsection{Photometric Catalog}

The primary photometric catalog used for selection of the spectroscopic
targets was drawn from the Supercosmos Sky Survey (SSS) 
server\footnote{See http://www-wfau.roe.ac.uk/sss/} using the object catalog
extraction function. We include only objects within a 15\arcmin\, 
($\sim 4$\,Mpc) radius centered at R.A.=$00^{\rm h}14^{\rm m}19^{\rm s}.5$, 
decl.=$-30^{\circ}23\arcmin 19\arcsec.0$ 
\citep[i.e. the center defined by ][]{abell1989} and brighter than a magnitude 
of \rf\,=21. We use the SSS classification flag to remove those objects with 
\rf$<19$ and not classified as galaxies \citep[the SSS image classification is 
most reliable for \rf\, $< 19$][]{hambly2001b} and do not apply any 
image-classification-based filtering for fainter objects. We cross-match our 
SSS catalog with the photometric catalog of \citet{busarello2002} and 
include those objects which are not detected in the SSS, but appear in the 
\citet{busarello2002} catalog and meet our magnitude limit. We exclude a 
number of objects which are not cluster members based on VLT VIMOS spectroscopy 
\citep[Filiberto Braglia, priv. comm.; see also][]{braglia2009}. The resulting
catalog contains 1443 targets.

\subsubsection{AAOmega Observations}
Observations were taken during the nights of 2006 September 12--18 using the 
AAOmega MOS on the Anglo-Australian Telescope  
\citep[AAT;][]{saunders2004,smith2004,sharp2006}. AAOmega consists of 392 2\arcsec\, 
diameter fibers available for target allocation across a two degree diameter field of 
view. During our observations, we used the medium resolution 
(R~1300) 580V (blue arm) and 385R (red arm) gratings which provide a wavelength 
coverage of 3700-8800\AA\, and spectral resolutions of 3.5\AA\, and 5.3\AA\, 
at 4800\AA\, and 7250\AA, respectively.

The target density greatly exceeds the maximum fiber packing density, requiring
multiple configurations in order to obtain a high level of spectroscopic 
completeness, particularly in the dense core regions of the cluster. 
We utilize the simulated annealing algorithm \citep{miszalski2006} option within 
the AAOmega specific 
{\it CONFIGURE}\footnote{See http://www.aao.gov.au/AAO/2df/aaomega/aaomega\_SAconfigure.html} 
software to generate the fiber allocation files. This software operates most 
efficiently with an input catalog of $\sim 500-800$ objects, particularly when
configuring clustered fields. These requirements are met by partitioning the 
$\sim 1500$ object primary target catalog into subsamples based on the target 
magnitude. The magnitude bins used were \rf$<19.63$, $19.63 < $\rf$<20.05$, 
$20.05 < $\rf$<20.37$, $20.37 < $\rf$<20.75$ and $20.75 < $\rf$<21.00$. 

The limiting magnitudes of the subsamples serve two purposes: 1) they allow a 
fairly random subsampling of the densest regions, thereby reducing the 
computation time for the {\it CONFIGURE} software and 2) allow a shorter 
observation time to reach the desired S/N limit for the brighter subsamples, 
enabling a more efficient observing strategy. The limits are chosen 
to allow approximately two 
configurations per subsample, without under-sampling fiber allocations. The 
configurations are done in sequence, starting with the brightest magnitude 
subsample and continuing to fainter subsamples. The unallocated objects in each 
subsample were propagated through to the next subsample in the configuration 
sequence. During the fiber configuration process, the {\it CONFIGURE} software 
attempts to allocate objects of higher priority ahead of lower priority objects,
where prioritization is based on a user-defined rank assigned to each object. 
We required a high level of spectroscopic completeness in 
the central regions, so objects within 500kpc of the cluster center were ranked 
highest. Rankings were decreased as a function of radius in 500\,kpc radial bins 
out to a limit of 4\,Mpc.

This method produced nine configurations and we summarize the observations of 
these configurations in Table~\ref{aao_obs}. For each configuration, 
$\sim 40$ fibers were allocated to blank sky regions for sky subtraction and a 
further 8 fibers were allocated to bright stars for acquisition and guiding. 
Differential atmospheric refraction limits the length of contiguous
exposures for any one configuration to about three hours, 
therefore configurations requiring exposures longer than 3 hours were observed 
across several nights. For each configuration, we took a 5s dome flat and 40s 
FeAr arc lamp exposures for the purpose of flat fielding and wavelength 
calibration of the spectra (Note: for configurations observed across several 
nights, night-specific calibrations were taken). The data were reduced using 
the AAO {\it 2dFDR} pipeline 
software\footnote{See http://www.aao.gov.au/AAO/2df/aaomega/aaomega\_software.html\#2dfdr} 
which reduces the red and blue arm data separately, extracting sky subtracted, 
flat fielded and wavelength calibrated data for each exposure frame. The 
reduced frames are then co-added (using a weight derived from the total flux in 
each frame), at which point cosmic rays are identified and rejected. After 
co-addition, the red and blue-arm spectra are spliced together. At this point, 
the data were processed using an IDL code which reduces the residuals left 
from sky subtraction via a Principal Component Analysis (PCA) process which is 
adapted from software developed by 
\citet[][Emily Wisnioski priv. comm. See \citealp{drinkwater2010} for a more detailed explanation]{wild2005}. 

The reduced, combined, PCA corrected data are then assigned redshifts using the 
{\it RUNZ} code, which was written by Will Sutherland for the 2dF Galaxy
Redshift Survey \citep[2dFGRS][]{colless2001}. The {\it RUNZ} code has been
adapted for use with AAOmega spectra and significantly enhanced from its earlier
versions. In particular, emission line redshifts measurements have been
significantly improved for the WiggleZ survey \citep{drinkwater2010}. The code 
measures redshifts using two methods: 1) for absorption line spectra the 
cross-correlation method of \citet{tonry1979} is employed with a suitable 
library of template spectra, and 2) for emission line galaxies the redshifts are
measured for each line by fitting a Gaussian, with the final emission line 
redshift being determined from the variance weighted mean of the redshifts 
determined for each emission line. The quality of the assigned redshift
was determined by visual inspection and an integer value classification, $Q$, 
was assigned, where $Q$ ranges from 1 to 6, using the same scheme adopted for 
the 2dFGRS, as described in \citet{owers2009b,owers2009a}. Briefly, 
objects with $Q \leq 2$ are assigned no redshift or an unreliable redshift, 
objects with $Q=3, 4$ or 5 have reliable redshifts (with $Q=$5 having a 
template-like quality spectrum) and objects with $Q=$6 are stars or 
non-extragalactic objects. The AAOmega observations produced reliable ($Q=$3,4 or
5) redshifts for 983 extragalactic objects and 276 stars.

\begin{deluxetable*}{ccccccc}
\tabletypesize{\scriptsize}
\tablecolumns{3}
\tablewidth{0pc}
\tablecaption{Summary of the AAOmega observations.\label{aao_obs}}
\tablehead {\colhead{Field} & \colhead{Date} & \colhead{\rf\ Limit} & \colhead{$T_{exp}$}
& \colhead{$N_{objects}$} &  $N_{Redshift}$ & \colhead{Seeing}}
\startdata
1 & 2006 Sep 12 & 19.63 & $5\times 1800$s & 192 & 180 &1\arcsec.2\\
2 & 2006 Sep 13, 14 & 19.63 & $4\times 1800 + 3\times 1800$s & 134 & 126 &1\arcsec.4\\
3 & 2006 Sep 12 & 20.05 & $10\times 1800$s & 172 & 161 & 1.2--1\arcsec.4\\
4 & 2006 Sep 13, 14, 15 & 20.37 & $4\times 1800 + 6\times 1800 + 5\times 1800$s & 181 & 147 & 1.0--1\arcsec.5\\
5 & 2006 Sep 13, 14, 15 & 20.37 & $1\times 1800 + 5\times 1800 + 4\times 1800$s & 130 & 114 & 1.0--1\arcsec.5\\
6 & 2006 Sep 15, 16 & 20.75 & $5\times 1800 + 7\times 1800$s & 190 & 152 & 1.0--1\arcsec.5\\
7 & 2006 Sep 16, 17 & 20.75 & $5\times 1800 + 5\times 1800$s & 133 & 109 & 1.\arcsec0\\
8 & 2006 Sep 17, 18 & 21.00 & $(10\times 1800 + 1\times 1200) + (6\times 1200+ 1\times800)$s & 191 & 157 & 1\arcsec.0\\
9 & 2006 Sep 18 & 21.00 & $(12\times 1800)$s & 149 & 113 & 1\arcsec.0\\
\enddata
\end{deluxetable*}

\subsubsection{Redshift uncertainties}

Of the 983 $Q=3,4$ or 5 AAOmega observations, there were 38 galaxy spectra with 
repeat observations. These repeat observations were used to check the 
uncertainties assigned by {\it RUNZ} to redshift measurements.
We use a robust biweight estimator 
to determine the mean and dispersion of the distribution of redshift 
differences, where the redshift measurement with the lower quality (defined by 
ether its lower $Q$ value, or higher error measurement) is subtracted from the 
higher quality measurement. We find a mean difference of 
$\overline{\Delta cz}=1\pm13$\kms\ and a dispersion of $87\pm21$\kms\ from which 
we can infer a single redshift measurement uncertainty of $61\pm15$\kms. The 
median value of the individual redshift uncertainties assigned by {\it RUNZ} to 
non-stellar objects was 66\kms\ (after excluding 6 outliers with error values 
greater than 1000\kms), which is consistent with the uncertainty derived from 
the repeat observations.

Independent checks on the redshifts and their uncertainties can be made by 
cross-correlating the AAOmega catalog with the \citet{couch1987}, 
\citet{couch1998}, \citet{boschin2006} and \citet{braglia2009} redshift 
catalogs. Further to these catalogs, we add archival 2dF data, which is 
reduced using 2dFDR and redshifted using the RVSAO package in IRAF \citep{kurtz1998}.
The results of the comparisons are presented in Table~\ref{czdiffs} 
where the columns are: Origin of external redshift catalog for comparison 
(Col. 1); number of redshifts in common with the AAOmega catalog, $N_{match}$,
where catalog members are deemed to be matches if their positions differ by 
less than 3\arcsec (Col. 2), the mean redshift difference, 
$\overline{\Delta cz}=\overline{cz_{ours}-cz_{ext.}}$, as determined by the 
biweight estimator (Col. 3), the dispersion in the $\Delta cz$ distribution, 
$\sigma ({\Delta cz})$, determined with the biweight estimator (Col. 4), the 
redshift uncertainty for the external redshift measurements (Col. 5), the 
quadrature difference between $\sigma ({\Delta cz})$ and the external redshift 
error which gives an external estimate of the uncertainty on the AAOmega 
measurements (Col. 6) and the ratio of the external uncertainty measurement to 
the internal uncertainty measurement derived above (Col. 7). We note that the results in 
Table~\ref{czdiffs} show our internal redshift uncertainties are systematically 
underestimated by a factor of $\sim1.8$. We conclude that the redshift precision
of the AAOmega measurements are $\sim 110$\kms.

\begin{deluxetable*}{ccccccc}
\tabletypesize{\scriptsize}
\tablecolumns{3}
\tablewidth{0pc}
\tablecaption{Comparison of AAOmega redshift measurements with literature 
values. The errors presented for $\overline{\Delta cz}$ and $\sigma (\Delta cz)$
are determined from the $68\%$ confidence limits based on 10000 bootstrap 
resamplings of the data.\label{czdiffs}}
\tablehead {\colhead{Literature source} & \colhead{$N_{match}$ }
&\colhead{$\overline{\Delta cz}$} & 
\colhead{$\sigma (\Delta{cz})$}& \colhead{Literature redshift}& \colhead{Quadrature difference}&ratio\\
&   &  \colhead{\kms}&&\colhead{uncertainty \kms}&\colhead{\kms}&\colhead{\kms}}
\startdata
\citet{braglia2009} & 60 & $88^{+39}_{-38}$ & $289^{+46}_{-40}$ & 276 & $86\pm156$&$1.4\pm2.6$\\
\citet{boschin2006} & 56 & $-18^{+23}_{-24}$ & $172^{+19}_{-18}$ & 90 & $147\pm22$&$2.4\pm0.7$\\
\citet{couch1998} & 29 & $177^{+24}_{-24}$ & $125^{+23}_{-18}$ & 90 & $87\pm32$&$1.4\pm0.6$\\
\citet{couch1987} & 28 & $-89^{+27}_{-27}$ & $151^{+24}_{-26}$ & 100 & $113\pm32$&$1.5\pm0.7$\\
2df 2003          & 24 & $-53^{+27}_{-26}$ & $136_{-23}^{+20}$ & 50 & $127\pm 21$& $2.1\pm0.6$\\
\enddata
\end{deluxetable*}

\subsubsection{Combined Redshift Catalog}
As can be seen from Table~\ref{czdiffs}, there are significant offsets from 
a mean of zero for all of the $\Delta cz$ distributions, aside from the 
comparison with the \citeauthor{boschin2006} results. In order to ensure 
the external redshift measurements are consistent with the AAOmega ones, 
prior to combining the catalogs we add the pertinent value of  
$\overline{\Delta cz}$ to each redshift measurement in the 
\citeauthor{braglia2009}, \citeauthor{couch1998} and \citeauthor{couch1987} 
catalogs. When combining the catalog, for consistency we always use 
AAOmega redshifts where an object has multiple redshift measurements.
Where an object has multiple AAOmega measurements, we choose the measurement with
the highest $Q$ value, and where the $Q$ values are equal, we choose the
measurement with the lower redshift uncertainty. Our final combined catalog
contains 1237 extragalactic objects with robust redshift measurements
within a 15\arcmin\, radius of the cluster. The position, \rf\ magnitude,
redshift, redshift uncertainty and redshift source for all galactic and
extragalactic objects in our catalog are tabulated in Table~\ref{cztab}.

\begin{deluxetable*}{ccccccc}
\tabletypesize{\scriptsize}
\tablecolumns{3}
\tablewidth{0pc}
\tablecaption{Combined redshift catalog.\label{cztab}}
\tablehead{\colhead{R.A. (J2000)} & \colhead{decl. (J2000)} & \colhead{\rf\ magnitude} & 
\colhead{($cz$)} & \colhead{$cz$ uncertainty} & \colhead{$cz$ Quality} & \colhead{$cz$ source}\\
    deg.      &     deg.       &          &   \kms     &    \kms  &            }
\startdata
3.296875  & -30.391694 & 20.69 &  400971.27 &  314.78    &    4   &           AAT/AAOmega \\
3.303542  & -30.358722 & 19.62 &      99.13 &  113.92    &    6   &           AAT/AAOmega \\
3.303875  & -30.404806 & 20.55 &   56010.02 &   20.99    &    5   &           AAT/AAOmega \\
3.304125  & -30.367528 & 19.47 &     234.52 &   65.95    &    6   &           AAT/AAOmega \\
3.305000  & -30.431861 & 20.75 &     103.98 &   65.95    &    6   &           AAT/AAOmega \\
3.305750  & -30.351361 & 20.48 &   28873.99 &   20.99    &    3   &           AAT/AAOmega \\
3.305833  & -30.339167 & 20.27 &   92533.37 &   23.98    &    5   &           AAT/AAOmega \\
3.306250  & -30.380945 & 22.20 &   63794.34 &  276.00    &    3   & Braglia et al. (2009) \\
3.308583  & -30.363083 & 19.57 &   78171.58 &   23.98    &    5   &           AAT/AAOmega \\
3.310292  & -30.324250 & 19.25 &   93436.37 &   74.95    &    4   &           AAT/AAOmega \\
3.313542  & -30.398889 & 18.91 &   91532.63 &   23.98    &    5   &           AAT/AAOmega \\
3.314292  & -30.344611 & 18.90 &   92477.01 &  113.92    &    4   &           AAT/AAOmega \\
3.313542  & -30.398889 & 18.91 &   91532.63 &   23.98    &    5   &           AAT/AAOmega \\
3.314292  & -30.344611 & 18.90 &   92477.01 &  113.92    &    4   &           AAT/AAOmega \\
3.314583  & -30.472861 & 19.14 &   92390.06 &   86.94    &    4   &           AAT/AAOmega \\
\enddata
\tablecomments{This table is available in its entirety in a machine-readable form in the online
journal. A portion is shown here for guidance regarding its form and content.}
\end{deluxetable*}

\subsubsection{Spectroscopic Completeness}\label{sec_compl_map}

\begin{figure*}
{\includegraphics[angle=-270,width=.48\textwidth]{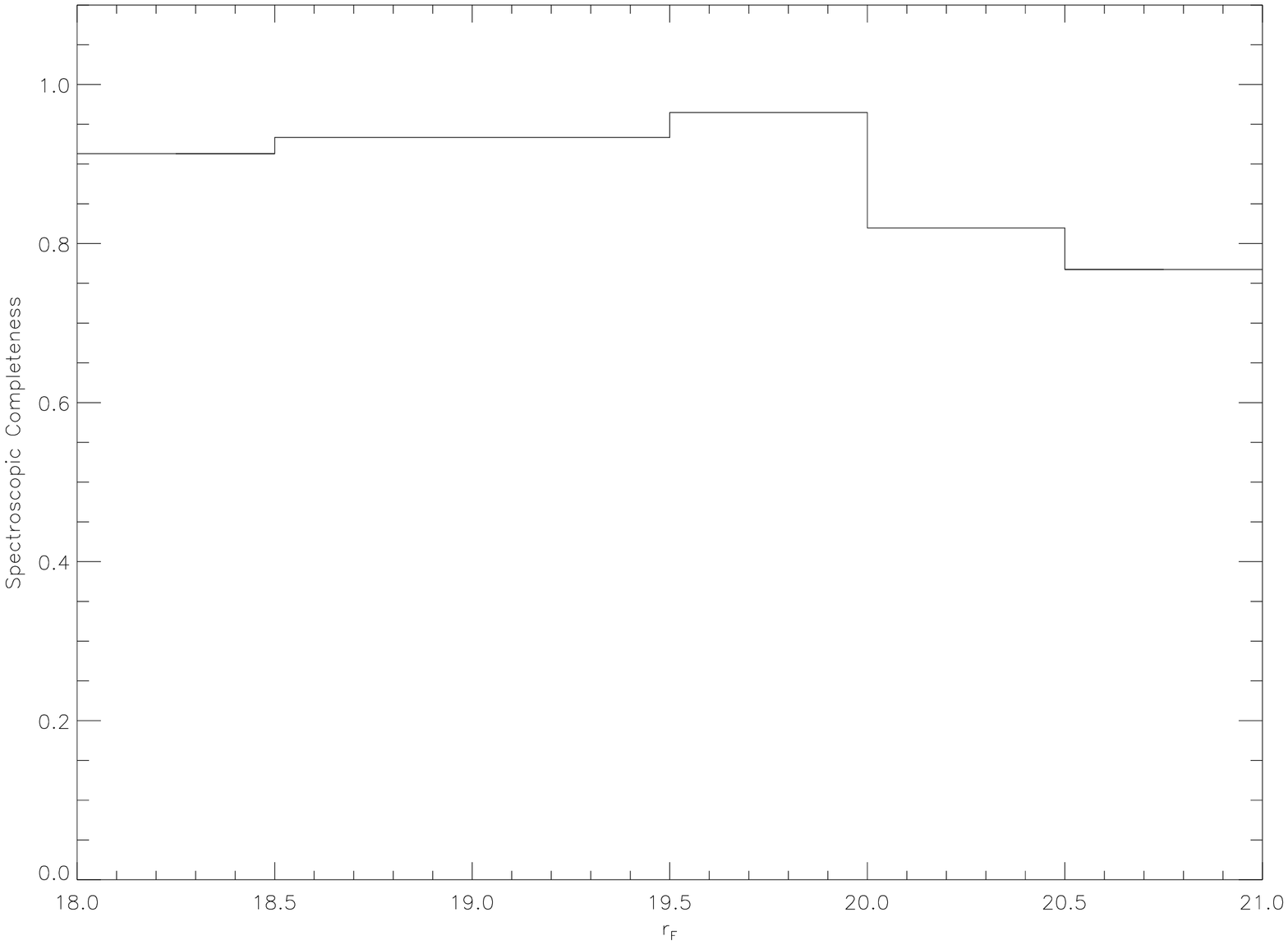}}
{\includegraphics[angle=-270,width=.48\textwidth]{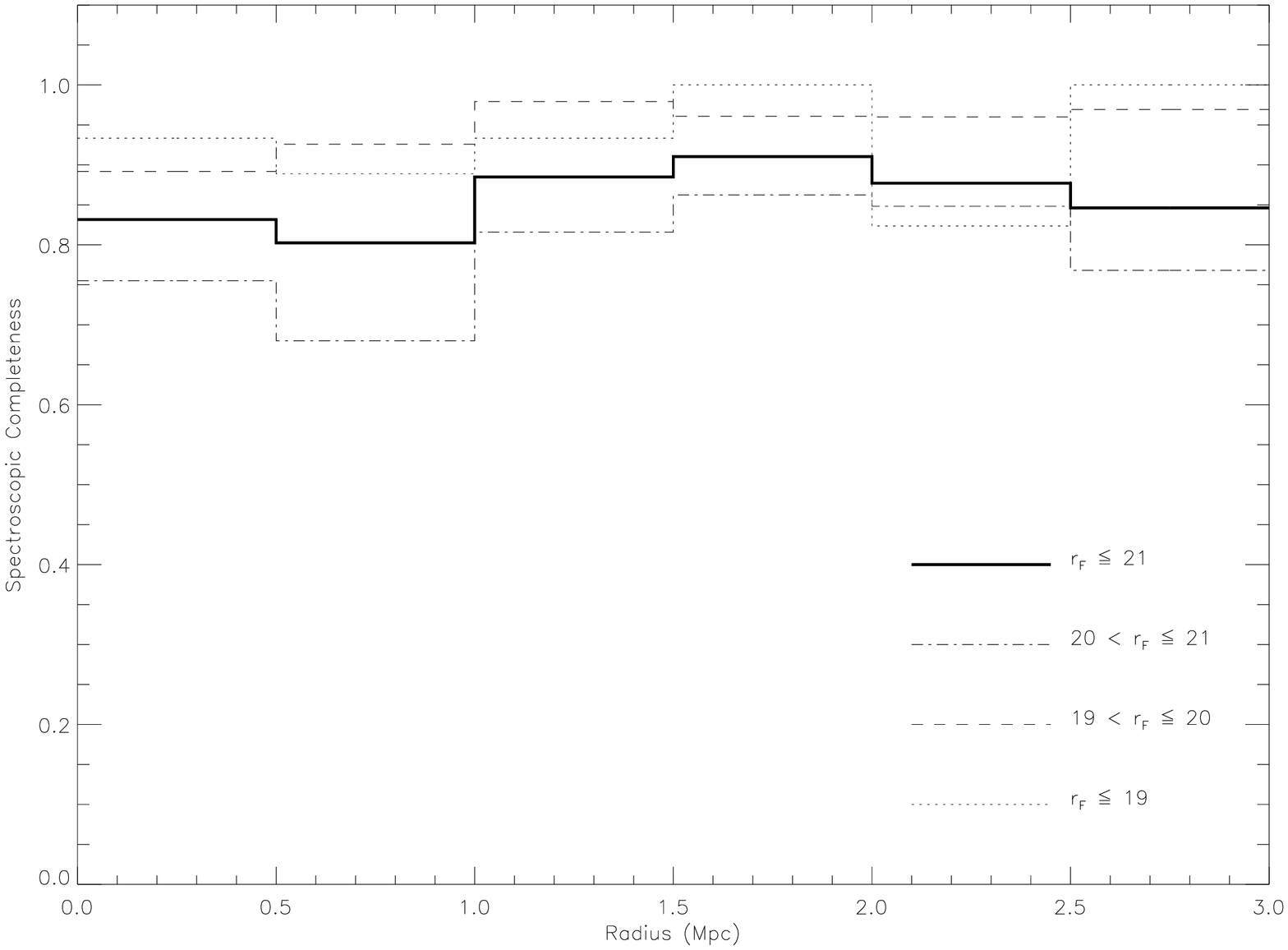}}
\caption{Spectroscopic completeness as a function of magnitude (left) and as a 
function of radius (right). Also plotted in the right panel is the 
spectroscopic completeness as a function of radius for the magnitude ranges 
indicated at the lower right of the figure.}
\label{compl}
\end{figure*}

In the forthcoming sections, we will use the redshift catalog to search for 
substructure in Abell~2744. Therefore, in order to be confident in the 
results, it is important that the spectroscopic completeness is well understood,
particularly for the interpretation of maps of the projected density of the
member galaxies. We use the combined redshift catalog 
to determine the spectroscopic completeness, which is defined as the ratio of 
the number of objects with reliable redshift determinations to the number of 
objects in the parent photometric catalog. In Figure~\ref{compl} we present 
the spectroscopic completeness as a function of radius, \rf\ magnitude and a 
combination of radius and \rf\ magnitude. The plots show that we have obtained
a high level of completeness at all cluster-centric radii and for all magnitude
ranges. However, the spectroscopic completeness is systematically lower within 
the central $\sim 1$\,Mpc (although we note it is still excellent and remains 
above $\sim 70\%$ for all magnitude ranges). This is due to a combination of 
fiber separation limits ($\sim 30$\arcsec) and the high galaxy surface density 
in the central regions, which means a small number of objects could not be 
observed.

\begin{figure*}
\vspace{10pt}
{\includegraphics[angle=-0,width=.48\textwidth]{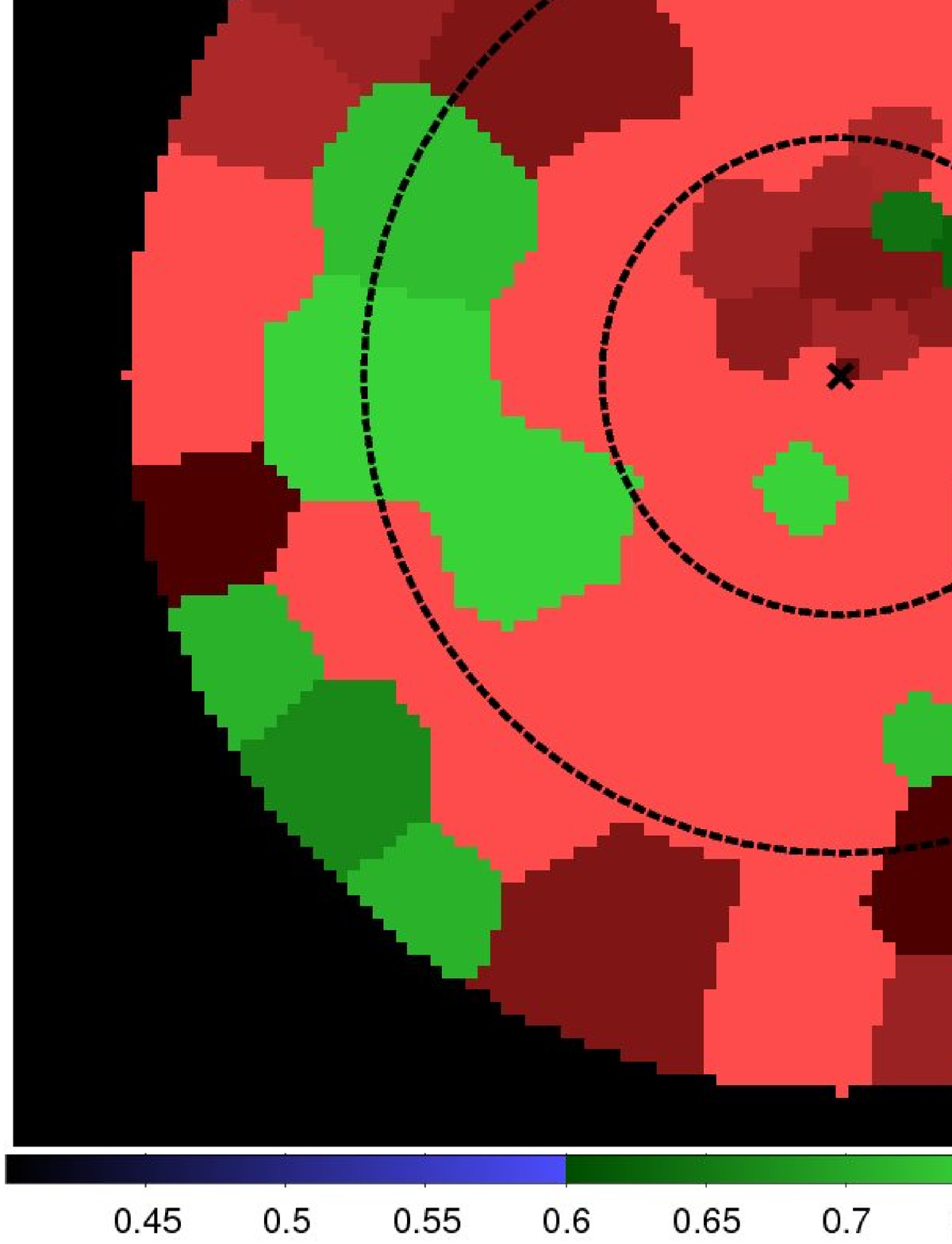}}
{\includegraphics[angle=-0,width=.48\textwidth]{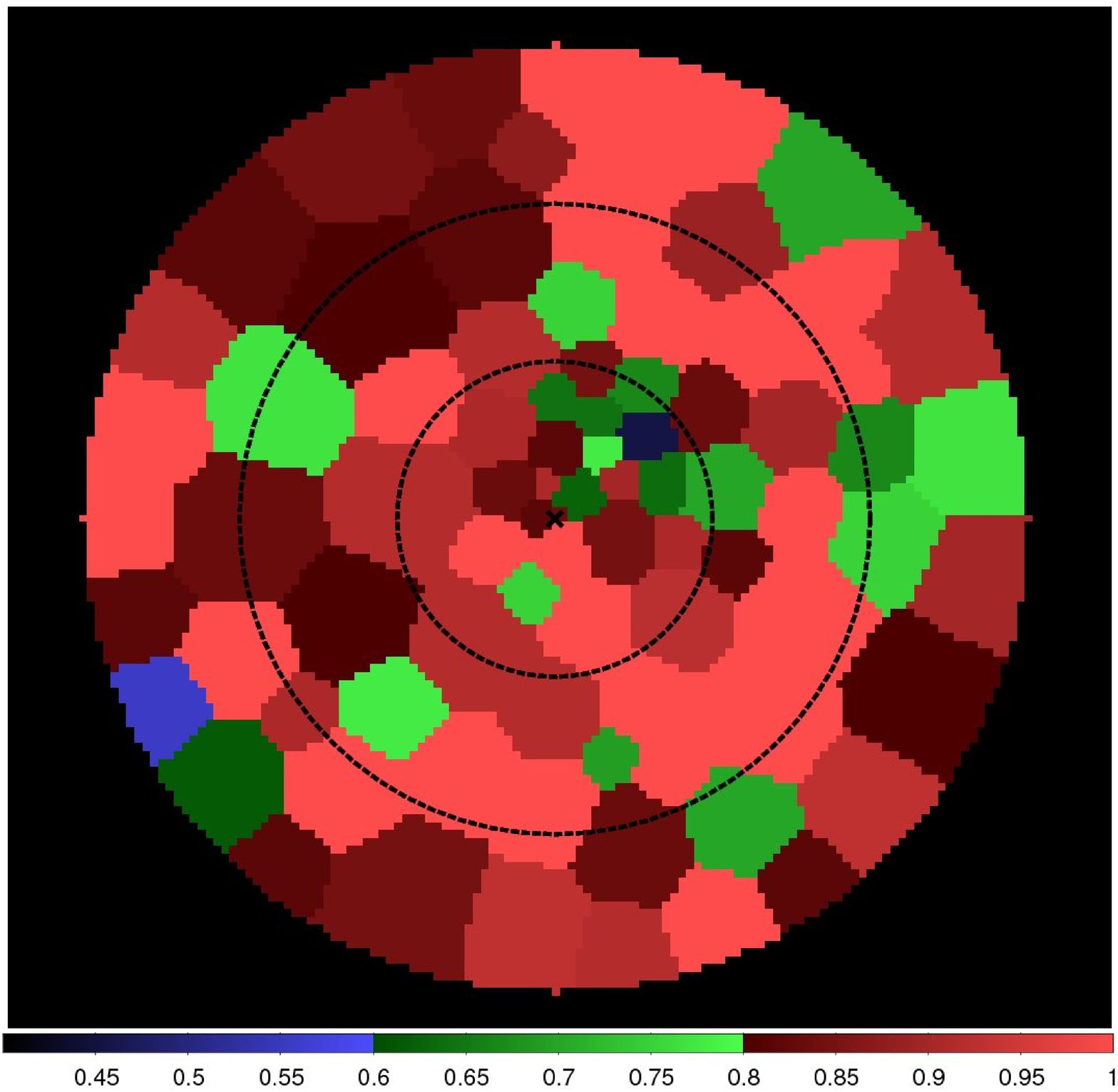}}
\caption{Left panel: Spectroscopic completeness map for $r_{\rm F}<20.5$. Right 
panel: Spectroscopic completeness map for $r_{\rm F}<21$. The black dashed 
circles have radii 1 and 2\,Mpc while the black cross marks the position of the 
BCG.}
\label{compl_map}
\end{figure*}

Figure~\ref{compl_map} shows two images of the spatial distribution of the 
spectroscopic completeness for the magnitude limits \rf $< 20.5$ (left 
panel) and \rf $< 21$ (right panel). The images were generated by first 
producing an adaptively binned image of the spatial distribution of the galaxies
in the photometric catalog where each spatial bin contains $\sim 10$ galaxies.
We made use of the WVT binning algorithm by \citet{Diehl2006}, which is a 
generalization of  \citet{Cappellari2003}'s Voronoi binning algorithm. 
The same spatial binning was then applied to the galaxies in the spectroscopic 
catalog to produce a second image. The ratio of the binned spectroscopic to 
photometric images gives the spectroscopic completeness map in 
Figure~\ref{compl_map}. The maps reveal the high level of spectroscopic 
completeness we have obtained with our observations and, in particular, for 
\rf $< 20.5$ we obtain $\gtrsim 90\%$ completeness over the majority of the 
6\,Mpc diameter field. For \rf $< 21$ we obtain $\gtrsim 80\%$  completeness
over the majority of the 6\,Mpc diameter field. The systematically lower 
spectral completeness within 1\,Mpc noted in the radial profile shown in 
the right panel of Figure~\ref{compl} is due to a region of low completeness
confined to the north west of the center. This patch of lower spectral 
completeness needs to be kept in mind in the forthcoming sections.

\subsubsection{Cluster Membership}\label{clust_member}

\begin{figure*}
{\includegraphics[angle=0,width=0.48\textwidth]{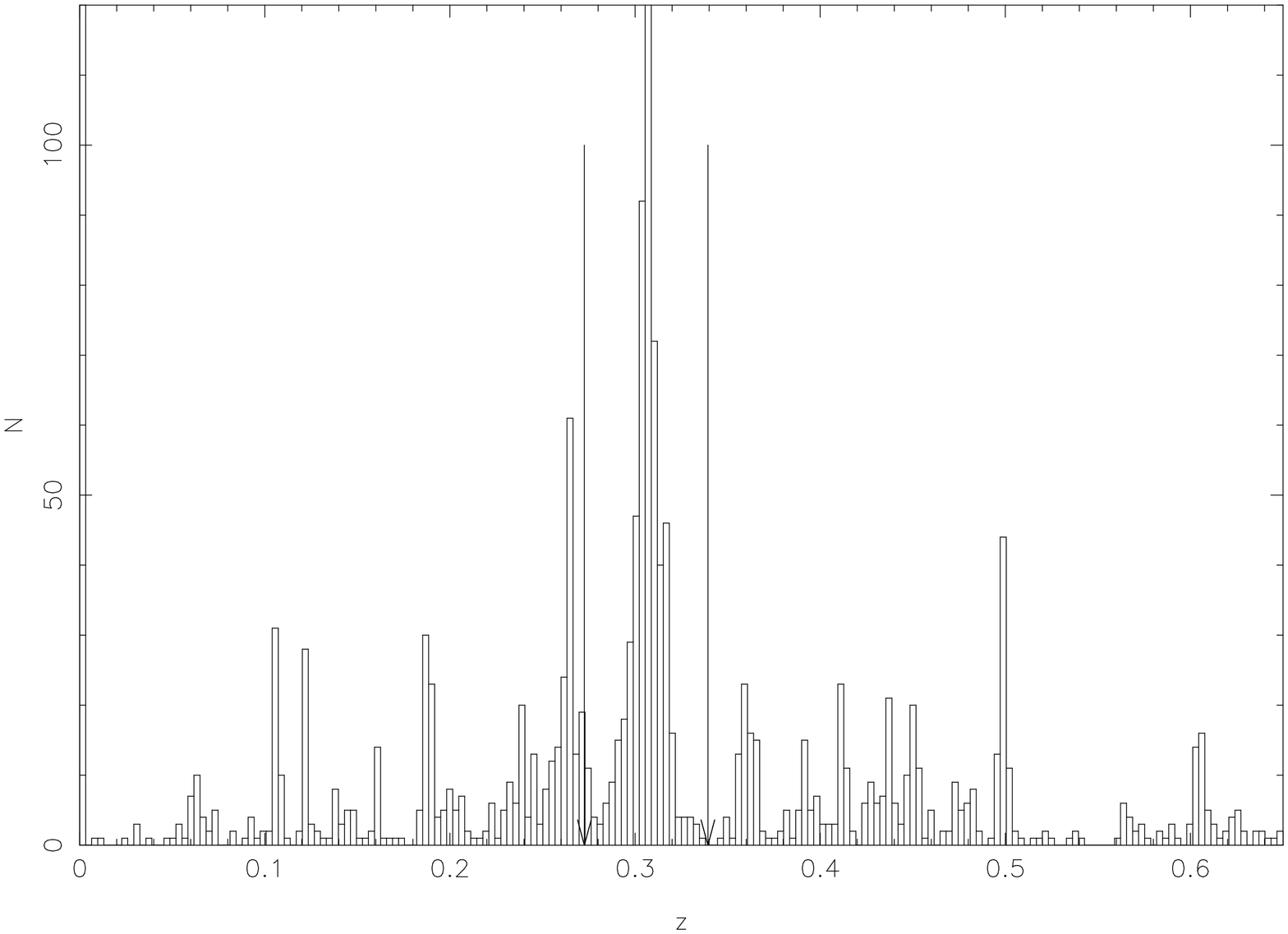}}
{\includegraphics[angle=0,width=0.48\textwidth]{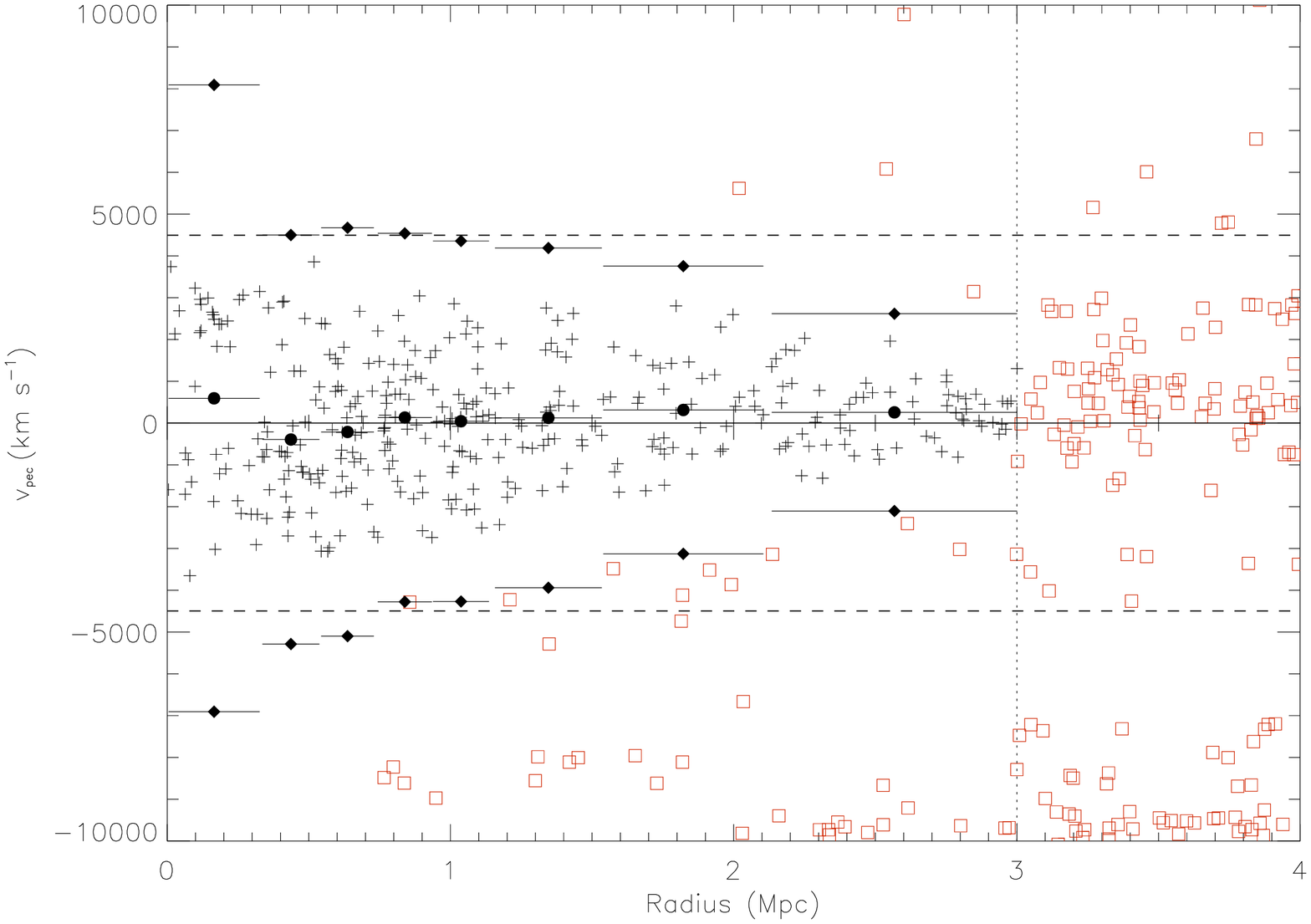}}
\caption{{\it Left panel:} A histogram of all the reliable redshifts measured 
within the Abell~2744 field. Arrows mark the initial cut in velocity used to 
crudely isolate the cluster in redshift-space. {\it Right panel:} Shift gapper 
plot for refined member allocation. The {\it black} crosses represent galaxies 
allocated as cluster members while the {\it orange} open squares are rejected 
foreground and background galaxies lying close to the cluster in redshift 
space.}
\label{allhisto}
\end{figure*}

Robustly distinguishing cluster members from foreground and background 
interlopers is a critical first-step when conducting a dynamical analysis. 
Membership allocation is a two-step process where the initial step involves 
crudely separating cluster members and projected interlopers in redshift-space 
only. This is achieved by selecting galaxies within the redshift window of 
$cz_{clus} \pm 10000$\kms\ \citep[where $z_{clus}=0.306$;][]{boschin2006} which, 
as the left panel in Figure~\ref{allhisto} shows, defines a peak in redshift 
space associated with Abell~2744. The second part of the process refines the 
membership allocation
by using a slightly modified version of the ``shifting gapper'' first employed 
by \citet{fadda1996}. This method utilizes both radial and peculiar velocity 
information to separate interlopers from members as a function of 
cluster-centric radius (defined with respect to the bright cluster galaxy 
[BCG] closest to the X-ray peak at R.A.=$00^{\rm h}14^{\rm m}20^{\rm s}.738$, decl.=$-30^{\circ}23\arcmin 
59\arcsec.90$) and is described in detail in \citet{owers2009a}. 
Briefly, the data are binned radially such that each bin contains at least 40 
objects. Within each bin galaxies are sorted by their peculiar velocity and 
interlopers are rejected based on the size of the gap in velocity between 
adjacent (in peculiar velocity) objects. The limit of 40 objects per bin is somewhat 
subjective and is chosen to minimize both the radial annulus used and the
likelihood of the algorithm rejecting either high velocity bona fide cluster members or 
coherent substructures associated with merger activity. Binning by 40 objects 
produces the best results, as shown in the right panel of 
Figure~\ref{allhisto} where we present the results of the shifting gapper
membership allocation procedure. Figure~\ref{allhisto} 
shows that within a cluster-centric radius of 3\,Mpc the cluster members are 
well separated from the surrounding interlopers. Beyond 3\,Mpc the separation
is less-clear, so we exclude the galaxies at these larger radii during the 
analysis. After culling the interlopers, 343 bona fide cluster members remain
from which we determine a cluster redshift of $z_{clus}=0.3064\pm0.0004$ and
a velocity dispersion of $\sigma_{clus}=1497\pm57$\kms\ using biweight estimators
of location and scale \citep{beers1990}, where the quoted errors are $1\sigma$
values determined using the jackknife resampling technique.

Also plotted in the right panel of Figure~\ref{allhisto} are the means (filled 
circles) and 
$3\sigma$ limits (filled diamonds) for each radial bin, determined using 
biweight estimators, showing that all excluded interlopers have peculiar 
velocities exceeding the $3\sigma$ limit at the radial bin of interest. The 
inadequacy of a simple $3\sigma$ clipping using only velocity information is 
highlighted in the right panel of Figure~\ref{allhisto}, where the 
dashed-lines show that the $3\sigma_{clus}$ limits fail to reject several 
interlopers at larger cluster-centric radii.

\section{Substructure Detection}\label{substructure_detection}

The primary motivation for this paper is to detect and characterize substructure
in order to understand the merger history of Abell~2744. While substructure tests that
use all available information (position on the sky and the line of sight velocity)
are more decisive \citep{dressler1988,west1990,pinkney1996, burgett2004}, the
results of simpler analyses are easier to comprehend. In
this section, we begin with a one-dimensional analysis of the velocity 
distribution, which is most sensitive to detecting substructure caused by 
line-of-sight mergers where the velocity difference is large. We then search for
substructure in the two-dimensional galaxy distribution, which is most sensitive
to detecting well-separated mergers occurring in the plane of the sky. Finally,
we combine the spatial and velocity information to search for local deviations
from the global velocity distribution which indicate dynamical substructures.

\subsection{Substructure in the Velocity Distribution}\label{veldist}

Cluster mergers can induce significant distortions in the observed velocity
distribution which, when a cluster is dynamically relaxed, is well approximated
by a Gaussian. Therefore, searching for evidence of substructure using 
velocity information is generally accomplished by testing for departures 
from a Gaussian shape. Here, we use the method first outlined by 
\citet{zabludoff1993} which approximates the velocity distribution by the 
summation of three Gauss-Hermite functions. The method is described in detail in
\citet{owers2009a}. The coefficient of the first term, $h_0 \approx 1$, 
multiplies the best-fitting Gaussian contribution to the distribution while the 
second and third terms, with coefficients $h_3\ {\rm and}\ h_4$, approximate the
third- and fourth-order asymmetric and symmetric deviations from a Gaussian 
shape, similar to skewness and kurtosis measurements but less sensitive to 
outliers in the distribution. For Abell~2744, we find the coefficients have 
values of $h_3 = 0.07$ and $h_4 = 0.06$. The significance of these values is 
determined by generating 10,000 Monte Carlo realizations of Gaussians with
$N=343$ data points having mean and standard deviation equal to the best-fitting
value derived from the data. We find that values of $|h_3|>0.07$ occur in only
$6\%$ of the realizations, while  values of $|h_4|>0.06$ occur $8\%$ of the 
time. Thus, there appear to be mildly significant symmetric and asymmetric 
deviations from Gaussianity present in Abell~2744's velocity distribution. 

The positive value for the $h_4$ term is likely to be due to the contribution
of galaxies in the cluster outskirts, which follow more radial orbits and hence
produce a more peaked velocity distribution \citep{merritt1987}, rather than the
presence of substructure. Because we are interested in 
detecting structure related to a cluster merger and not due to orbital 
anisotropy, we re-determine the Gauss-Hermite coefficients using only
data from the central 1.5\,Mpc region. There are 238 galaxies within this region
in our sample and for these we measure $h_3 = 0.12$ and $h_4 = -0.03$. The
observed $h_3$ coefficient is highly significant with $|h_3|>0.12$ occurring 
in less than $1\%$ of the realizations, while the $h_4$ term is not significant 
and $|h_4|>0.03$ occurs in $47\%$ of the realizations, verifying that the 
mildly significant positive $h_4$ value measured above is indeed due to the 
contribution of the galaxies at larger cluster-centric radii. 

\begin{figure*}
{\includegraphics[angle=0,width=0.33\textwidth]{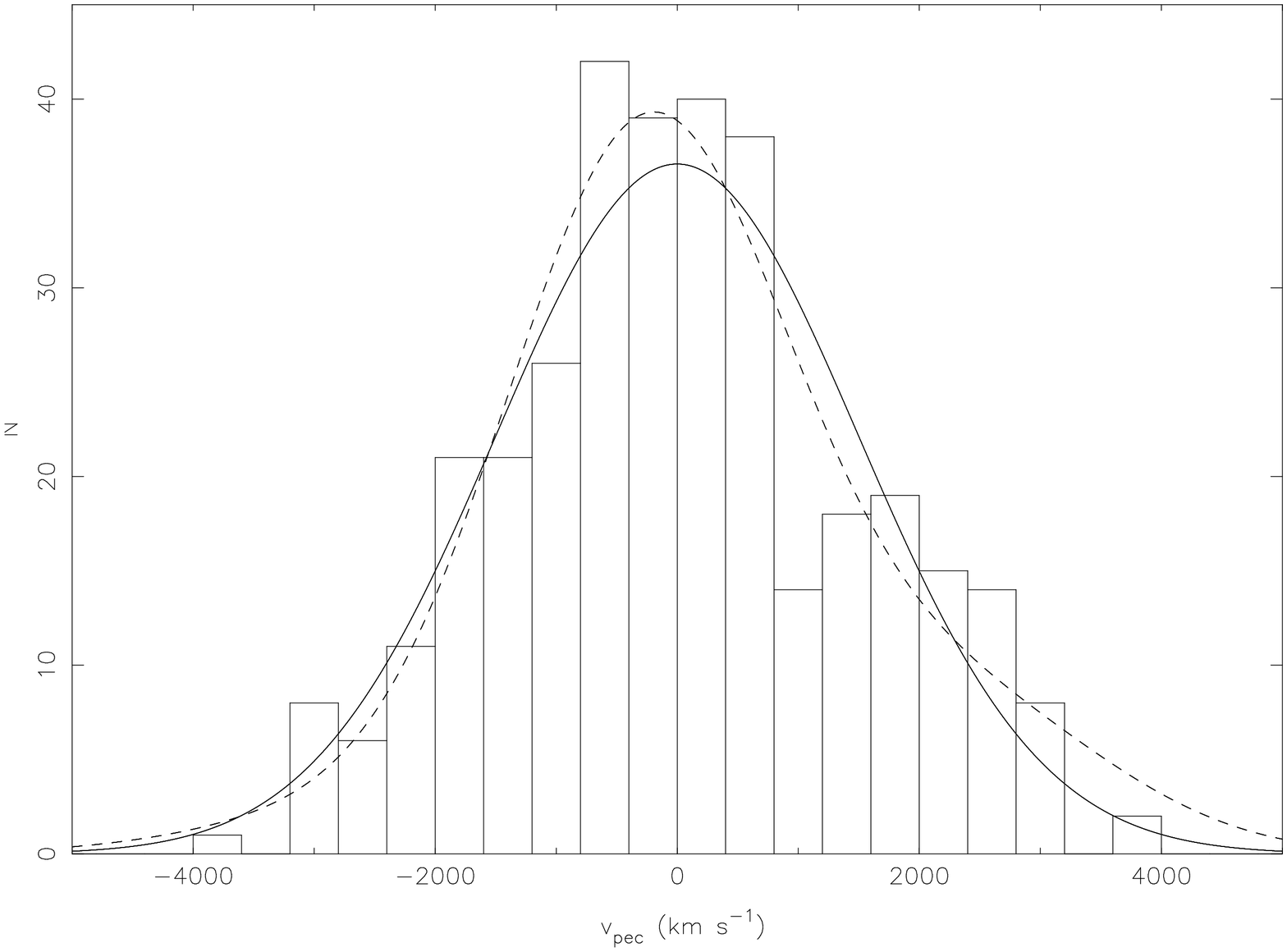}}
{\includegraphics[angle=0,width=0.33\textwidth]{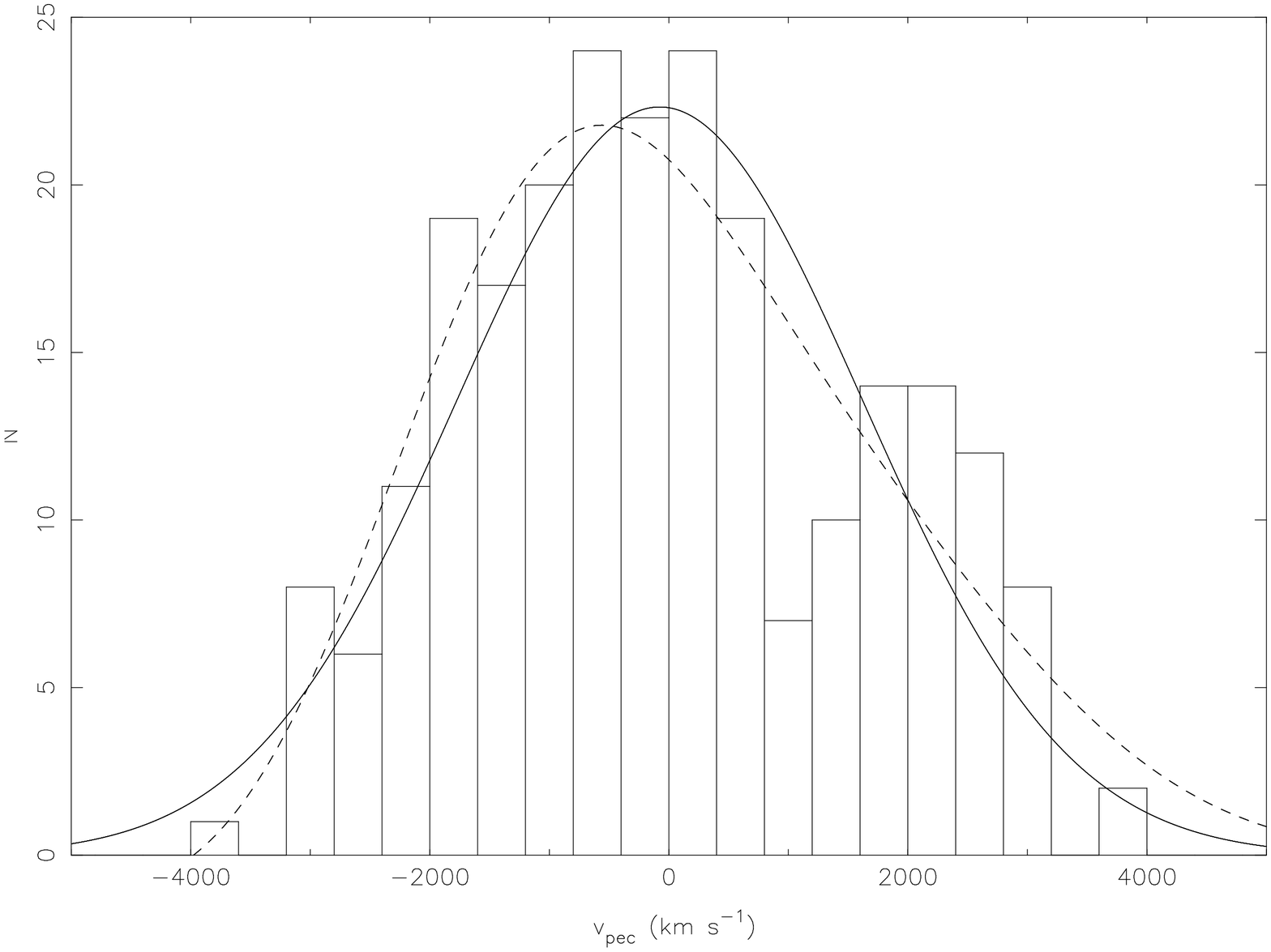}}
{\includegraphics[angle=0,width=0.33\textwidth]{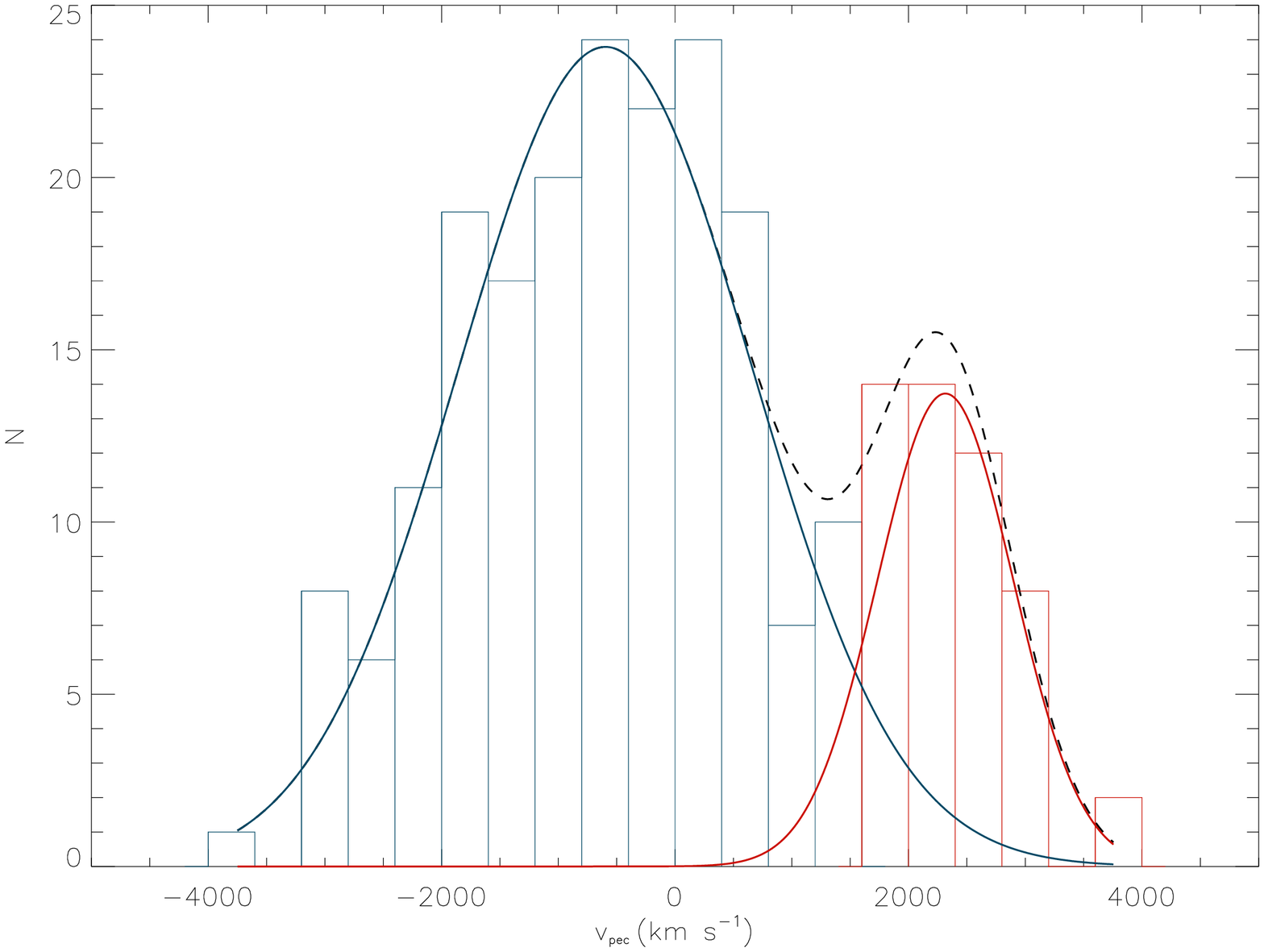}}\\\\
\caption{{\it Left panel:} Peculiar velocity histogram for all 343 members. The 
solid line shows a Gaussian with $\overline{v_{pec}}=0$km/s and $\sigma=1497$\kms\
and the dashed-line shows the Gauss-Hermite reconstruction of the velocity
distribution. {\it Middle panel:} Same as the {\it left panel} for the 238 
members within a 1.5\,Mpc radius of the cluster center. Here, the plotted 
Gaussian (solid line) has $\sigma=1701$\kms.{\it Right panel:} The results of 
the KMM partition for the members within 1.5\,Mpc. The blue Gaussian has 
$\overline{v_{pec}}=-596$km/s and $\sigma=1261$\kms\ and the red Gaussian has  
$\overline{v_{pec}}=2316$km/s and $\sigma=581$\kms. The dashed line shows the
combined red and blue Gaussians.}
\label{velhisto}
\end{figure*}

As noted previously, the velocity distribution in Abell~2744 is bi-modal 
\citep{girardi2001,boschin2006,braglia2007}. This bi-modality causes the
significantly positive $h_3$ term and Figure~\ref{velhisto} clearly shows a 
secondary substructure with a peak at $v_{pec} \sim 2000$\kms. To further 
quantify the significance of this bi-modality, we use the KMM algorithm 
\citep[Kaye's Mixture Model;][]{ashman1994} to fit two Gaussians to the velocity
distribution of the members residing within a cluster-centric radius of 
1.5\,Mpc. As input for the main component, we set a mean $\mu_{1,init}=0$,
dispersion $\sigma_{1,init}=1300$\kms\ and estimate the fractional contribution 
of the number of galaxies belonging to the main component with respect to the 
entire distribution is $f_{1,init}=0.7$. Likewise, for the secondary component 
we set $\mu_{2,init}=2000$\kms, $\sigma_{2,init}=500$\kms\ and $f_{2,init}=0.3$. As
output, the algorithm returns $\mu_{1,out}=-596$\kms, $\sigma_{1,out}=1261$\kms\ 
and $f_{1,out}=0.79$ for the main component and $\mu_{2,out}=2316$\kms, 
$\sigma_{2,out}=581$\kms\ and $f_{2,out}=0.21$ for the secondary component. It is
noted that the outputs generated by the KMM algorithm are robust to changes in
the initial input parameters. The resulting Gaussians are overplotted on the 
observed velocity distribution in right panel of Figure~\ref{velhisto}. Since 
we are using the algorithm for a heteroscedastic (i.e., where the variances 
differ) case, the estimate of the significance of 
the likelihood ratio test statistic (LRTS) comparing a bi-modal fit to the 
null-hypothesis uni-modal fit given by the algorithm is unreliable 
\citep{ashman1994}. To reliably determine the significance of the LRTS, we 
produce 10,000 randomly sampled Gaussian distributions with 238 data points, 
mean -38\kms\ and dispersion 1632\kms\ (i.e., mean and standard deviation which 
best describes the observed velocity distribution). For each random dataset,
two Gaussians are fitted using the KMM algorithm, as was done for the 
observations, where the initial inputs are set to the best-fitting outputs for 
the observed velocity distribution listed above and the LRTS is computed. 
Comparison of the observed LRTS to the distribution of 10,000 LRTS produced by 
the random uni-modal Gaussian datasets allows the significance of the observed
LRTS to be reliably determined. The observed LRTS is larger than $99.99\%$ of
the LRTSs produced in the 10,000 realizations and we conclude that a bi-modal 
fit is significantly better than a uni-modal one.

\subsection{Substructure in the Spatial Distribution of Member Galaxies}\label{2dsubstructure}

\begin{figure*}
{\includegraphics[angle=90,width=\textwidth]{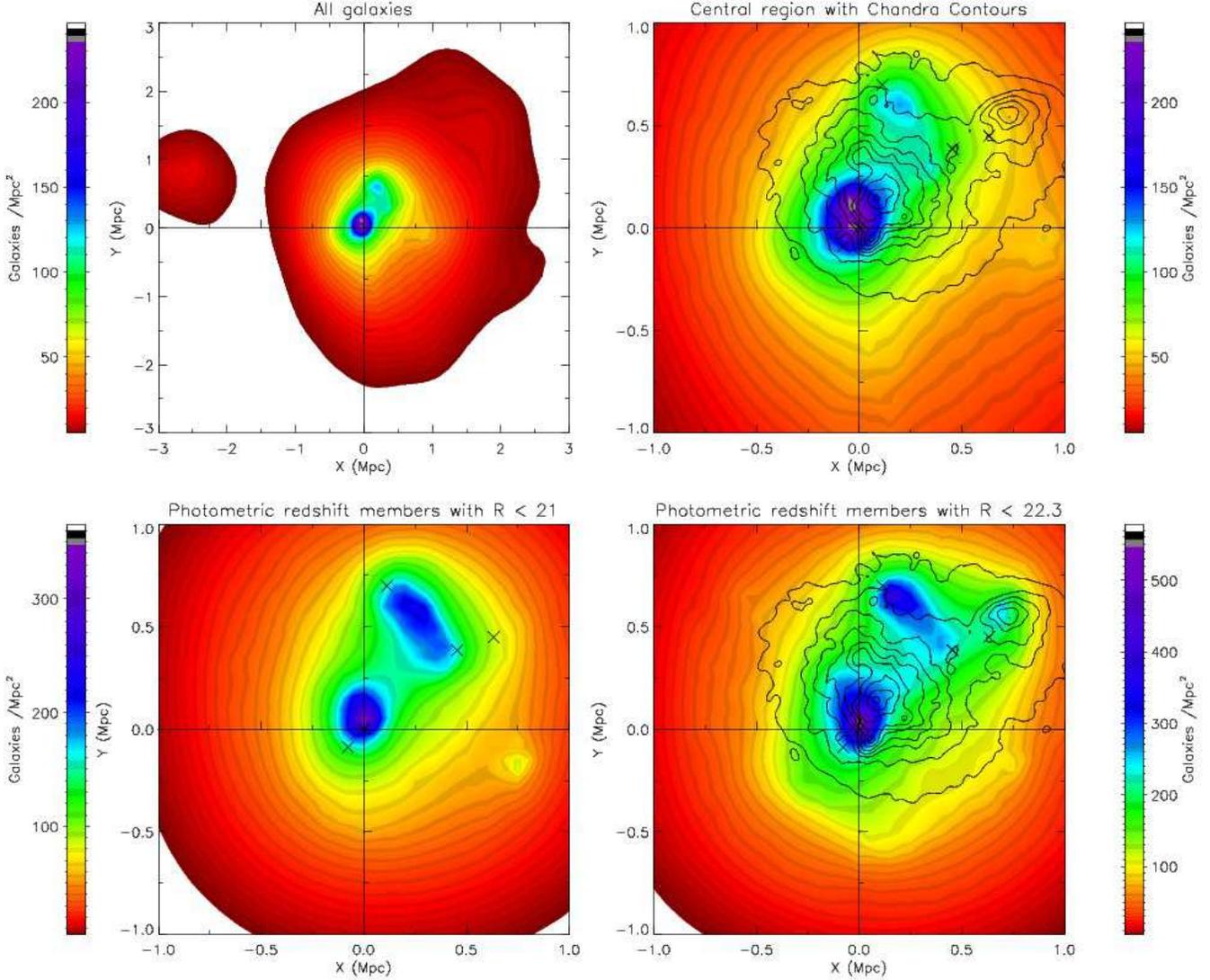}}
\caption{{\it Top left panel:} This panel shows the adaptively smoothed galaxy
surface density distribution for all spectroscopically confirmed cluster 
members. {\it Top right panel:} Close-up of the image in the {\it Top left 
panel} with \chan\ isointensity contours overlaid. Crosses mark the positions of
the 5 bright cluster members. {\it Bottom left panel:} Adaptively smoothed 
galaxy surface density for photometrically defined cluster members with 
magnitudes brighter than R=21 (see text). {\it Bottom right panel:} Same as 
{\it Bottom left panel} but for galaxies brighter than R=22.3. Note the
emergence of a substructure coincident with the X-ray substructure revealed by 
the \chan\ contours. In all panels, the color bars show the projected galaxy 
surface density scale in units of galaxies/${\rm Mpc}^2$.}
\label{galdense}
\end{figure*}

Having shown that the velocity distribution in Abell~2744 is bi-modal, we now
search for further evidence of merger-related substructure using 2D spatial 
information. To that end, we have used the positions of the 343 
spectroscopically confirmed members to produce maps of the smoothed  
galaxy surface density (Figure~\ref{galdense}).
We used a variable width Gaussian smoothing, where the width, 
$\sigma$, at each pixel is the projected distance to the tenth 
nearest galaxy, limited to a minimum of 90\,kpc and a maximum of 500\,kpc in
high and low density regions, respectively.
As detected previously \citep{boschin2006,braglia2009}, two 
distinct substructures inhabit the central regions of the cluster---the most
significant one is centered close the cluster center (which was arbitrarily 
defined in Section~\ref{clust_member} to be the position of the BCG nearest 
the peak in the X-ray surface brightness) and a second substructure 
$\sim 600$\,kpc to the north. At larger scales the distribution of galaxies 
appears smooth, apart from a structure $\sim 2.5$\,Mpc to the west and a 
mild asymmetry to the north-west.

\begin{figure*}
{\includegraphics[angle=-270,width=0.48\textwidth]{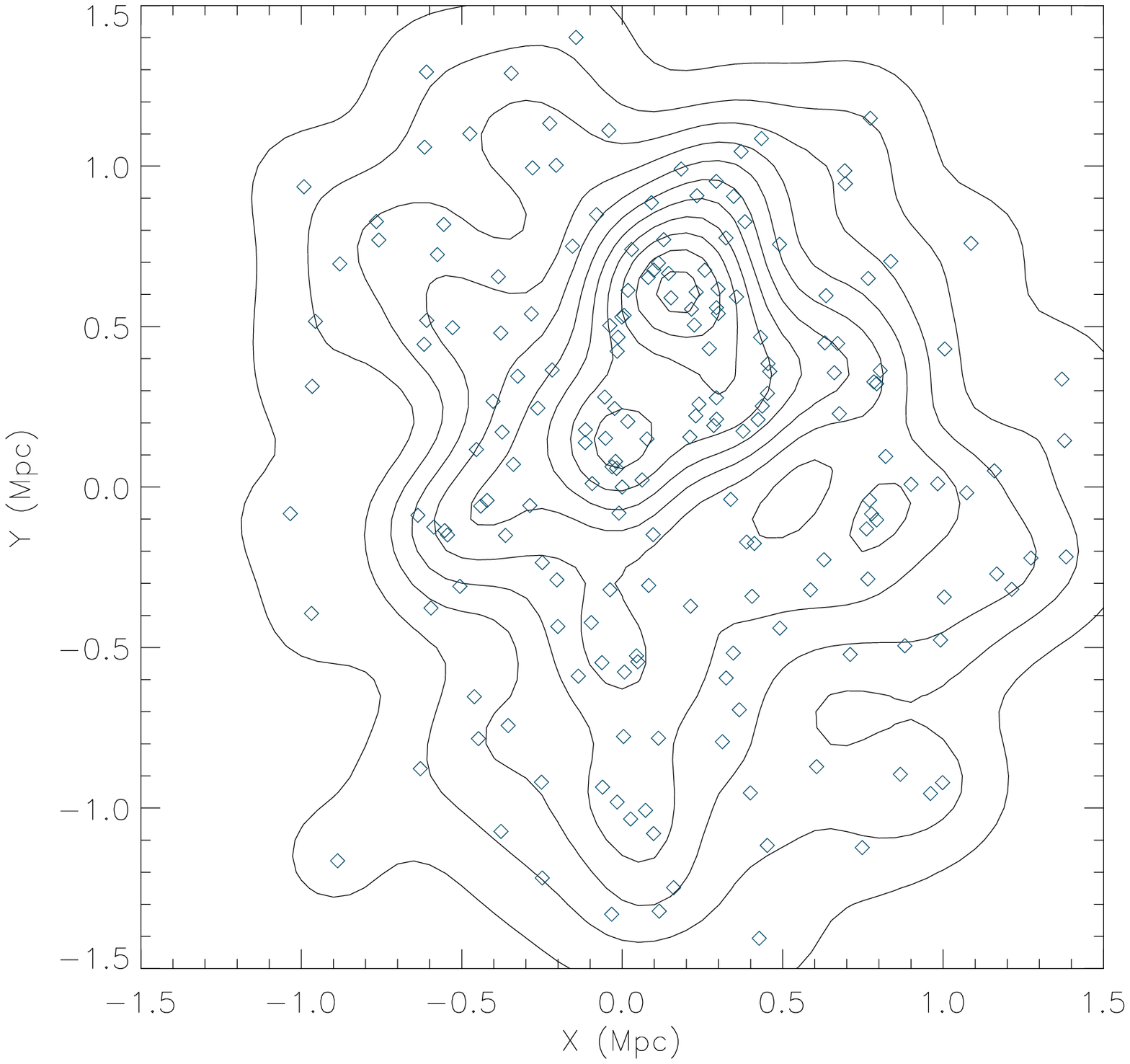}}
{\includegraphics[angle=-270,width=0.48\textwidth]{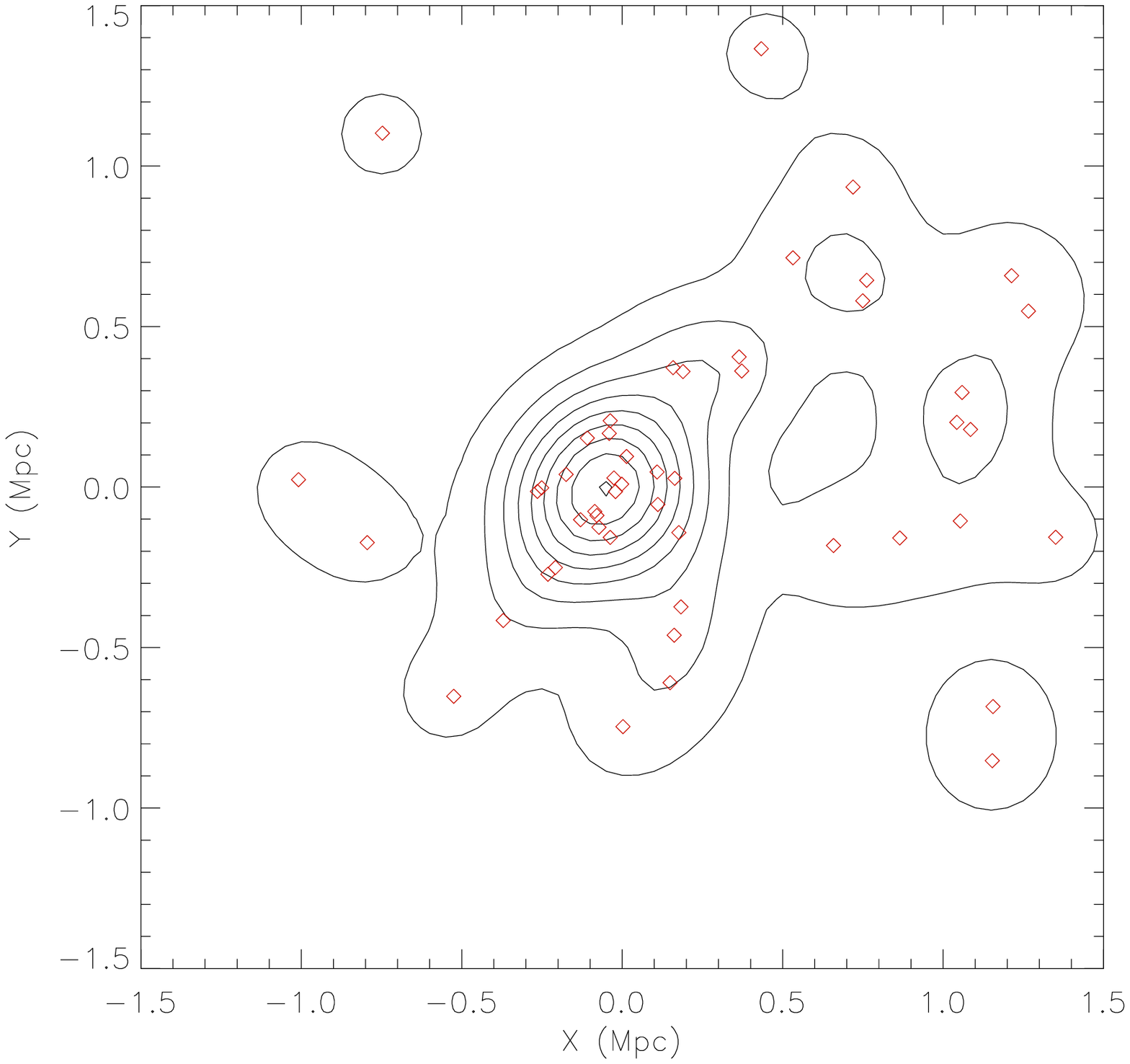}}
\caption{Spatial distribution of the two KMM velocity partitions. The positions
of the galaxies are marked by open diamonds which are color coded according to
the KMM partition they were allocated to (Figure~\ref{velhisto}). The
contours show the adaptively smoothed galaxy surface density levels, beginning
at 5 galaxies/Mpc and incrementing by 10 gals/Mpc.}
\label{kmmspatdist}
\end{figure*}

Given the bimodality detected in the velocity distribution, we search for
any correlation between the two velocity peaks and the two substructures
seen in the 2D galaxy density map for the spectroscopically confirmed members
(top panels of Figure~\ref{galdense}). To achieve this,
we plot separately the spatial distributions of the galaxies allocated
to the two KMM velocity partitions in Section~\ref{veldist} and smooth
the resulting distributions with an adaptive circular top hat filter
with radius set such that the smoothing region contains 10 galaxies. These
plots are shown in Figure~\ref{kmmspatdist} where the left and right panels
show the distributions of galaxies assigned to the lower and higher velocity 
partition, respectively. The spatial distribution of
the galaxies assigned to the dominant (both in terms of number of 
galaxies and velocity dispersion) lower velocity partition has two local 
peaks in galaxy surface density. The most significant of these two peaks
is located coincident with the northern substructure seen in 
Figure~\ref{galdense}, while the less significant one is approximately
coincident with the more central density peak seen in Figure~\ref{galdense}. 
The spatial distribution of the high velocity partition also has
its peak density approximately coincident with the central density 
peak seen in Figure~\ref{galdense}. This indicates that the density 
peak seen close to the center in Figure~\ref{galdense} is not the 
center of the cluster, but is caused by the alignment of two substructures 
separated by a large velocity and projected along our line of sight.

Inspecting the completeness map Figure~\ref{compl_map} reveals that galaxies 
in the region of the northern peak are underrepresented in our spectroscopic 
sample. Presumably, improving the sample completeness in this region would
enhance the local density peak. Thus the northern peak is likely to
be even more significant than it appears to be in Figure~\ref{kmmspatdist}.

The top right panel in Figure~\ref{galdense} shows a close up of the smoothed
member galaxy surface density in the central $1 \times 1$\,Mpc region with 
\chan\, X-ray contours overplotted. Neither of the local peaks in
galaxy surface density coincides with a peak in X-ray surface
brightness. Of particular concern is the lack of any significant 
overdensity coincident with the northwestern interloper where
\citet{boschin2006} and \citet{braglia2009} both  detected substructure using 
photometrically selected cluster members. As noted in Section~\ref{sec_compl_map}, 
the region just northwest of the center has the lowest spectroscopic 
completeness and this may be a contributing factor in the non-detection of 
substructure in this region. To test this, we obtain the photometric
catalog
\footnote{http://vizier.cfa.harvard.edu/viz-bin/VizieR?-source=J/A+A/389/787/}
of \citet{busarello2002}, which covers the central $2.3 \times 2.3$\,Mpc region,
and utilize their publicly available photometric redshifts to define cluster 
membership as per the definition in \citet{busarello2002}. The lower left
panel of Figure~\ref{galdense} shows the adaptively smoothed galaxy surface
density for the photometric members with $R \leq 21$ (where $70 \leq \sigma 
\leq 400$\,kpc). At this magnitude limit, which approximately matches our 
spectroscopic limit of \rf =21, there is qualitative agreement with the 
structures detected in the galaxy surface density map for the spectroscopically 
confirmed members. Therefore, the non-detection of a significant local 
overdensity coincident with the northwestern interloper, as
detected by \citet{boschin2006} and \citet{braglia2009}, is not due to the lower
spectroscopic completeness, but rather is due to the fact that the majority of
the galaxies contributing to the overdensity there are fainter than our limiting
magnitude. This can be seen in the lower right panel of Figure~\ref{galdense} 
where we have plotted the adaptively smoothed galaxy surface density for the 
photometric members down to the completeness limit of  $R = 22.3$ (where $50 
\leq \sigma \leq 400$\,kpc). This map reveals the presence of an overdensity
approximately coincident with the northwestern interloper, 
consistent with the maps of \citet{boschin2006} and \citet{braglia2009}. This
result is consistent with \citet{andreon2001} who, using deep $K-$band imagery, 
find significantly more dwarf galaxies and significantly fewer bright galaxies 
in this northwestern region when compared to the central cluster region (which 
is roughly coincident with our defined cluster center). Confirmation that this 
overdensity is part of the Abell~2744 system requires deeper spectroscopic 
observations.

\subsection{Substructure from 3D spatial plus velocity information}\label{3dsubstructure}

We now combine the spatial and velocity information to determine if there
is any correlation between the substructures detected in the 2D spatial and 1D 
velocity distributions above. This is achieved by making use of the 
$\kappa$-test which was successfully employed by \citet{colless1996} to detect 
substructure in the Coma cluster. The $\kappa$-test searches for local 
departures from the global velocity distribution around each cluster member 
galaxy. This is accomplished by using the Kolmogorov-Smirnov (KS) test 
which determines the likelihood that the velocity distribution of the 
$n=\sqrt N$ (where $N$ is the number of cluster members) nearest neighbors 
around the galaxy of interest and the global cluster velocity distribution, 
which includes the remaining $N-n$ galaxies, are drawn from the same parent 
distribution. The likelihood is determined by measuring the maximum separation
of the two cumulative distribution functions, $D_{obs}$ and determining the
probability that the $D$-statistic is larger than $D_{obs}$ for the given
sample sizes, $P_{KS}(D > D_{obs})$. The global measure of the dynamical 
substructure present within the cluster, $\kappa$, is determined by summing the 
negative log-likelihood for each galaxy over the entire sample, i.e.
\begin{equation}
\kappa = \sum -{\rm log}P_{KS}(D > D_{obs}).
\end{equation}
The significance of $\kappa$ is determined by performing 10,000 Monte Carlo 
realizations where, for each realization, the observed positions of the
galaxies are retained, while the galaxy velocities are shuffled randomly,
removing any correlation between positions and velocities. The observed 
$\kappa$ value is then compared 
to the distribution of the 10,000 Monte Carlo realizations of $\kappa$. A value 
as high as the observed value of $\kappa=481$ does not occur in the 10,000 
realizations, which have a distribution that is well modeled by a log-normal
distribution with $\mu(\ln \kappa)=4.95$ and $\sigma(\ln \kappa)=0.19$. 
Thus, the observed $\kappa$ value lies $\sim 6.6 \sigma$ from the mean of the
realizations and the upper limit on the probability of observing a value this 
high by chance is $10^{-4}$. 

\begin{figure*}
{\includegraphics[angle=-270,width=.9\textwidth]{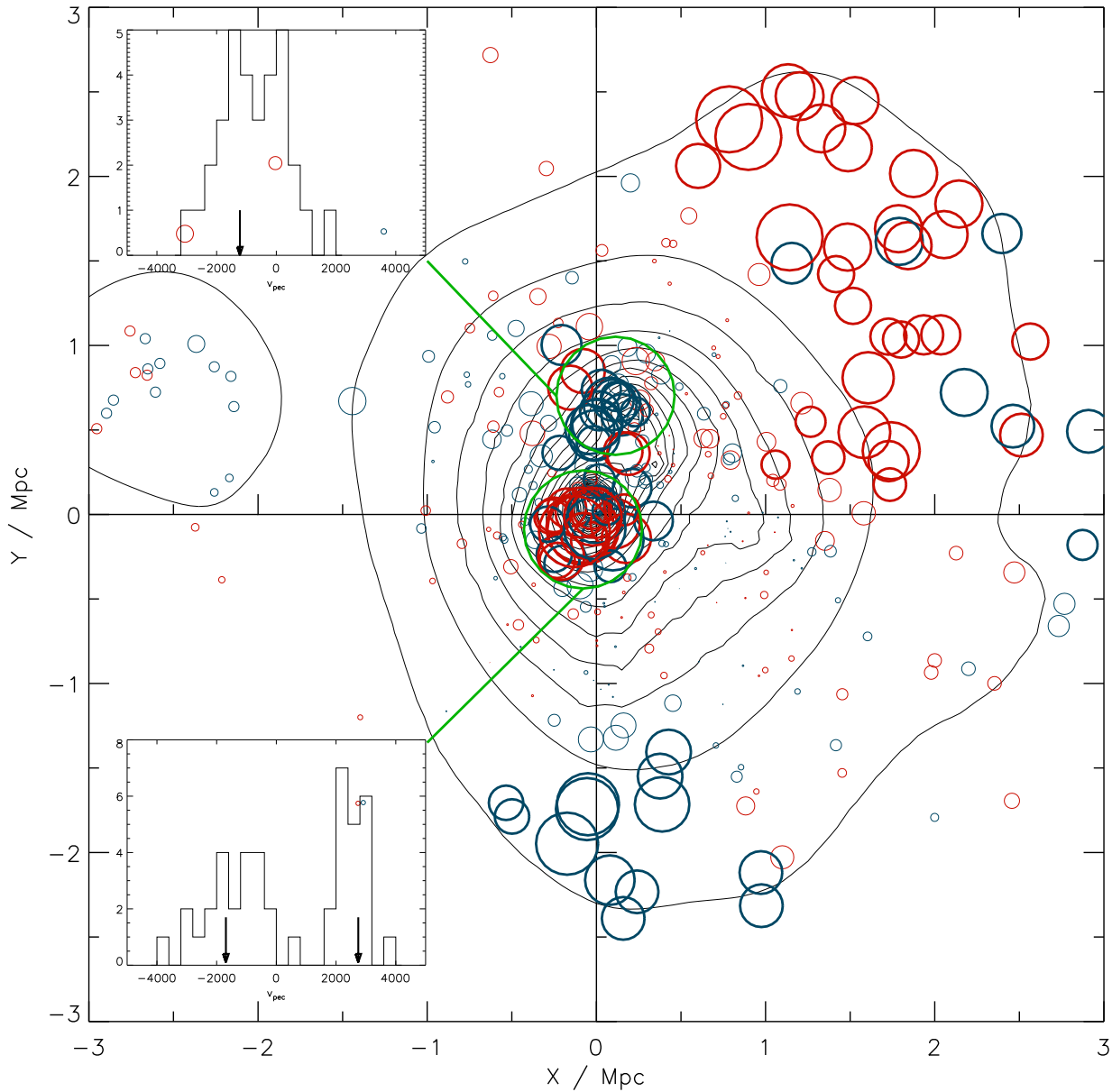}}\\\\\\
\caption{Bubble plot outputs from the $\kappa$ test. The {\it bold} bubbles are 
those deemed to be significant insomuch as they only occur in $1\%$ of 10,000 
realizations. The total $\kappa$ test is significant at the $6.6\sigma$ level 
and the observed value does not occur in all 10,000 realizations. {\it Blue}
and {\it red} bubbles have negative and positive $v_{pec}$, respectively. The 
contours are galaxy density contours generated from the adaptively smoothed 
image in the top left panel of Figure~\ref{galdense} and are linearly spaced by
10 in the interval 5-245 $\rm{gals\,Mpc}^{-2}$. Note the two clusterings of 
significant bubbles are coincident with overdensities in the projected galaxy 
density. The inset panels show the peculiar velocity histograms for the regions
indicated by the green circles, which have radii of 350\,kpc. The arrows mark 
the velocities of the bright (\rf$<17.8$) member galaxies within each green
region.}
\label{bubbleplot}
\end{figure*}

Figure~\ref{bubbleplot} shows the results of the $\kappa$-test in the
form of a ``bubble plot.'' At the position of each member galaxy, a circle
is plotted with radius $\propto -log P_{KS}(D > D_{obs})$. Clusters of large
bubbles reveal regions where the local dynamics are different from the global 
cluster dynamics; we define significantly large bubbles as those with radii that
occur in less than $1\%$ of the 10,000 realizations and these are emboldened in
Figure~\ref{bubbleplot}. The bubbles are also color-coded based on the sign
of the galaxy's peculiar velocity, where red and blue mark positive and negative
$v_{pec}$, respectively. Overplotted are the galaxy surface density contours
taken from the adaptively smoothed image shown in the top left panel of 
Figure~\ref{galdense}. There are two clear dynamical substructures detected in
the central $\sim 1$\,Mpc; one coincident with the cluster center and another 
to the north, coincident with the second projected galaxy overdensity. At larger 
radii, we also detect the two substructures to the northwest and south first
noted by \citet{braglia2007}.

Radial gradients in the cluster velocity dispersion are expected and, in 
particular, the velocity dispersion is expected to decrease at larger radii due 
to a combination of the radial orbits of infalling galaxies and the decreasing
escape velocity of the cluster. This can cause spurious results in 3D tests, 
such as the $\kappa$-test used here, which are not related to dynamical 
substructures \citep{pinkney1996}. For this reason, we re-ran the $\kappa$-test 
considering only those galaxies residing within a central 1.5\,Mpc radius 
region. Again, we find the observed value of $\kappa=200$ does not occur in any
of the 10,000 Monte Carlo realizations and lies $\sim 3.6 \sigma$ from the
mean of the simulated $\kappa$ distribution, which has $\mu(\ln \kappa)=4.58$
and $\sigma(\ln \kappa)=0.20$.

\begin{figure}
\vspace{10pt}
{\includegraphics[angle=0,width=.45\textwidth]{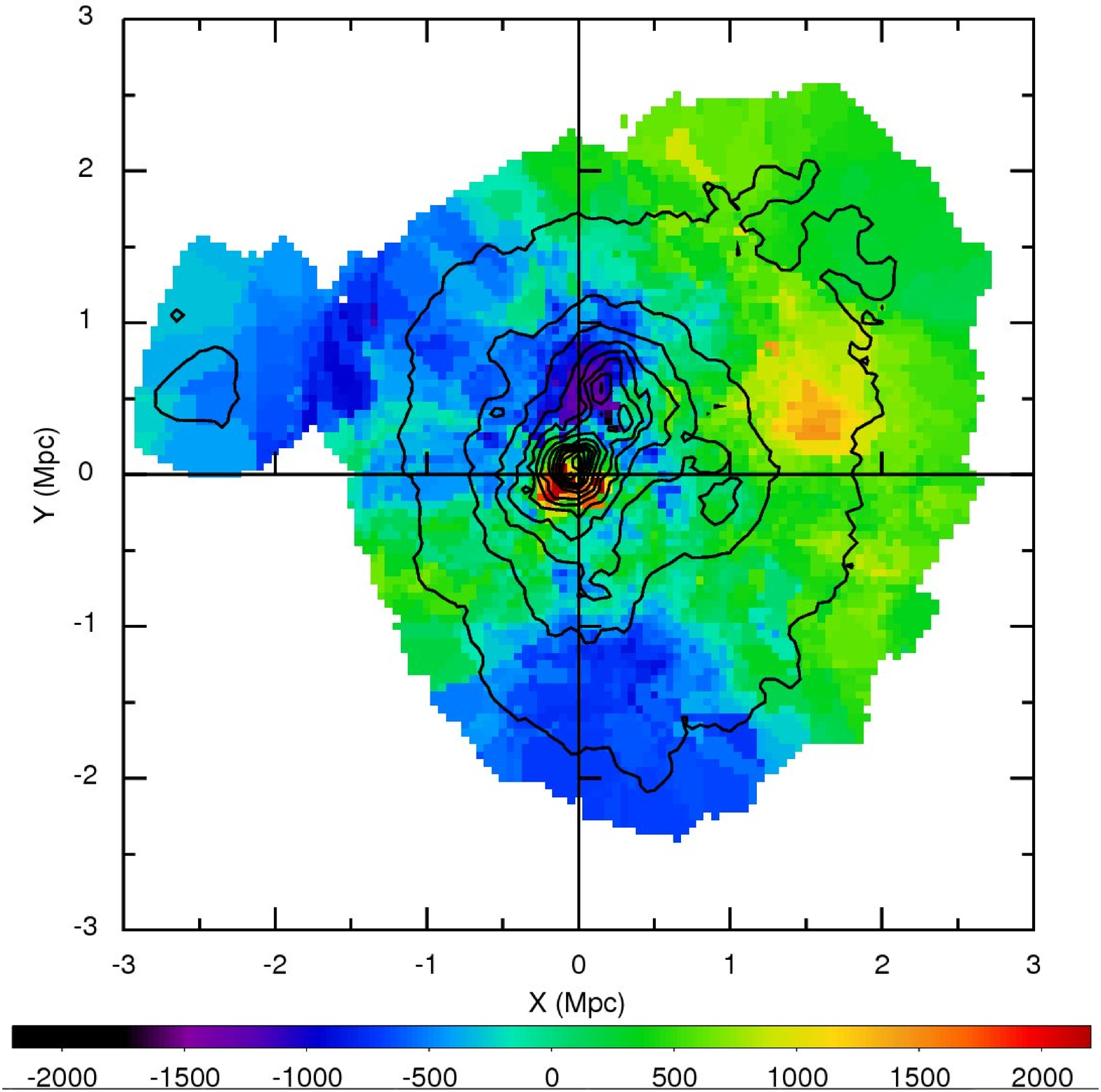}}
\caption{The line-of-sight velocity field determined at each
pixel using the trimmed mean of the velocity distribution of the 15 nearest
galaxies. The black contours show the galaxy density isopleths and the pixel
size is 50\,kpc.}
\label{velfield}
\end{figure}

The $\kappa$-test is an excellent tool for locating dynamically distinct 
substructure, however it provides no insight into the origin of the differences
in the kinematics of the galaxies within the substructures. As a first step 
towards visualizing and characterizing the substructure revealed by the
$\kappa$-test in Figure~\ref{bubbleplot}, we present Figure~\ref{velfield}
which shows the line-of-sight velocity field. The value of each pixel in the
line-of-sight velocity field map is the trimmed mean of the velocity 
distribution for the 15 nearest neighbors in projection. Only pixels where the
radius to the 15th nearest neighbor is $\le 1000$\,kpc are presented in 
Figure~\ref{velfield}. The trimmed mean is measured by first determining the 
median and median absolute deviation (which are insensitive to outliers)
as proxies for the mean 
and standard deviation for the distribution, trimming those points which are 
further than $3\sigma$ from the median and calculating the mean of the 
remaining points. The velocity field is complex and, most significantly, there 
is clear evidence for velocity structure within the central 1\,Mpc region where
a low velocity ($\sim -1300$\kms) structure is seen to the north, while a high 
velocity structure ($\sim 2400$\kms) is seen to the south. These two structures 
coincide with the two substructures revealed by the $\kappa$-test in 
Figure~\ref{bubbleplot}. 

We now attempt to disentangle these substructures and ascertain their dynamics.
To do this, we again utilize the KMM algorithm of \citet{ashman1994} as in
Section~\ref{veldist}, however, here we use both spatial and velocity 
information to partition the cluster into its dynamical 
substructures. Since we are mainly interested in the dynamics of the central 
region, where we have \chan\, data, and the KMM algorithm can be biased by 
outliers, we consider only members within a cluster-centric radius of 1.5\,Mpc 
for the KMM analysis. As input, the KMM algorithm requires the number of 
partitions needed to describe the data, as well as the approximate mixing
proportions and estimates of the means and covariance matrices for each 
partition. One of the major uncertainties when using this method is 
in determining the number of partitions required to adequately describe the 
data. We implement two strategies to overcome this uncertainty. First, we use 
the results from the $\kappa$-test, in combination with the galaxy surface 
density maps, to guide us in selecting the number of likely substructures. 
Second, we utilize the hypothesis testing capabilities of the KMM algorithm
by measuring the LRTS, which gives a measure of the improvement in the fit for a
complex, $g$-partition, model over a simpler, $g_0$-partition model.
In \citet{ashman1994}, the simpler model was a unimodal one (i.e., $g_0=1$) 
where the significance of the
LRTS, expressed as a $P$-value, was estimated by comparison with a $\chi^2$ 
distribution, which is only suitable when comparing two homoscedastic (i.e.,
the same variance), univariate models---clearly not the case here. Furthermore, 
we would like to assess how significantly the fit is improved by the addition of
one partition, i.e., the $g$-partition case over the $g_0=g-1$-partition case, in order 
to decide how 
many partitions to use. Therefore, we employ a parametric bootstrap resampling 
technique to determine the $P$-values as follows. We use KMM to fit the $g_0$- 
and $g$-partition 3D Gaussian models to the data. We then produce 5000 bootstrapped
samples by randomly sampling the best fitting $g_0$-partition 3D Gaussian mixture 
model. To each of these bootstrap samples, we re-fit the $g_0$- and $g$-partition
models, using the means and covariances of the models fitted to the data as
initial input estimates for the KMM algorithm and remeasure the LRTS. The
observed value for the LRTS is then compared to the null distribution of
bootstrapped LRTS values to determine the $P$-value.

\begin{figure*}
\begin{center}
\begin{tabular}{c}
{\includegraphics[angle=90,width=0.95\columnwidth]{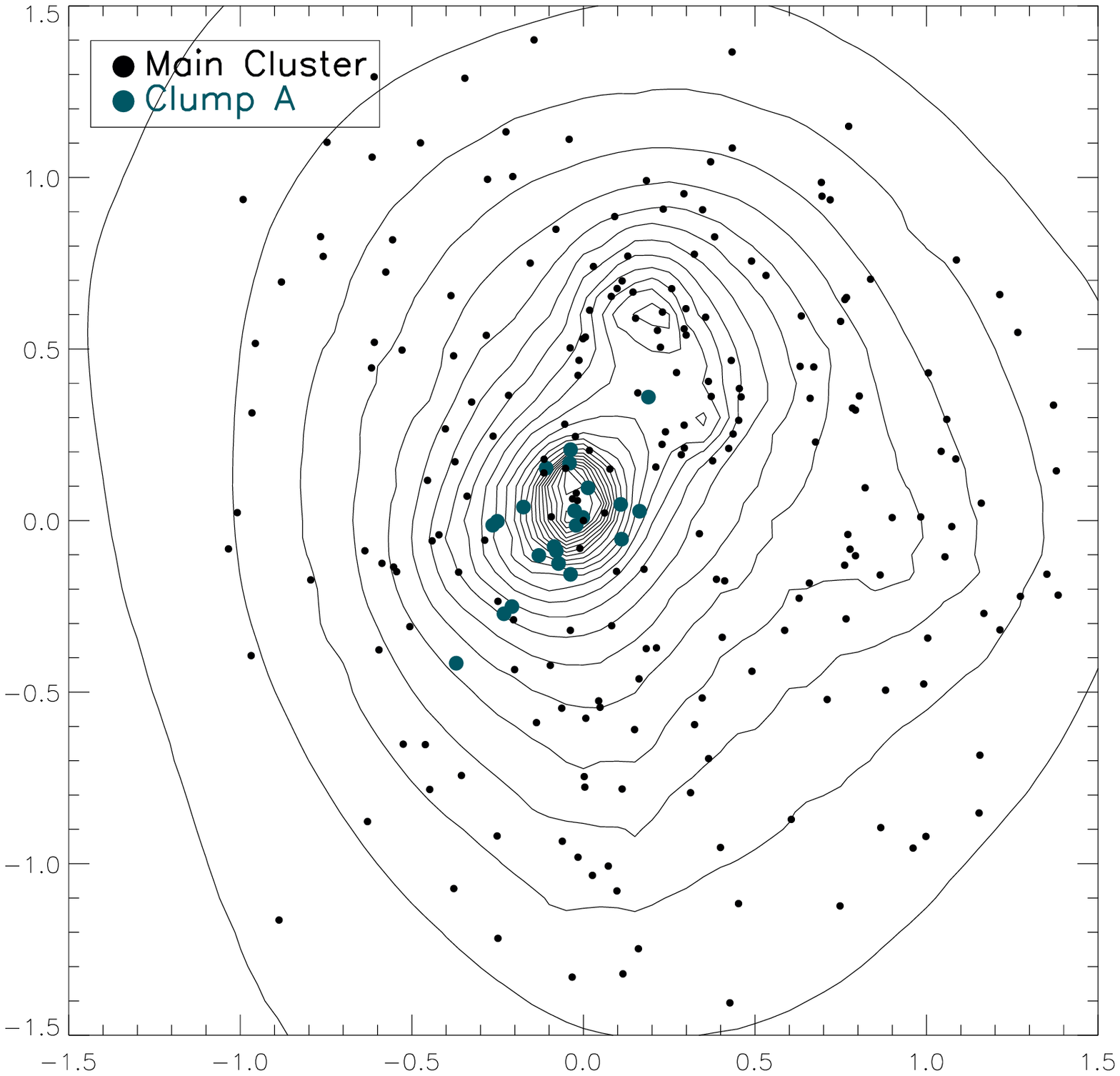}}
{\includegraphics[angle=90,width=0.95\columnwidth]{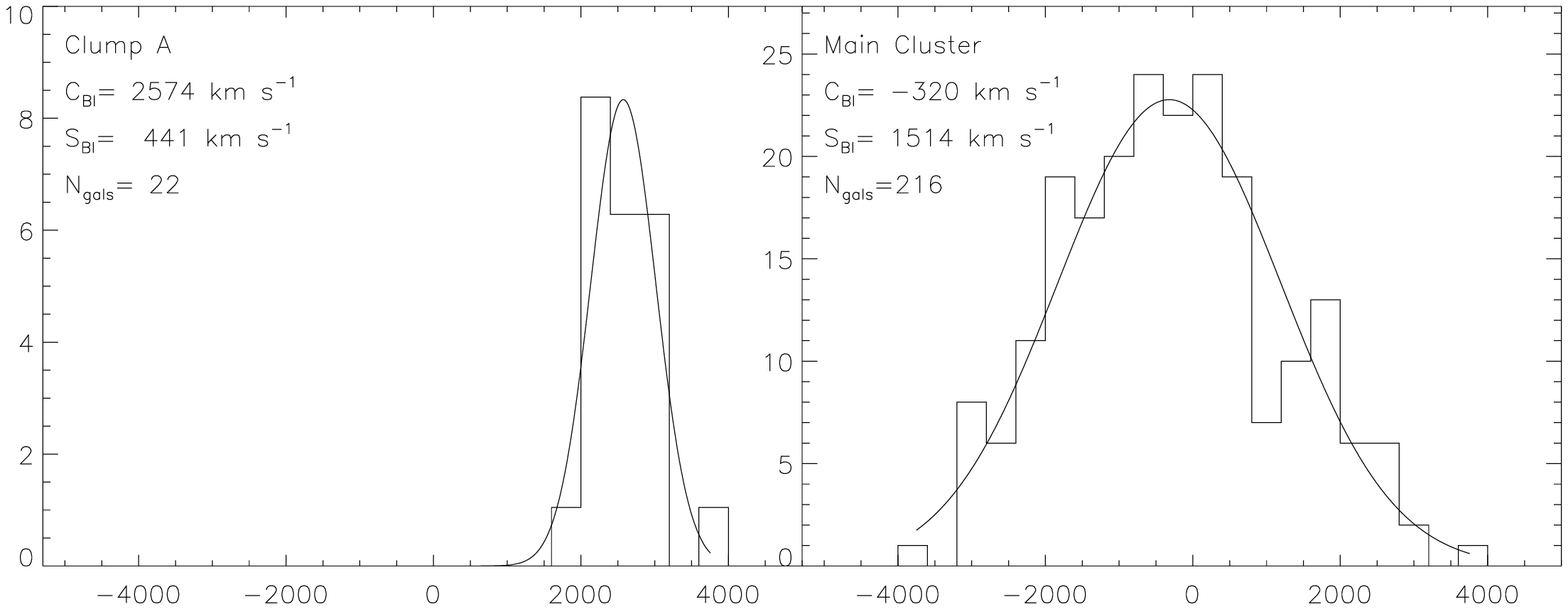}}\\
{\includegraphics[angle=90,width=0.95\columnwidth]{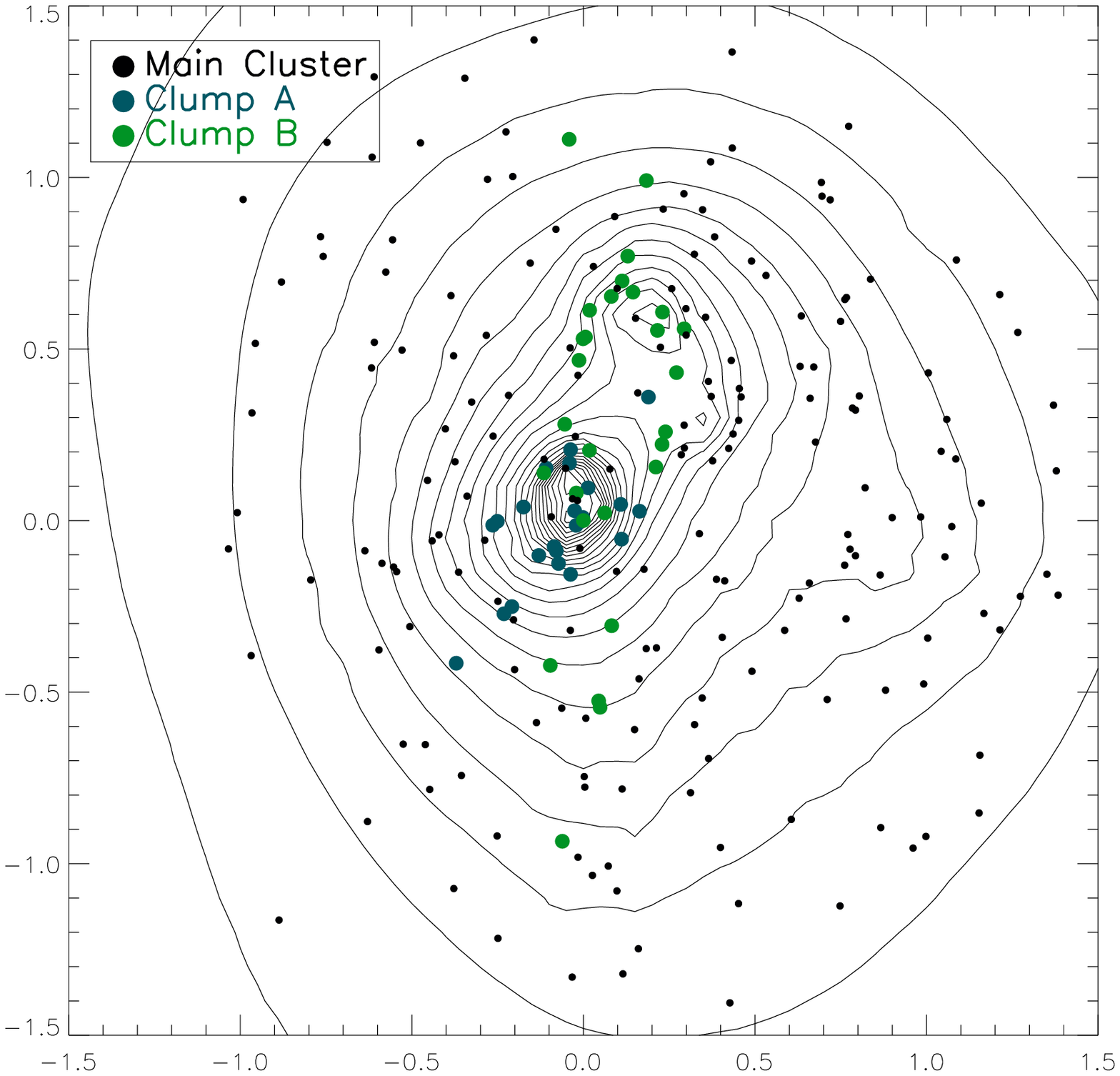}}
{\includegraphics[angle=90,width=0.95\columnwidth]{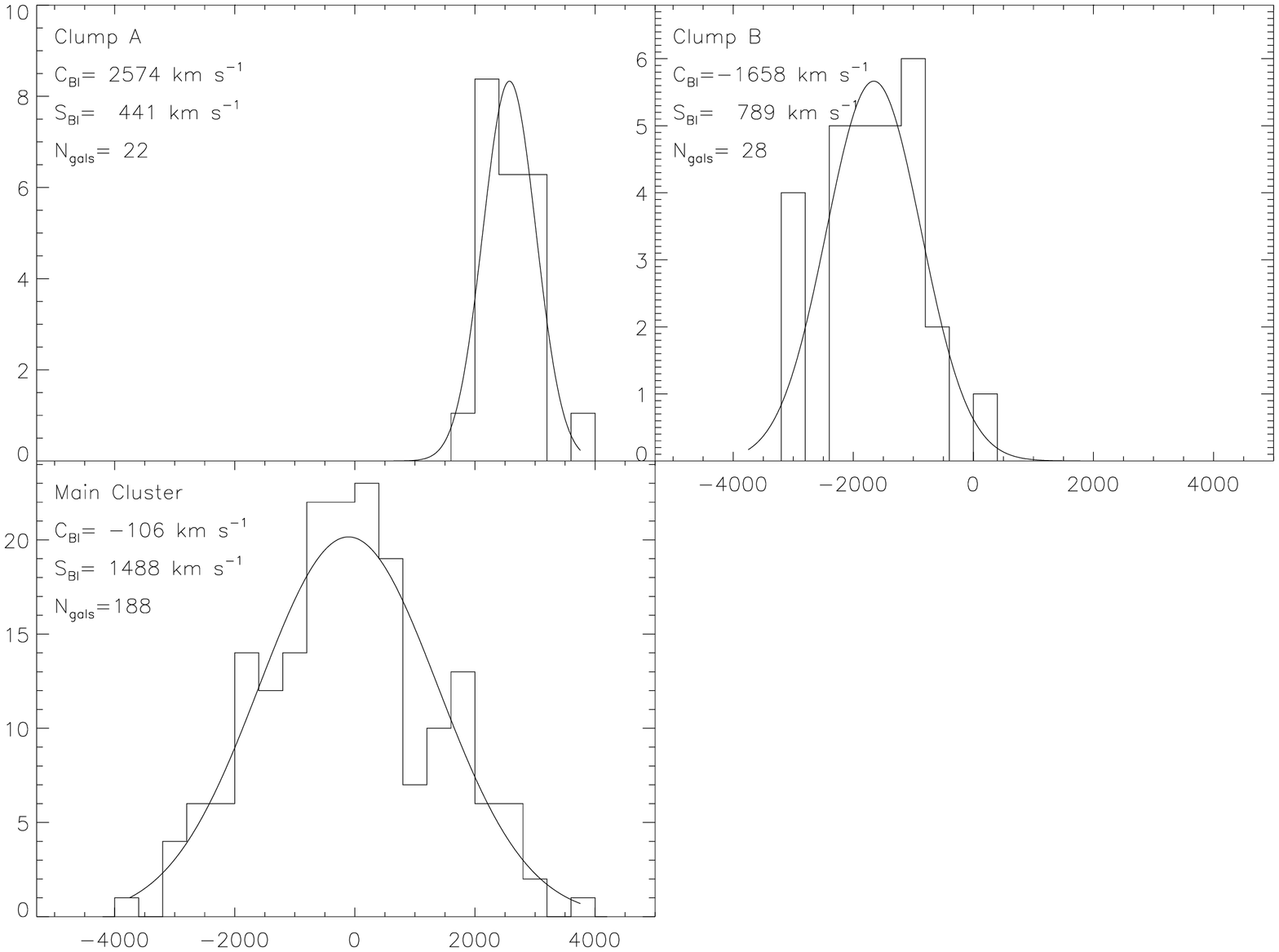}}\\
{\includegraphics[angle=90,width=0.95\columnwidth]{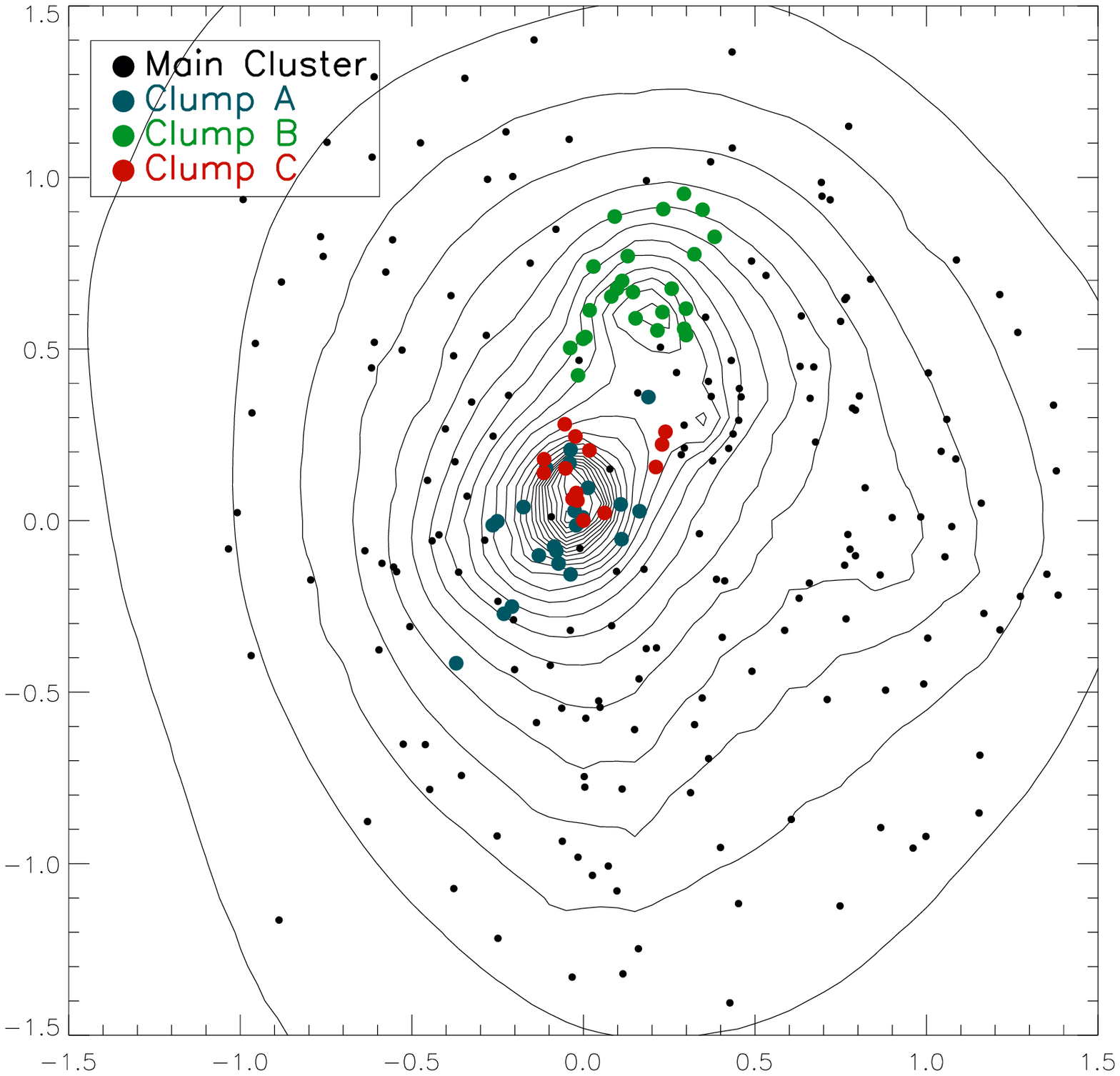}}
{\includegraphics[angle=90,width=0.95\columnwidth]{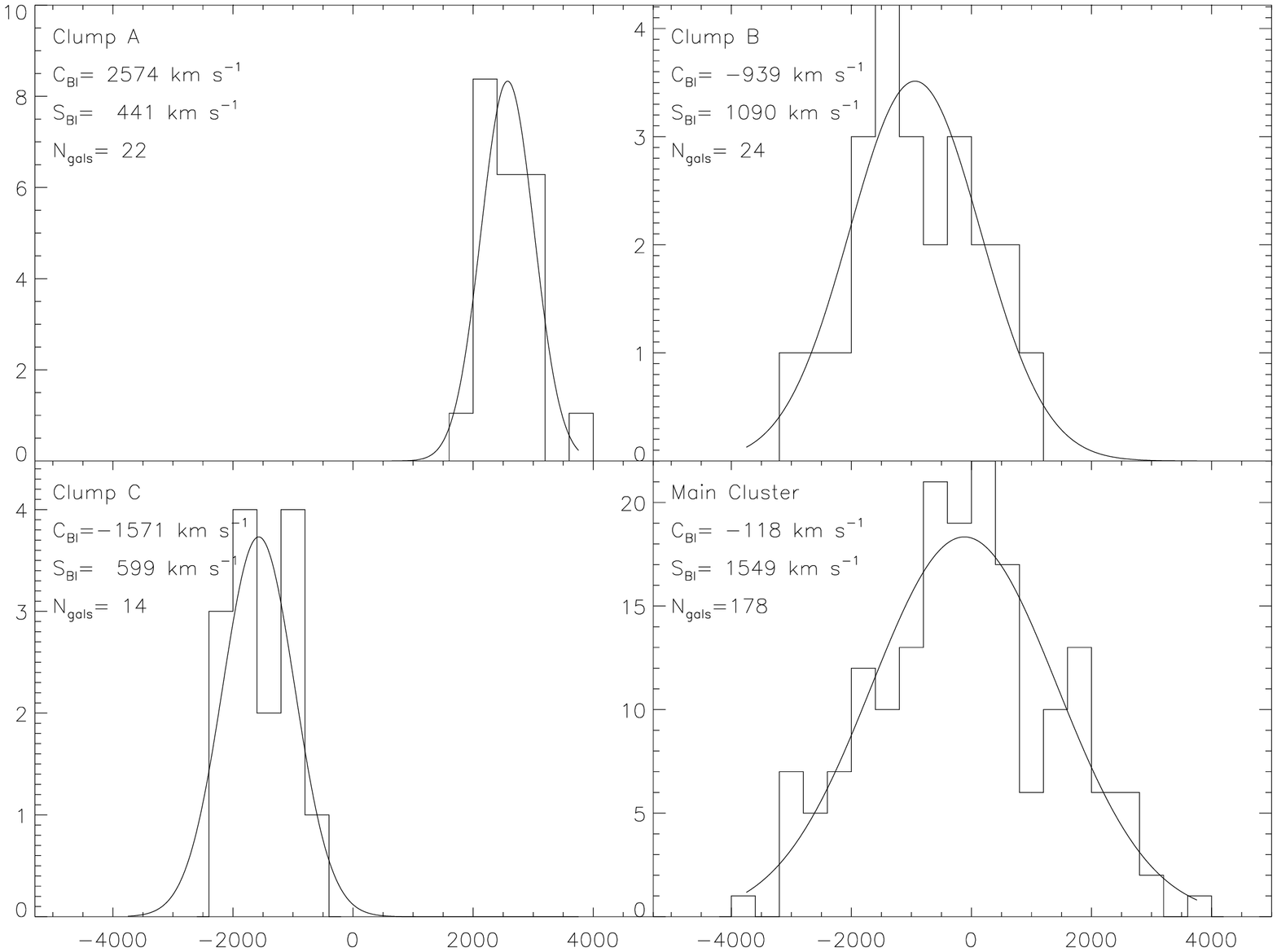}}
\end{tabular}
\caption{These plots show the results of the KMM partitioning of the data using the full spatial plus
velocity information.}
\label{kmmplot}
\end{center}
\end{figure*}

Figure~\ref{bubbleplot} reveals two clear substructures: 
one with $v_{pec} \sim 2300$\,\kms\, close to the cluster center, which we label
Clump~A and one with $v_{pec} \sim -1600$\,\kms\ and $\sim 700$\,kpc north of 
the cluster center, which we label Clump~B. There is also a possible 
substructure approximately spatially coincident with Clump~A, but with 
a $v_{pec}$ similar to that of Clump~B, which we call Clump~C. These partitions 
were used as input for the KMM algorithm. Thus, we fit three
models: a $g=2$-, $g=3$- and $g=4$-partition model and at each stage we determine 
the $P$-value for the null hypothesis $g_0=g-1$ fit, as outlined above.
To determine the initial estimates for input into the KMM algorithm, we
take advantage of the fact that each of these clumps houses a bright 
(\rf$<17.8$) galaxy which has $v_{pec}$ at the peak of the 
velocity distribution (Figure~\ref{bubbleplot}). We assume 
that these bright galaxies lie at the spatial and dynamical centers of each 
partition and crudely define partition membership as any galaxy within 
$v_{pec}=\pm1500$\,\kms\ and radius $\leq 350$\,kpc of the central galaxy. The
means and covariance arrays for these initial partitions serve as the initial
estimates for the KMM algorithm. The results of the $g=2,3\ {\rm and}\ 4$-partition 
KMM fits are presented in Figure~\ref{kmmplot}. The $P-values$ for these fits
are $P_{g=2}<0.0002, P_{g=3}=0.02,\ {\rm and}\ P_{g=4}=0.93$. Thus, the $g=3$ 
partition fit, with Clumps B and C combined, is the statistically preferred one,
despite a spatial separation of $\sim 700$\,kpc between the two bright galaxies
in these clumps. 

\section{The Nature of the Detected Substructures.}\label{SS_nat}

\begin{figure*}
\begin{center}
\begin{tabular}{c}
{\includegraphics[width=0.95\textwidth]{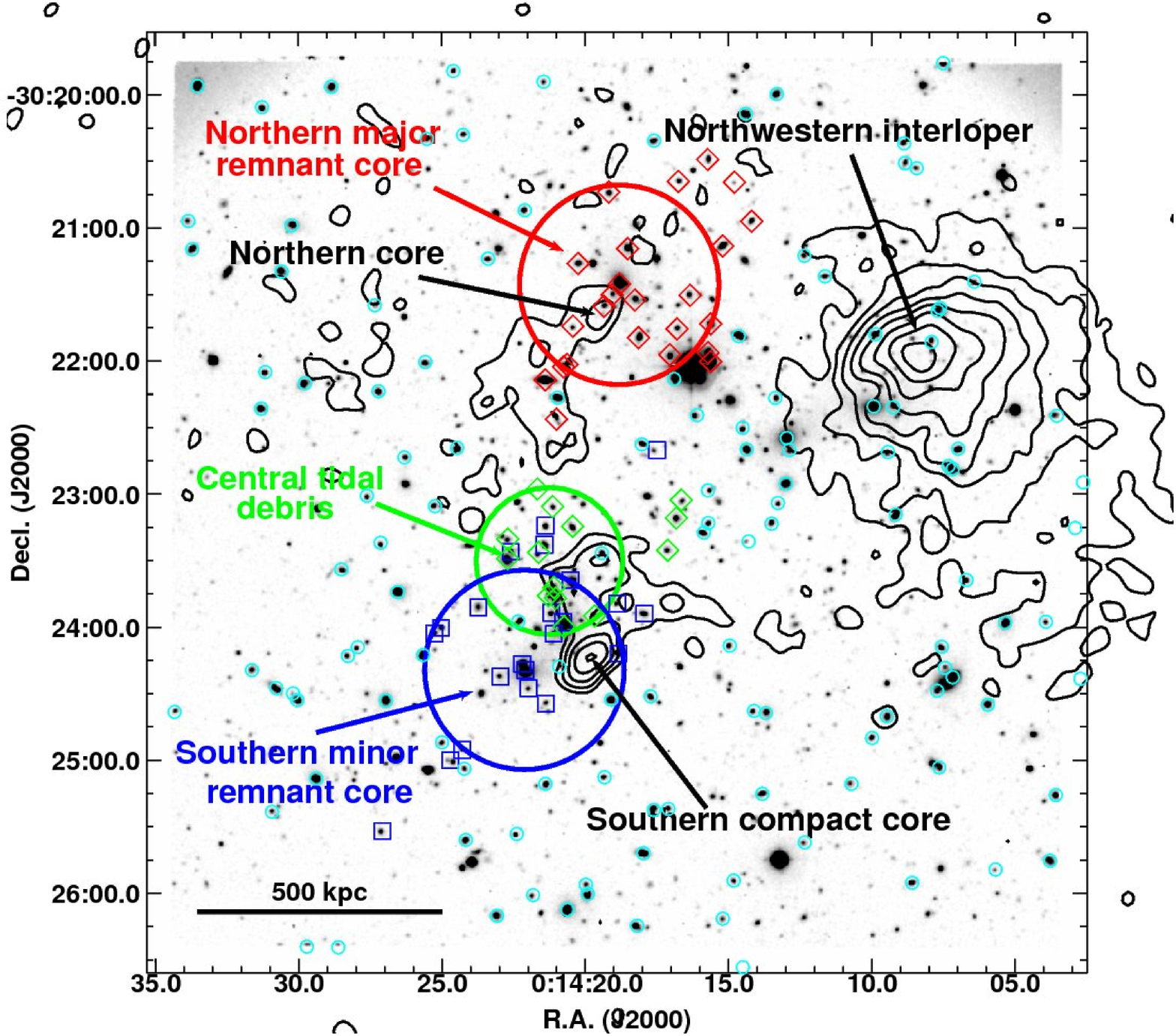}}
\end{tabular}
\caption{This image shows an AAT $R-$band image of Abell~2744. Contours from the X-ray 
residual map (Figure~\ref{residimage}) are shown in black. The regions show cluster 
members and are color/shape coded with red diamonds, green diamonds, blue squares and
cyan circles corresponding to members allocated to KMM partitions labeled in the lower 
left panel of Figure~\ref{kmmplot} as Clump B (Northern major remnant core), Clump C 
(Central tidal debris), Clump A (Southern minor remnant core) and the Main Cluster.}
\label{opt_xray_labels}
\end{center}
\end{figure*}

In this section, we discuss the relationship between the structures
detected in the X-ray and optical analyses and present interpretations
of the nature of these structures. As a visual aid to the discussion
below, we present Figure~\ref{opt_xray_labels} which is an $R-$band
image taken with the AAT. Overlaid are contours from the X-ray
residual map (Figure~\ref{residimage}) showing the relevant X-ray
substructures. The cluster member positions are overplotted and
color coded to match the allocations from the KMM $g=4$ 
partition (Section~\ref{3dsubstructure}). While this partition
was not found to be favored over a $g=3$ partition fit, it serves
the purpose of a visual aid by highlighting
the dynamical structures discussed below in a more coherent manner.

\subsection{The Northern core and northern major remnant core}\label{northcore}

In their analysis of the dynamics of Abell~2744, \citet{boschin2006} detect
the northern major remnant core in both the 2D and 3D analyses but discard the
hypothesis that it is the undisturbed remnant core of the main cluster.
The combination of the new \chan\ X-ray data and comprehensive AAOmega
spectroscopy reveal that the substructure north of the central
regions of Abell~2744 is more significant than previously thought.
The strongest evidence for this assertion comes from Figure~\ref{kmmspatdist}
which shows that when the contribution to the galaxy surface density from the 
high velocity southern component is removed, the most significant peak in the 
galaxy surface density is coincident with the northern substructure. In 
Section~\ref{3dsubstructure}, we show that this galaxy surface density 
enhancement is  detected as a dynamically distinct substructure with 
$v_{pec} \sim -1600$\kms\ and $\sigma_v \sim 800$\kms. Furthermore, the deeper 
\chan\, observations resolve a gas substructure and reveal a tail of gas 
trailing to the south which curves towards the west (Figure~\ref{wvtimage}). The 
abundance map presented in Section~\ref{thermomaps} 
(Figure~\ref{Zmap}) shows that this gas substructure has significantly 
higher metallicity than the global cluster value, while the analysis of detailed
spectra (NC1 in Table~\ref{reg_interest}) show that this gas structure has 
a significantly lower temperature ($kT\sim7$\,keV) than the global cluster
temperature. 

Observations of relaxed clusters show that the central core regions 
are characterized by a steep decline in gas temperature and a corresponding 
increase in metallicity \citep{vikhlinin2005}. Given these observations
and the thermodynamic properties outlined above, we interpret
the northern gas substructure as the remnant of a cool-core which has survived
the merging activity. In addition to this, the large velocity dispersion, the 
significant number of galaxies in this region, along with the 
fact that this structure harbors one of the bright cluster members (with 
$v_{pec} \sim -1200$\kms), we postulate that this northern component has a 
significant mass and is the core of the more massive system involved in the 
central merger in Abell~2744. 

\subsection{The central tidal debris}

This structure, located coincident in projection with the high velocity 
southern substructure (Section~\ref{bullet}), but separated in peculiar 
velocity by $\sim 4000$\kms\ as discussed in Section~\ref{3dsubstructure} 
(Figure~\ref{bubbleplot} inset), has previously been interpreted as 
being the core of the main negative $v_{pec}$ component in the Abell~2744 system 
\citep{kempner2004, boschin2006}. This interpretation is supported by the local 
peak in the surface density of the galaxies and the presence of a luminous
galaxy with $v_{pec}\sim -1600$\kms. However, as discussed in 
Section~\ref{2dsubstructure} (see also Figure~\ref{kmmspatdist}), the peak in 
the projected galaxy density here is artificially enhanced by the high velocity
southern substructure. When this is accounted for, it can be seen that the more 
significant peak in the galaxy surface density of the negative $v_{pec}$ 
component is in fact the northern substructure, which we interpret as the 
remnant core of the main cluster, as discussed in Section~\ref{northcore}. 

Despite not being the core of the negative $v_{pec}$ component, this
structure is a dynamically distinct component with a
$v_{pec} \sim -1600$\kms. In Section~\ref{3dsubstructure}, we found that adding a 
component representing this
structure to the KMM analysis failed to improve the fit significantly
over the simpler 3-partition fit.  Therefore, we tentatively identify
this structure as tidal debris stripped from the main cluster (the
remnant core of which now lies mainly to the north) during the core 
passage phase of the merger with the high velocity southern substructure.

\subsection{The Southern compact core and southern minor remnant core: A Bullet-like remnant viewed from nearer to the
merger axis}\label{bullet}

The southern compact core stands out as a prominent surface brightness
enhancement in the \chan\ images (Figures~\ref{wvtimage} and 
\ref{residimage}). It is conspicuous in the thermodynamic maps presented in 
Section~\ref{thermomaps} as a region with significantly cooler, 
higher metallicity and lower entropy gas (Figures~\ref{scottstmap} and 
\ref{Zmap}). This provides strong evidence for the southern compact core
being the remnant cool core of a merging substructure. In Section~\ref{veldist}
we found that the cluster peculiar velocity distribution was bimodal, containing
a high velocity component. In Section~\ref{2dsubstructure} we explored the 
spatial distribution of this high velocity component, finding that it is 
spatially compact, while the combination of spatial and velocity information in 
Section~\ref{3dsubstructure} confirmed the existence of this structure as a
dynamically distinct, spatially compact entity which we label as the southern minor
remnant core in Figure~\ref{opt_xray_labels}. The proximity of the
southern minor remnant core to the southern compact core indicates that they 
are associated; the  bright galaxy accompanying the southern minor
remnant core is located only 
$\sim 140$\,kpc to the southeast of the X-ray peak due to the southern compact core. The 
peculiar velocity of the southern minor remnant core, obtained with the KMM analyses in 
Sections~\ref{veldist} (for the 1D velocity analysis) and \ref{3dsubstructure} 
(for a full 3D velocity plus spatial information analysis), is 
$v_{pec}=2300-2500$km/s. We find evidence that the southern compact core also has a high 
peculiar velocity by considering the shock front to the southeast 
(Section~\ref{southern_ss}). The Rankine-Hugoniot shock jump conditions can be 
used to derive a Mach number for the shock from the density jump and thus a 
shock velocity \citep[e.g., Eqn 14 from][]{markevitch2007,landau1959}.
For the measured density jump of $1.60_{-0.11}^{+0.17}$, we determine
a Mach number of $M=1.41_{-0.08}^{+0.113}$. We can check this result
by using Eqns 4 and 14 from \citet[][]{markevitch2007} and the temperature 
measurements taken on either side of the front to obtain $M=1.81_{-0.37}^{+0.49}$,
consistent within the errors. The sound speed for a $\sim 8.6$\,keV gas is 
$\sim 1523$\,km/s, therefore the shock velocity is $\sim 2150$km/s, using the 
Mach number derived from the density jump. 

We note that this method does not give the total velocity of the shock, rather 
it gives the shock velocity in the plane of the sky. This is due to the 
following. In order that the shock front be observable as an edge in the surface
brightness, our line of sight must be tangent to some point on the shock front 
surface. Assuming we know the radius of curvature, we measure the density jump 
at the point which is tangent to our line of sight. The Rankine-Hugoniot shock 
jump conditions relate the measured density jump to the Mach number of the 
component of gas velocity travelling perpendicular to the shock surface. Since 
our line of sight is tangent to the point at which we measure this Mach number, 
the measured velocity component must be in the plane of the sky. We can use this
fact, in combination with the line of sight velocities of the northern major 
remnant core and the southern minor remnant core, to constrain the inclination 
of the merger axis along the line of sight and to approximate the shock velocity. 
To do this, we assume that the southern compact and northern cores have the 
same line of sight velocities as their respective galaxy components, i.e., the
southern minor and northern major remnant cores (measured from the 3D KMM partition in 
Section~\ref{3dsubstructure}). The shock leading the southern compact core is 
meeting the unshocked gas associated with the northern core with a line of sight 
velocity $v_{los} \simeq 2574+1658 = 4232$\kms\ which, assuming the unshocked gas has
kT$\sim 8.6$\,keV, gives a line of sight Mach number of 
$M_{los} \simeq 4232/1523=2.78$. The inclination angle of the velocity vector 
to our line of sight is then $\arctan (1.41/2.78) \simeq 27^\circ$ and the total
Mach number is $M_{tot}=3.1$ leading to a shock speed of $\sim 4750$\kms, similar
to that observed in the Bullet cluster \citep{markevitch2002,markevitch2006}. Using the
Mach number derived from the temperature jump gives a similar value for the
inclination angle of $\sim 33^\circ$, the total Mach number, $M_{tot}=3.31$ and 
a shock speed of $5041$\kms. We note that, as in the Bullet cluster 
\citep{springel2007,mastropietro2008}, the 
shock velocity is measured relative to the
surrounding gas which will have high bulk motions due to the merger. These 
measurements also rely critically on the assumption that the velocities of the
gaseous and galactic components is the same. Future X-ray telescopes, such as 
the International X-ray Observatory, will have high enough spectral resolution 
to directly probe the line of sight gas kinematics and will enable us to 
test this assumption.

The measured velocity dispersion of the southern minor remnant core ranges from 
581 km/s, measured from the 1D KMM partition, to 441 km/s, measured from the 
3D KMM partition. This dispersion is likely to be an underestimate of the real 
dispersion due the KMM partitioning which allocates objects which are probably
physically associated with the southern minor remnant core, but in the tails of the 
velocity distribution, to the main cluster. This effect is seen in the 
right-most panel of Figure~\ref{velhisto}, which shows that objects with 
peculiar velocities overlapping with the main structure have been allocated to 
the main structure, thereby truncating the  southern minor remnant core's velocity 
distribution. The act of merging will also affect the measured velocity dispersion, 
since the merger acts to disperse the southern minor remnant core's galaxies 
\citep{pinkney1996}. The galaxies most likely to be stripped are those that are 
least tightly bound to the southern minor remnant core, e.g., those in the 
outskirts, or those in the tails of the velocity distribution. This stripping, 
compounded by the lack of certainty in the allocation of galaxies as members of 
the  southern minor remnant core, mean 
it is extremely difficult to gauge the  southern minor remnant core's initial velocity 
dispersion. In any case, the measured velocity dispersion indicates the 
 southern minor remnant core contains significant mass, as does the measured 
temperature of $\sim 7$\,keV for the southern compact core.

\begin{figure*}
{\includegraphics[angle=-0,width=0.9\textwidth]{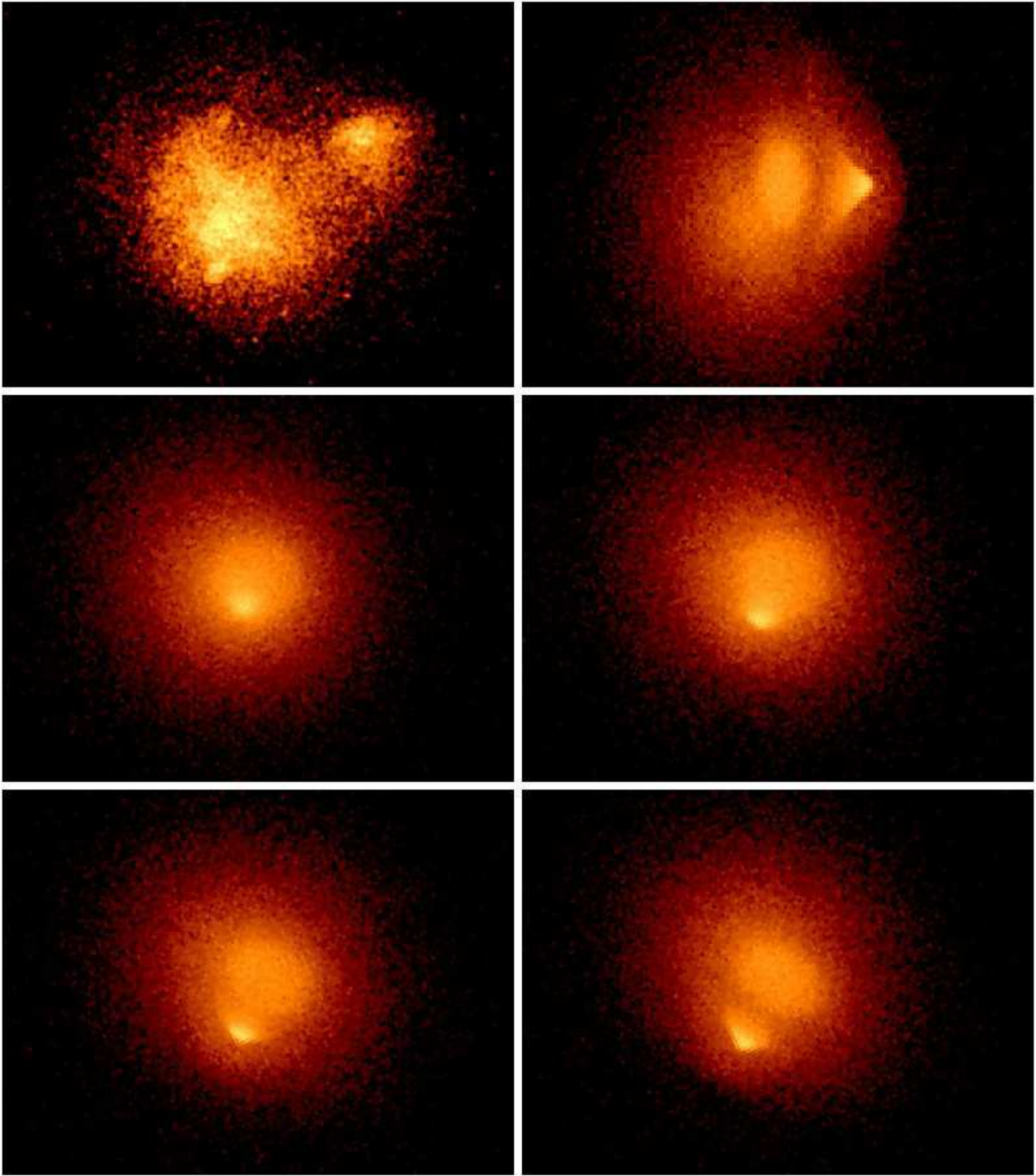}}
\caption{{\it Top left:} A2744 \chan\ image. {\it Top Right:} Simulated 100ks
\chan\ ACIS-I observation of the Bullet cluster gas model (at z=0.308) viewed
in the same projection as the Bullet cluster, i.e., $\theta_{inc}=90$ where 
$\theta_{inc}$ gives the angle of the merger axis with respect to the line of 
sight. {\it Middle left:} As for {\it top right} with $\theta_{inc}=15$. 
{\it Middle Right:} As for {\it top right} with $\theta_{inc}=35$. {\it Bottom 
Left:} As for {\it top right} with $\theta_{inc}=48$. {\it Bottom Right:} As for 
{\it top right} with $\theta_{inc}=60$. The bottom four panels are rotated in the
plane of the sky to best match the orientation of Abell~2744 shown in the top 
left panel.}
\label{bulletproj}
\end{figure*}

Given the above evidence, we interpret the southern compact core as a 
Bullet-like system \citep{markevitch2002} merging along an axis that lies well 
out of the plane of the sky. To support this interpretation, we present  
Figure~\ref{bulletproj}, which shows a model for the X-ray appearance of 
the Bullet cluster \citep[as used in][]{bradac2006} viewed from a range of 
directions. Briefly, the \chan\ X-ray image of the Bullet
cluster is fitted with a density model which consists of a shuttlecock with
concave sides (for the Bullet substructure), power-law profiles across the
cold and shock fronts, a flattened pancake (for the main cluster's emission)
and a beta-model at larger radii. The {\sf XIM} software \citep{heinz2009} is 
used in combination with this density model to simulate 0.6-5 keV \chan\ images 
with 100\,ks exposures and projected so that the merger axis is inclined
to our line of sight with angles of $\theta_{inc}=90,\, 15,\, 35,\, 
48,\, {\rm and}\, 60$. To best match the orientation of the southern 
compact core in Abell~2744, the images shown in the lower 4 panels of 
Figure~\ref{bulletproj} are again rotated in the plane of the sky.

Comparing the simulated images in Figure~\ref{bulletproj} with the
X-ray morphology of the southern compact core, we surmise that
an inclination angle in the range $\theta_{inc}=35 - 60^\circ$ provides 
the best qualitative fit. This is based primarily on the observability
of the shock front (Section~\ref{southern_ss}) and on the 
blunted cone appearance of the southern compact core in the \chan\ 
images (Figure~\ref{wvtimage}). These inclination angles compare well 
with that derived above from the combination of the shock and kinematic 
parameters, particularly when considering uncertainties inherent 
in both techniques.

\subsection{The northwestern interloper}

In their analysis of the northwestern interloper, \citetalias{kempner2004}
presented tentative evidence for a cold front and a bow shock
on the eastern side of the northwestern interloper (their Figure 3.). In our deeper
observations, we find no convincing evidence for the existence of a shock front
at the position and orientation indicated in \citetalias{kempner2004}. However,
both of the putative fronts appear to be part of a larger front 
on the northeastern side of the northwestern interloper, as indicated in 
Figure~\ref{wvtimage}. The analysis in 
Section~\ref{NW_edge} reveals that this is a cold front. 
In Figure~\ref{residimage} we noted that the emission surrounding the 
northwestern interloper extends towards the south. The analyses conducted 
in Sections~\ref{thermomaps} and \ref{roi} indicate that this extension is cooler 
and has lower entropy than the ICM immediately to the east and west. We interpret
the low entropy gas as a trail of gas stripped from the northwestern interloper by
ram pressure.

While we detect no dynamical system associated with 
the northwestern interloper, the pair of bright galaxies lying $\sim 140$\,kpc 
and $\sim 320$\,kpc southeast of the X-ray peak have peculiar velocities of 561\kms\, 
and 663\kms, respectively. The velocity field (Figure~\ref{velfield}) has values 
of $\sim 400$\kms\, in the region of the northwestern interloper, consistent 
with the velocities of the pair of bright galaxies. This hints at a
dynamically distinct system.  However, because the velocities are
close to those of the main cluster and because of the
small number of spectroscopically confirmed cluster members in this region, 
dynamical substructure is more difficult to detect at a significant level 
\citep{pinkney1996}. We note also that if the northwestern interloper has
passed pericenter, as we argue in Section~\ref{NWinfall}, it is plausible
that a large fraction of the galaxies have been tidally stripped. This will
further inhibit the detection of dynamical substructure associated with the 
northwestern interloper.

Using the photometric redshift catalog of \citet{busarello2002} to define cluster 
members, probing $\sim 1.3$ mag fainter than our
spectroscopy, we do detect a local peak in the galaxy surface density
coincident with the northwestern interloper's X-ray peak. We showed that the 
non-detection of a local peak in the 2D galaxy distribution of the 
spectroscopically confirmed cluster members is not due to spectroscopic 
incompleteness, but due to the majority of galaxies associated with the 
overdensity detected using the photometrically defined members having magnitudes 
fainter than our limiting magnitude for the spectroscopy. Thus, spectroscopic 
observations probing to fainter magnitudes are required to confirm the existence
of a substructure associated with the northwestern interloper. 

\section{A merger scenario}\label{merger_scen}

Here we present a merger scenario based on our new observations. Our
interpretation of the data reveal a merger scenario which agrees with
previous authors in the broadest sense, i.e., we are observing a major
merger in the central regions of the cluster and a more minor infall 
of a group to the northwest. However, the details of our merger scenario
differ from those outlined previously in \citetalias{kempner2004} and
\citet{boschin2006}. In the following, we sketch an outline for this 
major-plus-minor-merger scenario, but stress that confirmation of our 
hypotheses require rigorous testing in the form of planned Abell~2744 specific 
merger simulations.

\subsection{The central major merger}

The central region of the cluster has extremely complex X-ray morphology,
temperature and metallicity structure, as well as multiple dynamically distinct
components. These features can be reconciled with an off center, two body
merger after core passage, with the merger axis nearer to our line of
sight than the plane of the sky. Below, we provide evidence for these 
assertions.

During cluster mergers, the strong ram pressure felt by the
gas means that it can become decoupled from the collisionless galaxies which
will lead the gas 
\citep[e.g.,][]{barrena2002,markevitch2002,clowe2006,mahdavi2007,bradac2008}. 
From the observed offset between the galaxies and gas, the X-ray morphologies, 
and also the velocities of the subclusters, we can determine the direction of 
motion of the two subclusters. The northern major remnant core is blueshifted with 
respect to the systemic velocity of the cluster complex ($v_{pec} \simeq -1600$\kms), 
the galaxy substructure is offset from the northern core to the northwest, 
with the bright cluster galaxy associated with the remnant core leading by 
$\sim 100$\,kpc. This evidence indicates that some component of the northern 
subcluster's motion is directed towards the north-northwest, while a large portion 
is also directed towards us along the line of sight. Further evidence for the 
north-northwest motion is revealed by the enhanced X-ray emission to the 
southeast of the northern core (Figures~\ref{wvtimage} and \ref{opt_xray_labels}), which 
we interpret as a trail of stripped gas. The curvature of the gas trail 
implies significant westward acceleration, so that the northward moving 
core must have passed to the east of the core of the other major subcluster, 
which we associate with the southern compact core.

Using a similar argument, we can ascertain the direction of motion of the 
southern subcluster. The brightest galaxy associated with the southern 
minor remnant core is offset some $\sim 140$\,kpc south of due east of
the peak in the X-ray emission associated with the southern compact core
and assuming that this bright galaxy was
once at the center of the subcluster and coincident with the peak in
the X-ray emission, this offset indicates some 
fraction of the southern subcluster's motion is directed towards the 
south-east. Further evidence for motion towards the southeast comes from 
the \chan\ image in the form of the edge to the southeast 
(Figure~\ref{wvtimage}), which is part of the shock front 
(Section~\ref{bullet}), and the morphology of the southern compact core, 
which exhibits a blunted cone structure where the compact core is the head
of the blunted cone, and the tail of stripped gas to the northwest forms 
a wedge-like structure. The mean peculiar velocity of the 
southern minor remnant core galaxies ($\sim 2300-2500$\kms) indicates that a large 
fraction of the southern substructure's motion is directed away from us along 
our line of sight. However, the edge and blunted cone morphology of
the southern compact core show that the direction of motion cannot be entirely
directed along our line-of-sight.

These arguments require that the northern and southern subclusters
have passed one another and now are moving apart. The gaseous
envelopes of the two subclusters have largely been stopped and
stripped from their former hosts by shocks, separating them from the
remnant cool cores, which can survive the off-center merger due to
their higher densities and pressures. In our merger scenario, the 
central peak of emission is the gas which has been stripped from 
the outer parts of the two subclusters during the core passage phase 
of the merger. This interpretation
is supported by high temperatures in the central gas, consistent with
shock heating and adiabatic compression during the core passage.
Patches of low metallicity gas suggest that at least some of this gas
originates from outside the more highly enriched cores \citep{vikhlinin2005}, 
as described above.

According to the KMM partitioning in Section~\ref{3dsubstructure}, a majority 
(188 out of 238) of the galaxies are allocated to a partition with 
$\mu = -105$\kms\ and $\sigma =1438$\kms. This is also clearly seen in 
Figure~\ref{allhisto} where within a cluster-centric radius of 500\,kpc there 
is a clear segregation in velocity with essentially no galaxies with $v_{pec}=0$\kms, 
whereas at $R> 500$\,kpc the velocity distribution appears to be unimodal
and centered around $v_{pec}=0$\kms. Presumably the galaxies at larger radii 
were once part of the less-tightly-bound outskirts of the southern and northern
substructures. The outskirts of the two merging clusters decoupled from
the tightly bound cores during pericenter and are in the process of mixing
to form the outskirts of the new cluster. 

In our interpretation, the northern subcluster was the more massive of
the two merging systems and the southern subcluster is the remnant of the less
massive one, consistent with the higher velocity dispersion of the
northern major remnant core.  However, the southern compact core is brighter
now. This suggests that the smaller subcluster hosted a more robust
cool core, i.e. a greater mass of low entropy gas, prior to the merger.
The higher entropy northern core of the larger subcluster would then be more
prone to disruption during the merger. Indeed, the clumpy tail of 
gas trailing the northern core indicates that a significant fraction of 
its gas has been removed and it is in the process of being destroyed 
by the merger.

\subsection{The infall of the northwestern interloper}\label{NWinfall}

The deeper Chandra observations presented here fail to confirm
the proposal by \citetalias{kempner2004} that the northwestern interloper 
is traveling eastward on its first passage through the cluster. 
\citetalias{kempner2004} used the positions of two bright cluster galaxies 
$\sim 140$kpc and $\sim 320$kpc southeast of the northwestern interloper, 
along with tentative detections of cold and shock fronts on the eastern side 
of the interloper, to support this scenario. We find little evidence for 
the shock proposed to exist by \citetalias{kempner2004}, while their proposed 
cold front is part of a more extended cold front. Furthermore, it is difficult 
to explain the large separation of gas and galaxies prior to a core passage 
in a subcluster which appears to have both significant mass (indicated by its 
temperature kT$\sim 5$\,keV) and ICM density (indicated by its X-ray brightness).

Three main features in the deeper X-ray data provide support for an alternative 
interpretation. They are 1) the cold front to the north-east, 2) the extended region 
of enhanced X-ray surface brightness to the south and 3) the swirl of 
low entropy gas seen in Figure~\ref{scottstmap}, which is
associated with the southern extension and is curved in a southeasterly direction 
towards the cluster center. The cold front to the north-east, along with the 
extension of the emission towards the south, indicate that the northwestern interloper
is currently traveling towards the north-northeast. The curvature of the swirl of low 
entropy gas indicates significant angular momentum. Using these features to
guide our interpretation, we suggest that the northwestern interloper fell into 
the main cluster from the south or southeast, initially traveling roughly north or 
northwest and passing the main cluster core off-center to the southwest. The
gravity of the main cluster has subsequently deflected it onto its
current path towards the north-northeast \citep[see, e.g., the evolution of 
the 3:1 and 10:1 mass ratio, offcenter mergers in][]{poole2006}. The fact these 
features are observed at all, along with the non-detection of any significant 
dynamical substructure in this region, indicates the merger of the northwestern 
interloper is taking place fairly close to the plane of the sky. We note,
however, that the non-detection of dynamical activity here may be due to the 
limited depth and completeness of our sample. 

One issue with the above interpretation for the northwestern interloper
is the position of the two bright galaxies which are offset to the southeast 
from the peak in the X-ray emission. Assuming these bright galaxies trace the dominant
mass component (i.e., the collisionless dark matter core) of the northwestern
subcluster, we would expect to see the classic Bullet-like scenario
where the collisionless dark matter and galaxies lead the collisional gas
component and thus to see the bright galaxies offset to the
northeast of the gas clump. However, we note that during the later stages of
a merger, the simulations of \citet{mathis2005} showed that the cool, dense
cores of subclusters can lead the dark matter cores. Indeed, this phenomenon 
has been observed in the clusters Abell~168 \citep{hallman2004} and 
Abell~754 \citep{markevitch2003}. The physical mechanism producing this 
offset is the rapid changes in ram pressure felt by the subcluster gas as it 
traverses the main cluster; during closest approach, the ram pressure peaks
and the gas and dark matter separate, with the dark matter leading. After the 
core passage phase, the ram pressure drops significantly due to the decrease in
the surrounding ICM density and the cool core begins to fall back towards its 
parent dark matter core, eventually overshooting it in a 
``ram pressure slingshot'' as explained in \citet{markevitch2007}. This effect 
is also simulated in \citet{ascasibar2006} and their Figure~12 shows 
qualitative agreement with what is proposed above and with the temperature 
morphology of the northwestern substructure. High resolution weak lensing
maps are required to determine the location of the local peak in the
mass density corresponding to the northwestern interloper and to test the
scenario outlined above.

While the northwestern subcluster is due to a more minor merger than
the central merger, there should still be observable effects due to
the core passage, particularly in the ICM. Although the central major merger
complicates this matter and may erase any effects due to the northwestern 
subcluster, we tentatively point out the bridge of emission in
Figure~\ref{wvtimage} along with the slight asymmetry of the central cluster 
emission in the direction of the northwestern interloper, as evidence of the 
previous interaction with the central ICM. We also suggest that the 
hot regions lying between the main cluster and the northwestern
interloper (Figure~\ref{scottstmap}) were produced by shocks and/or 
compression of the ICM during the subclusters passage.

\subsection{Comparison to previous interpretations}

The interpretations outlined above for the different structures revealed by our 
new observations give cause to re-visit the hypotheses outlined in
\citet{boschin2006} and \citetalias{kempner2004} for the merging history of 
Abell~2744. Considering the central regions, \citetalias{kempner2004} propose
a merger scenario where Abell~2744 is seen just prior to a core passage and the 
merger is occurring along a north-south direction with the majority of the 
motion directed along our line-of-sight. \citet{boschin2006}, using a slightly 
larger sample of spectroscopically confirmed cluster members, agreed with 
\citetalias{kempner2004} that the central merger is occurring along a 
north-south direction with the majority of the motion directed along our 
line-of-sight, but suggested that the merger is in a more advanced 
post-core-passage phase. Both \citetalias{kempner2004} and \citet{boschin2006} 
suggest that the northwestern subcluster is infalling onto the main cluster from 
the west and can be treated as separate from the major merger occurring in the 
central regions.

With regard to the southern subcluster, our interpretation of the new 
observations is mostly in accordance with the conclusion of \citet{boschin2006}
that it is viewed after a core passage, has a large fraction of 
its motion directed along the line-of-sight and some component to the south. We
slightly modify this scenario based on the detection of the shock front and the
blunted cone X-ray morphology, which indicate a more southeasterly direction
of motion. The detection of the shock front also indicates that the component of
the velocity perpendicular to our line-of-sight must be larger than previously
suggested. Indeed, our observations of the shock front, combined with the 
dynamical analyses, allow us to constrain the direction of motion of the
southern subcluster to be inclined at around $30^\circ$ to the line of sight.

Our scenario differs most from that of \citet{boschin2006} in
the importance of the northern 
subcluster. \citet{boschin2006} interpret this structure as an additional 
feature in their scenario due to the lack of a counterpart in the X-ray images. 
Our deeper \chan\ data reveals the northern core as a cool, high metallicity 
structure which is trailing the dynamically distinct northern major remnant
core galaxies. Furthermore, the larger sample of spectroscopically confirmed
cluster members provide strong evidence that, when separated by velocity, the 
most significant projected galaxy overdensity is the northern major remnant core.
Taking these new results in concert provides strong evidence in support of our 
view that the northern subcluster is the surviving core of the main component 
involved in the central merger. Thus, while we agree with \citet{boschin2006} 
that we are observing a post-core passage merger in the central regions, we 
suggest a somewhat later phase, given the large ($\sim800$\,kpc) projected 
separation of the main northern subcluster from the high velocity southern 
subcluster.

Turning to the interloping northwestern subcluster, the deeper \chan\ data have muddied 
the waters somewhat in terms of a simple interpretation and indicate that 
the direction of motion of this subcluster is currently to the 
north/northeast---perpendicular to the direction of motion suggested by both
\citetalias{kempner2004} and \citet{boschin2006}. Both \citet{boschin2006} and
\citet{braglia2009} detect an overdensity of red galaxies associated with the 
bright cluster galaxy lying $\sim 140$\,kpc to the southeast of the X-ray peak.
We propose that this offset has occurred due to a ram pressure slingshot which 
caused the gas to ``overshoot'' the galaxies after pericentric passage. This 
interpretation provides a better fit to the X-ray data and we note that the
galaxy overdensities found here and in the \citet{boschin2006} and 
\citet{braglia2009} studies rely on photometric cluster member allocation, rather
than the more robust spectroscopic identification of cluster members, which is 
less prone to projection effects. A more rigorous investigation involving
deeper spectroscopy and detailed simulations are required to test the different
infall scenarios for this substructure.

\section{Summary and Conclusions}\label{summary}

We have presented an analysis of the merging cluster Abell~2744 based on new 
\chan\ X-ray data and AAOmega optical spectroscopy. The deeper \chan\
data reveal a plethora of substructure including:

\begin{enumerate}
\item A significant structure to the north with a trail of X-ray
emission to the south and curving westward. This northern core
harbors cooler gas with higher metallicity than its surroundings, indicating it 
is a remnant cool core.
\item The Bullet-like southern compact core which is surrounded by
hot, shocked gas. There is an edge to the southeast of the southern compact core, 
which has the characteristics of a shock front with a Mach number $M=1.41_{-0.08}^{+0.13}$.
\item A global peak in the X-ray emission with no corresponding galaxy
overdensity, which is interpreted as the stripped atmospheres of the merging 
subclusters.
\item The northwestern interloper has an edge to the northeast and a tail of
emission extending towards the south. Analysis of the spectra of these features 
indicate that the edge is a cold front and that the tail of gas harbors cooler,
lower entropy gas when compared to the surrounding gas. 
\end{enumerate}

From the new AAOmega optical spectroscopy, we have found:
\begin{enumerate}
\item From 343 spectroscopically confirmed members within a 3\,Mpc radius, the 
redshift and velocity dispersion of Abell~2744 are, respectively, 
$z_{clus}=0.3064\pm 0.0004$ and $\sigma=1497\pm 47$\kms. 
\item The velocity distribution is significantly skewed and the KMM analysis 
reveals the velocity distribution in the central region is characterized by two 
distinct structures in velocity space, separated by $\sim 3000$\kms.
\item Examination of the spatial distribution of the galaxies within the two
partitions found in the KMM analysis of the velocity distribution reveal that
for the negative peculiar velocity component the more significant galaxy
concentration lies further to the north, approximately coincident with the 
cool, high metallicity northern core revealed by the \chan\ data.
\item Applying the KMM algorithm to the combined velocity and spatial 
information, we find the statistically favored fit is one where the sample is
partitioned into three. Two of the partitions correspond to the
remnant cores of the two merging clusters in the central regions, while the third 
likely contains the galaxies which were in the outskirts of the two subclusters
and are currently in the process of mixing to form the new cluster. The third 
substructure also contains those galaxies which belong to the northwestern interloper.
\item The northwestern interloper detected in the \chan\ image is not detected as
a dynamical substructure, nor as a substructure in local galaxy surface density 
when considering only spectroscopically confirmed members. An increase in local 
galaxy surface density is detected when considering cluster members defined 
using photometric redshifts, although only when including galaxies down to 
magnitude R=22.3.
\end{enumerate}

In combination, the observations presented here have been used to outline a 
skeleton hypothesis for the merger history of Abell~2744. In this scenario, we
have identified two significant subclusters in the central region which have
undergone a violent core passage and are now moving away from each other along 
a roughly north-south axis, with a large line-of-sight component. For the
northwestern subcluster, we propose a scenario where the structure is 
traveling to the north/northeast after pericenter.
Confirmation of this merger hypothesis requires detailed simulations and deeper
optical spectroscopy in the central regions in order to probe the dynamics of 
the northwestern subcluster.

The evidence for a post-core passage merger phase reported here, along with the 
significant enhancement of blue starburst and poststarburst galaxies residing 
within Abell~2744 \citep{couch1998}, provides further evidence that major 
cluster mergers can act as catalysts for rapid, environmentally driven  
evolution in the star forming properties of cluster galaxies and that this 
evolution occurs when the cluster galaxy environment is violently rearranged 
during the core passage phase of a major merger. Our large spectroscopic sample,
in combination with deep radio data, can now be used to define a large sample
of actively transforming galaxies which will further test this hypothesis of
merger-induced galaxy transformation.

\acknowledgments

We thank Gregory Poole for useful discussions and sanity checks. We are 
grateful to Will Saunders and the staff at the 
Anglo-Australian Observatory for their support during AAT observations. We thank
Lee Spitler for his assistance in preparing the AAT image and Emily Wisnioski for
supplying code for the PCA sky subtraction.
We acknowledge the financial support of 
the Australian Research Council (via its Discovery Project Scheme) throughout 
the course of this work. PEJN was partly supported by NASA grant NAS8-03060. 
This work was partly supported through Chandra GO7-8127X.
This research has made use of software provided by the Chandra X-ray Center 
(CXC) in the application packages CIAO, ChIPS, and Sherpa and also of data 
obtained from the Chandra archive at the NASA Chandra X-ray center 
(http://cxc.harvard.edu/cda/). This research has made use of the NASA/IPAC 
Extragalactic Database (NED) which is operated by the Jet Propulsion Laboratory,
California Institute of Technology, under contract with the National Aeronautics
and Space Administration. 

{\it Facilities:} \facility{CXO (ACIS)}, \facility{AAT (AAOmega)}

\end{document}